\documentclass{article} % For LaTeX2e
\usepackage{iclr2026_conference,times}

\usepackage{hyperref}
\usepackage{url}

\usepackage[T1]{fontenc}
\usepackage[utf8]{inputenc}
\usepackage[warn]{textcomp}

\usepackage{siunitx}

\usepackage{enumitem} %< control spacing in itemize/enumerate/...
\usepackage{overpic} %< add raw math symbols to figures
\usepackage{color}
\usepackage{xspace}
\usepackage{xfrac}
\usepackage{colortbl}
\usepackage{bm}

\usepackage{rotating}

\usepackage{caption}
\usepackage{subcaption}
\usepackage{wrapfig}
\usepackage{multirow}

\usepackage{graphicx}
\usepackage{amsmath}
\usepackage{amssymb}
\usepackage{booktabs}
\usepackage{mathtools}

\usepackage{algorithm}
\usepackage{algpseudocode}

\usepackage{epigraph}

\usepackage{ragged2e}

\usepackage{tikz}
\usepackage{pgfplots}
\pgfplotsset{compat=1.14}

\usetikzlibrary{shapes}
\usetikzlibrary{calc}
\usetikzlibrary{backgrounds}
\usetikzlibrary{positioning}
\usetikzlibrary{arrows}
\usetikzlibrary{arrows.meta}
\usetikzlibrary{decorations.text}    
\usetikzlibrary{fit}
\usetikzlibrary{spy}
\usetikzlibrary{3d}
\usepgfplotslibrary{groupplots}
\usepgfplotslibrary{colorbrewer}
\usepgfplotslibrary{statistics}

\usepackage{tikz-3dplot}

\usepackage[skins]{tcolorbox}

\definecolor{turquoise}{cmyk}{0.65,0,0.1,0.3}
\definecolor{purple}{rgb}{0.65,0,0.65}
\definecolor{dark_green}{rgb}{0, 0.5, 0}
\definecolor{orange}{rgb}{0.8, 0.6, 0.2}
\definecolor{red}{rgb}{0.8, 0.2, 0.2}
\definecolor{darkred}{rgb}{0.6, 0.1, 0.05}
\definecolor{blueish}{rgb}{0.0, 0.3, .6}
\definecolor{light_gray}{rgb}{0.7, 0.7, .7}
\definecolor{pink}{rgb}{1, 0, 1}
\definecolor{greyblue}{rgb}{0.25, 0.25, 1}

\definecolor{bestcol}{RGB}{254,196,79}
\newcommand{\best}[1]{\cellcolor{bestcol} \textbf{#1}}
\definecolor{secondbestcol}{RGB}{255,247,188}
\newcommand{\secondbest}[1]{\cellcolor{secondbestcol} #1}

\newcommand{\expnumber}[2]{{#1}e{#2}}

\newcommand{\fig}[1]{Fig.~\ref{fig:#1}}

\newcommand{\Table}[1]{Table~\ref{tab:#1}}

\makeatletter
\DeclareRobustCommand\onedot{\futurelet\@let@token\@onedot}
\def\@onedot{\ifx\@let@token.\else.\null\fi\xspace}

\def\eg{\emph{e.g}\onedot} 
\def\ie{\emph{i.e}\onedot}

\def\etal{\emph{et al}\onedot}

\makeatother

\newcommand{\inlinesection}[1]{\vspace{0.05cm} \noindent {\bf #1}}
\newcommand{\titlecaption}[2]{\caption{\textbf{#1.}\xspace#2}}

\newcommand{\titlecaptionof}[3]{\captionof{#1}{\textbf{#2.}\xspace#3}}

\usepackage{pifont}
\usepackage{blindtext}

\definecolor{mpurple}{RGB}{106,27,154}
\definecolor{mpurplelight}{RGB}{206,147,216}

\definecolor{mblue}{RGB}{40,53,147}
\definecolor{mbluelight}{RGB}{159,168,218}

\definecolor{mteal}{RGB}{0,105,92}
\definecolor{mteallight}{RGB}{128,203,196}

\definecolor{morangelight}{RGB}{255,171,145}

\definecolor{mgrayblue}{RGB}{55,71,79}
\definecolor{mgraybluelight}{RGB}{176,190,197}

\definecolor{mamber}{RGB}{255,143,0}
\definecolor{mamberlight}{RGB}{255,224,130}

\definecolor{mdeeporange}{RGB}{216,67,21}
\definecolor{morange}{RGB}{245,124,0}
\definecolor{myellow}{RGB}{253,216,53}

\definecolor{mgreen}{RGB}{85,139,47}
\definecolor{mgreenlight}{RGB}{174,213,129}

\definecolor{mred}{RGB}{198,40,40}
\definecolor{mredlight}{RGB}{239,154,154}

\definecolor{corange1}{RGB}{254,237,222}
\definecolor{corange2}{RGB}{253,190,133}
\definecolor{corange3}{RGB}{253,141,60}
\definecolor{corange4}{RGB}{230,85,13}
\definecolor{corange5}{RGB}{166,54,3}

\definecolor{cblue1}{RGB}{239,243,255}
\definecolor{cblue2}{RGB}{189,215,231}
\definecolor{cblue3}{RGB}{107,174,214}
\definecolor{cblue4}{RGB}{49,130,189}
\definecolor{cblue5}{RGB}{8,81,156}

\definecolor{cgray1}{RGB}{247,247,247}
\definecolor{cgray2}{RGB}{204,204,204}
\definecolor{cgray3}{RGB}{150,150,150}
\definecolor{cgray4}{RGB}{99,99,99}
\definecolor{cgray5}{RGB}{37,37,37}

\definecolor{cred1}{RGB}{254,229,217}
\definecolor{cred2}{RGB}{252,174,145}
\definecolor{cred3}{RGB}{251,106,74}
\definecolor{cred4}{RGB}{222,45,38}
\definecolor{cred5}{RGB}{165,15,21}

\definecolor{cgreen1}{RGB}{237,248,233}
\definecolor{cgreen2}{RGB}{186,228,179}
\definecolor{cgreen3}{RGB}{116,196,118}
\definecolor{cgreen4}{RGB}{49,163,84}
\definecolor{cgreen5}{RGB}{0,109,44}

\definecolor{cdiv11}{RGB}{166,97,26}
\definecolor{cdiv12}{RGB}{223,194,125}
\definecolor{cdiv13}{RGB}{245,245,245}
\definecolor{cdiv14}{RGB}{128,205,193}
\definecolor{cdiv15}{RGB}{1,133,113}

\definecolor{cdiv21}{RGB}{208,28,139}
\definecolor{cdiv22}{RGB}{241,182,218}
\definecolor{cdiv23}{RGB}{247,247,247}
\definecolor{cdiv24}{RGB}{184,225,134}
\definecolor{cdiv25}{RGB}{77,172,38}

\definecolor{cdiv31}{RGB}{230,97,1}
\definecolor{cdiv32}{RGB}{253,184,99}
\definecolor{cdiv33}{RGB}{247,247,247}
\definecolor{cdiv34}{RGB}{178,171,210}
\definecolor{cdiv35}{RGB}{94,60,153}

\newif\ifarxiv
\arxivtrue
\ifarxiv
\iclrfinaltrue
\fi

\newcommand{\ours}{ReSWD\xspace}

\title{\ours: ReSTIR‘d, not shaken. \\ Combining Reservoir Sampling and Sliced Wasserstein Distance for Variance Reduction}
\author{Mark Boss$^1$ \quad Andreas Engelhardt$^{1,2\footnotemark[2]}$ \quad Simon Donné$^1$ \quad Varun Jampani$^1$ \\ \\
$^1$Stability AI \quad $^2$University of Tübingen}

\begin{document}

\maketitle
\begin{center}
    \centering
    \captionsetup{type=figure}
    \begin{tikzpicture}[
  >=Latex,
  font=\footnotesize,
  pics/hourglass/.style args={#1/#2/#3/#4}{% width/height/neck
    code={
      \def\W{#1}
      \def\H{#2}
      \def\neck{#3}
      \coordinate (TL) at (-\W/2,\H/2);
      \coordinate (TR) at ( \W/2,\H/2);
      \coordinate (BR) at ( \W/2,-\H/2);
      \coordinate (BL) at (-\W/2,-\H/2);
      \coordinate (CU) at (0, \neck);
      \coordinate (CL) at (0,-\neck);
      \path[fill=#4!35,draw=#4,thick]
        (TL)--(CU)--(TR)--(BR)--(CL)--(BL)--cycle;
    }
  }
]

%%% General Matching
\ifarxiv
\node (general_matching_start) {\includegraphics[width=3cm]{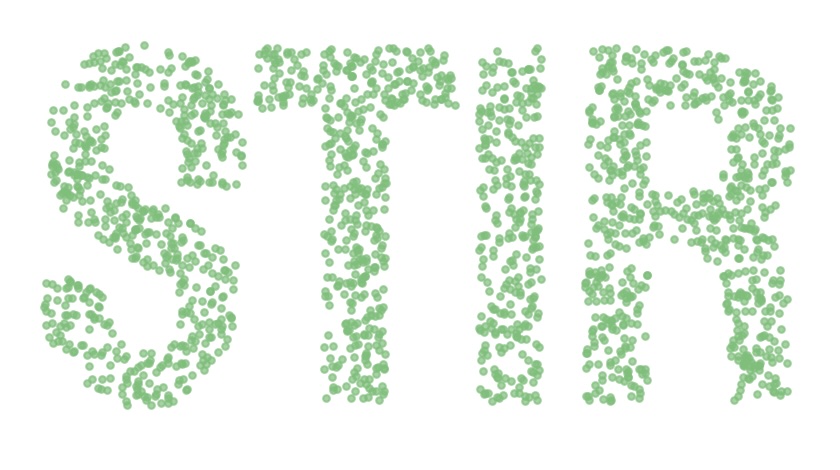}};
\else
\node (general_matching_start) {\includegraphics[width=3cm]{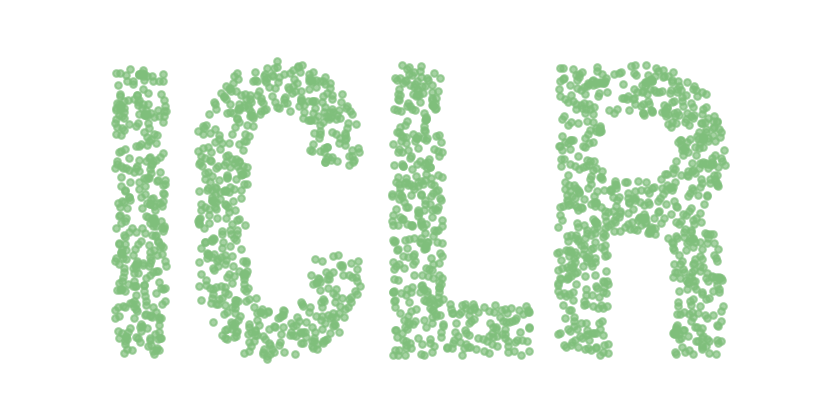}};
\fi

\ifarxiv
\node[below=0.1cm of general_matching_start] (general_matching_mid) {\includegraphics[width=3cm]{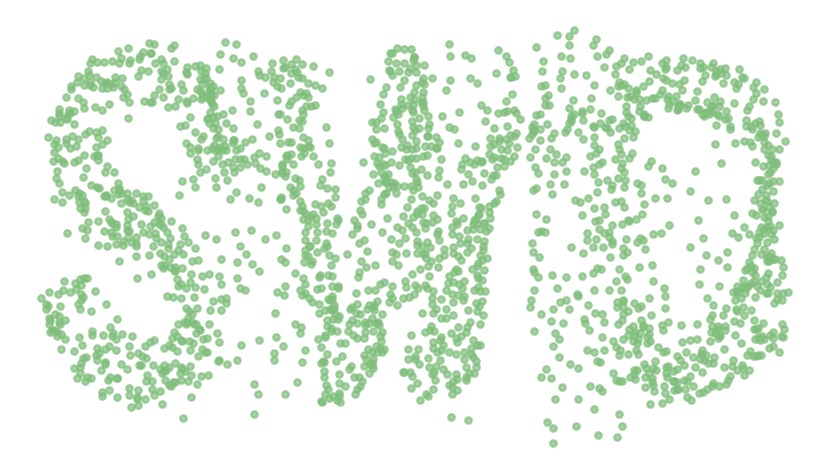}};
\else
\node[below=0.1cm of general_matching_start] (general_matching_mid) {\includegraphics[width=3cm]{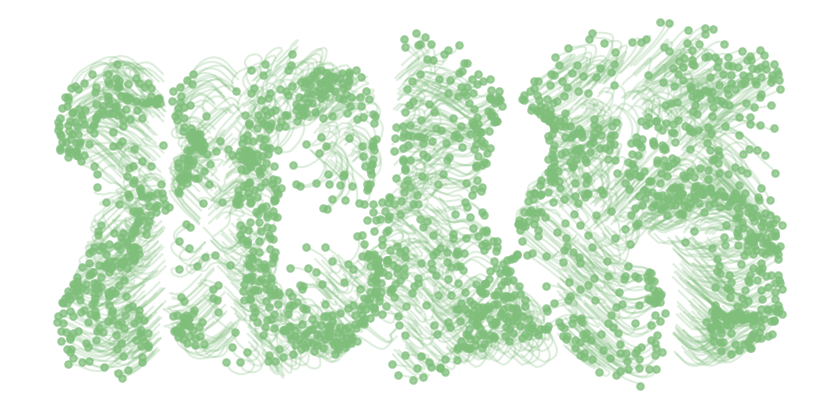}};
\fi

\ifarxiv
\node[below=0.1cm of general_matching_mid] (general_matching_end) {\includegraphics[width=3cm]{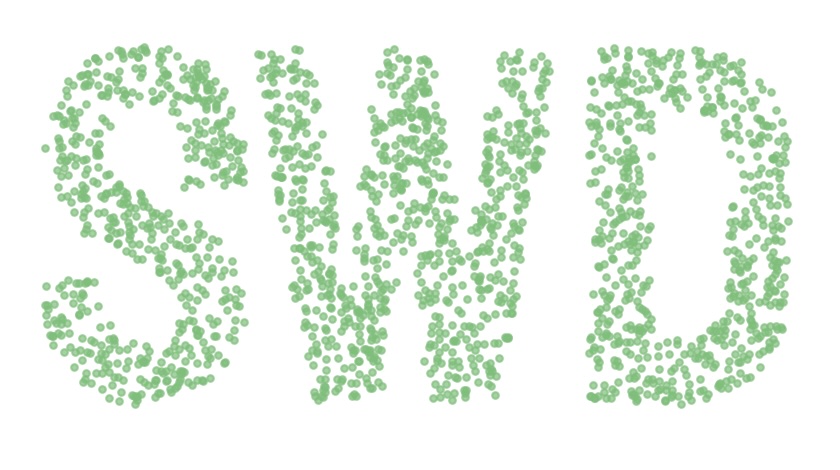}};
\else
\node[below=0.1cm of general_matching_mid] (general_matching_end) {\includegraphics[width=3cm]{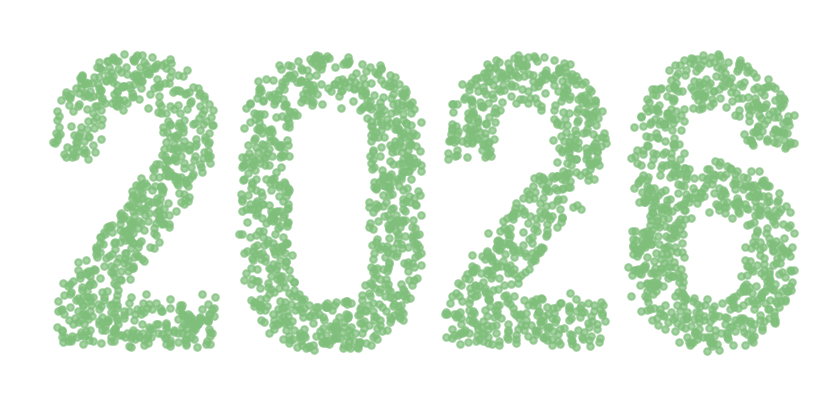}};
\fi

\draw[->] (general_matching_start.north west) -- (general_matching_end.south west);

%%% Diffusion Guidance

\node[right=0.5cm of general_matching_start.north east, text width=2cm, align=center, anchor=north west, yshift=0.25cm] (prompt) {\textit{Watercolor Hand Painted Fantasy City}};
\node[right=0.2cm of prompt.north east, anchor=north west] (reference_img) {\includegraphics[width=1.5cm,trim=0pt 0pt 0pt 100pt, clip]{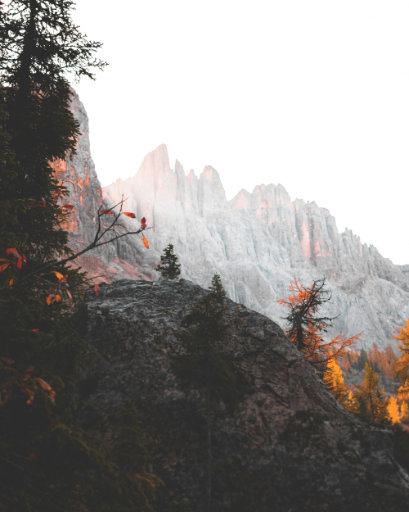}};
\node[above=-0.1cm of reference_img] (reference_lbl) {Reference Image};

\pic[below=2cm of $(prompt.west)!0.5!(reference_img.east)$] {hourglass={2.5/1.5/0.4/teal}};
\node[below=2cm of $(prompt.west)!0.5!(reference_img.east)$,yshift=12pt,text width=2.5cm, align=center] (network_label) {Image Generation + ReSWD}; 

\node (image_guidance) at (general_matching_end -| prompt) {\includegraphics[width=2cm]{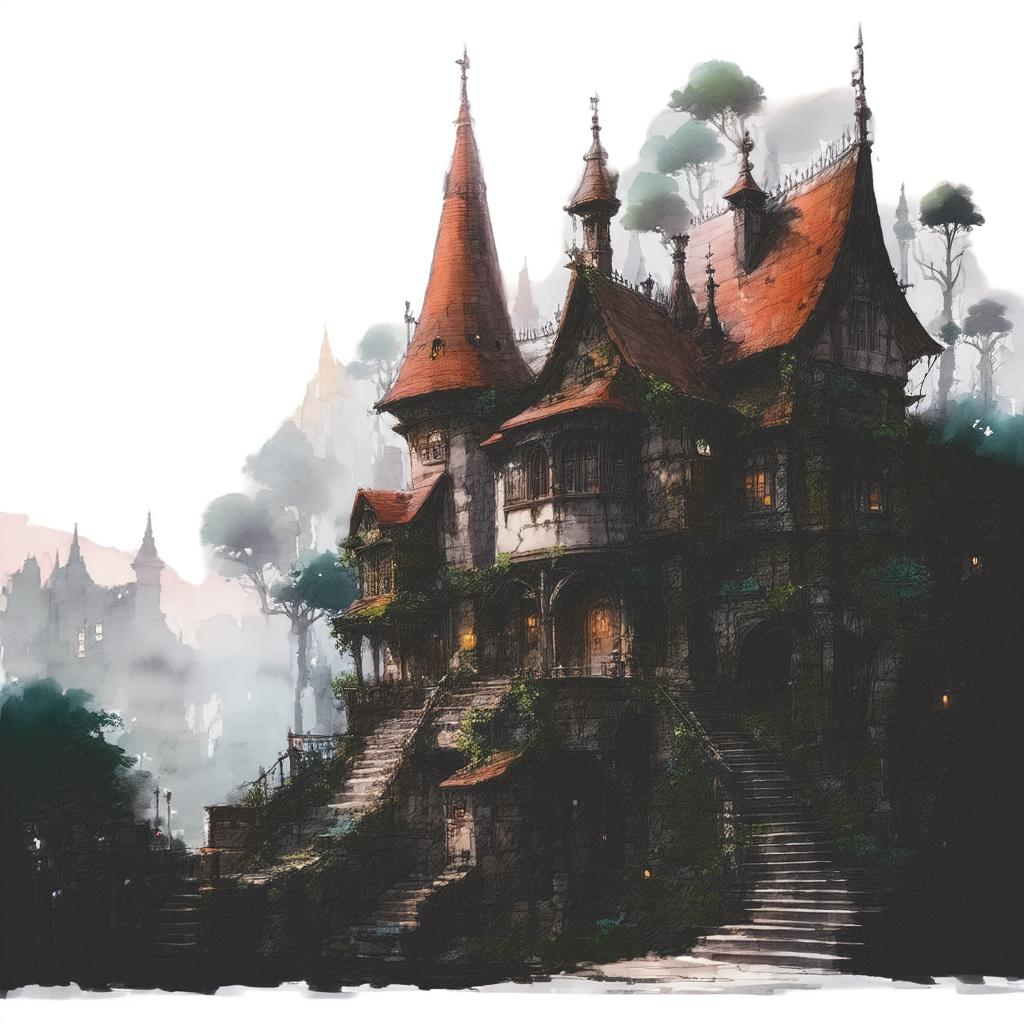}};
\node[right=-0.1cm of image_guidance.north east, anchor=north west] (guidance_1) {\includegraphics[width=1cm]{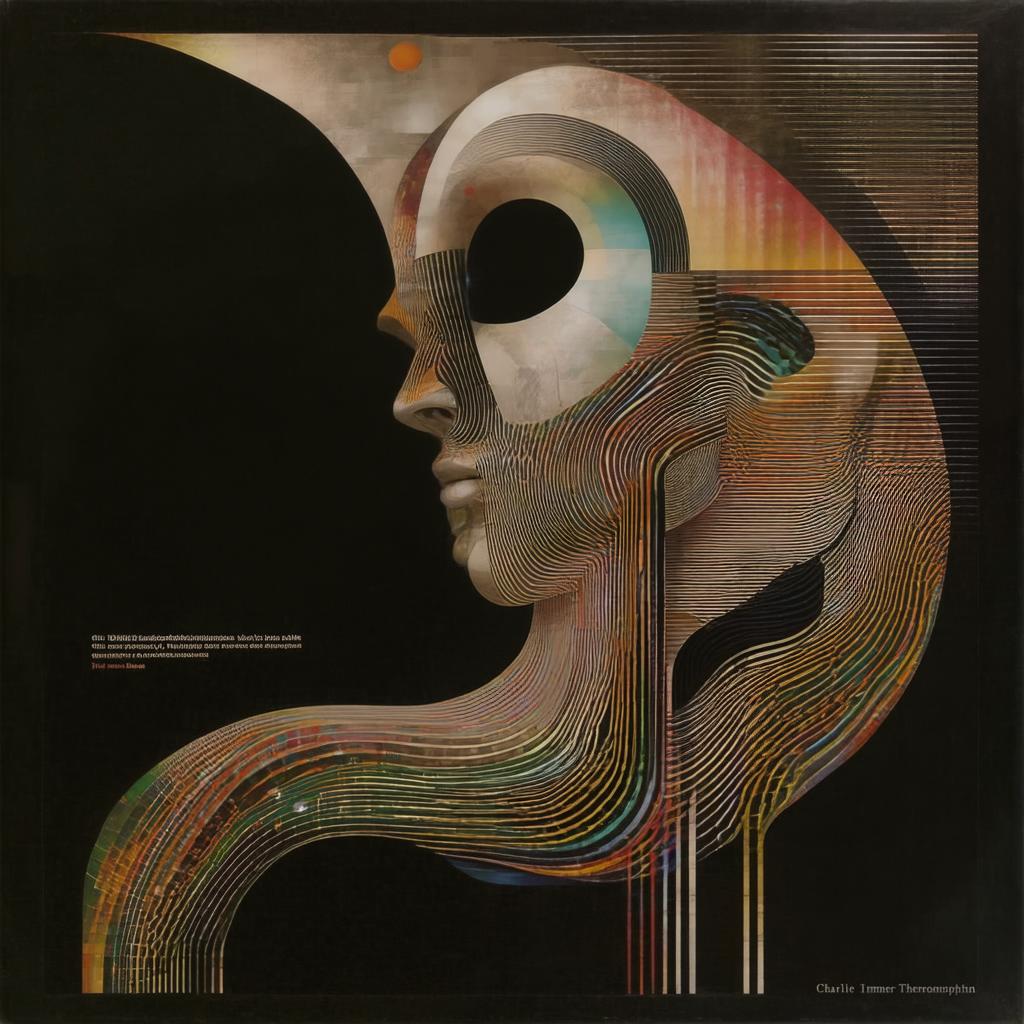}};
\node[right=-0.1cm of guidance_1.north east, anchor=north west] (guidance_2) {\includegraphics[width=1cm]{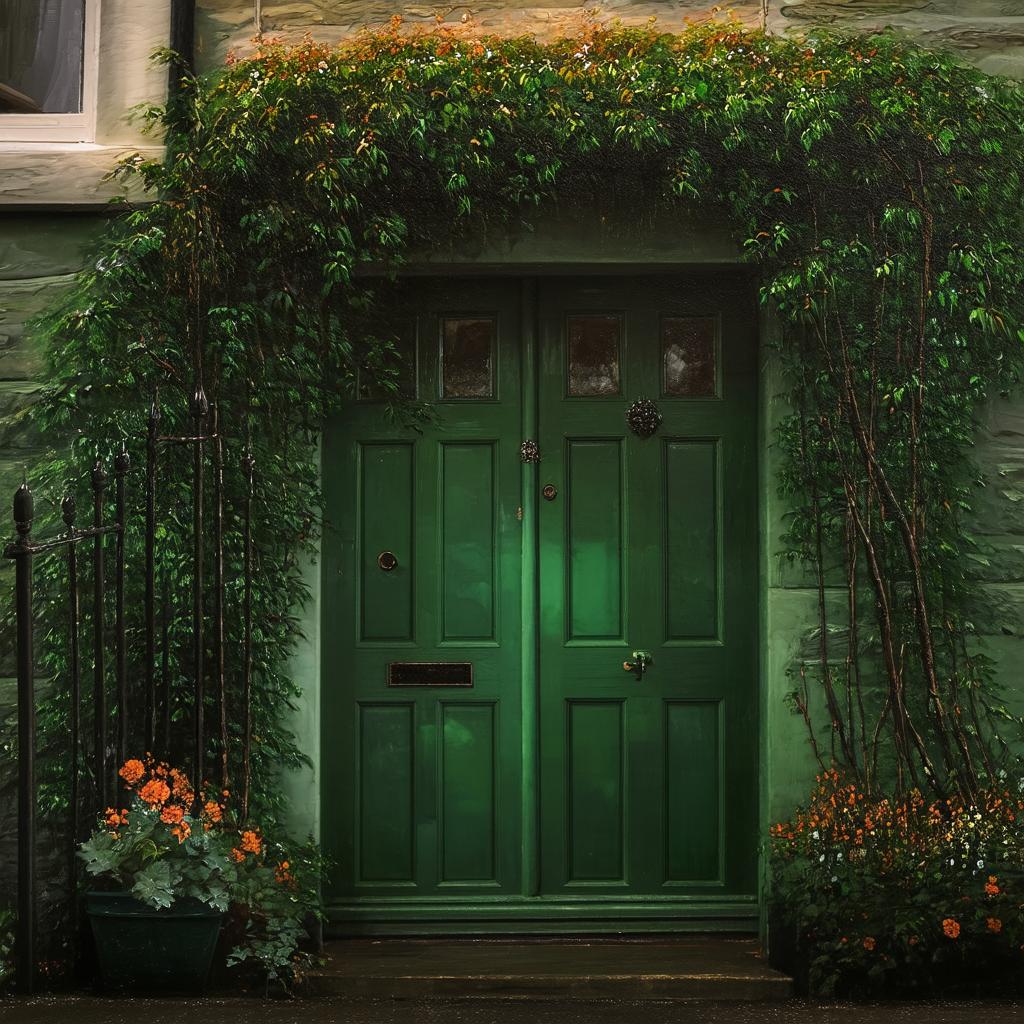}};
\node[right=-0.1cm of image_guidance.south east, anchor=south west] (guidance_3) {\includegraphics[width=1cm]{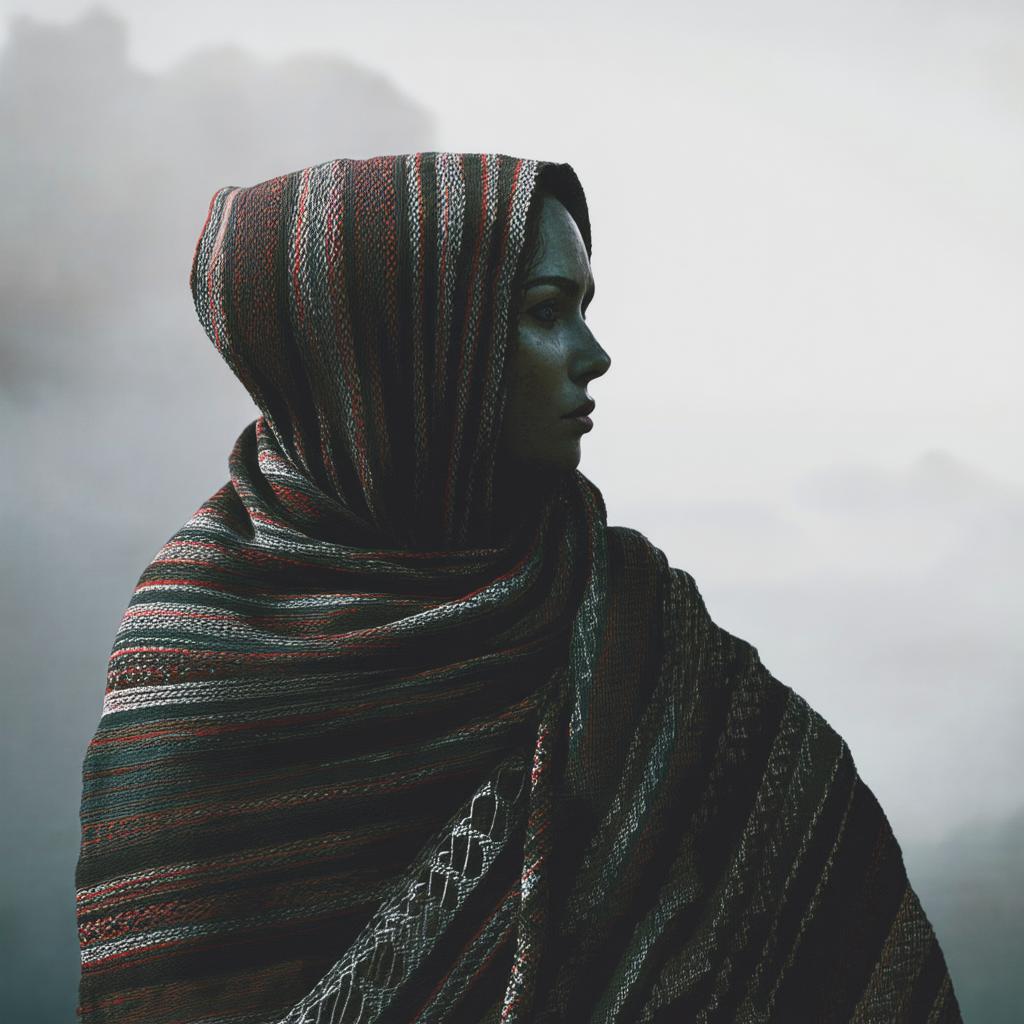}};
\node[right=-0.1cm of guidance_3.south east, anchor=south west] (guidance_4) {\includegraphics[width=1cm]{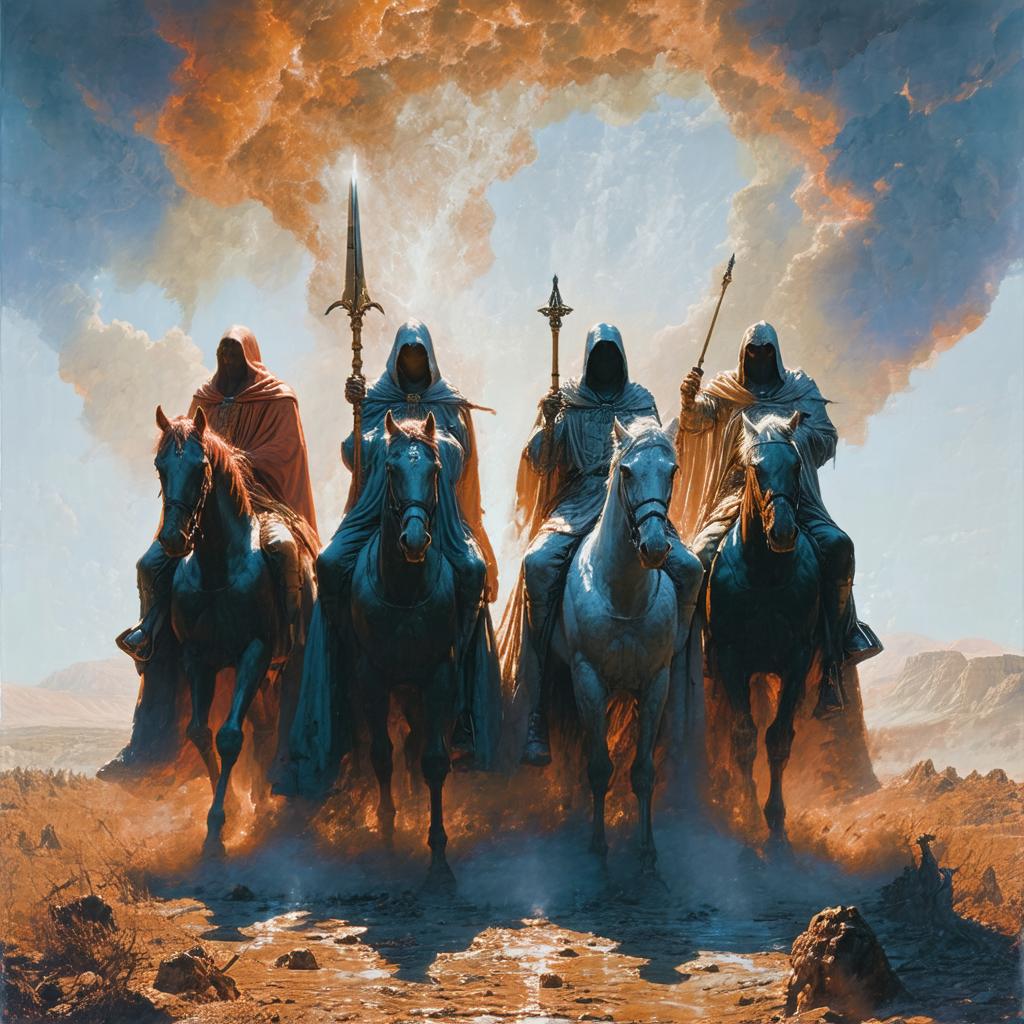}};

\draw[->, shorten >=3mm] (reference_img) -- (network_label);
\draw[->, shorten >=4mm] (prompt) -- (network_label);
\draw[->, shorten <=1mm, shorten >=-2mm] (network_label) -- (image_guidance);

%%% Color Correction

\node[right=of reference_img] (source_img_matching) {\includegraphics[width=2cm,height=2cm,trim=250pt 0pt 250pt 0pt, clip]{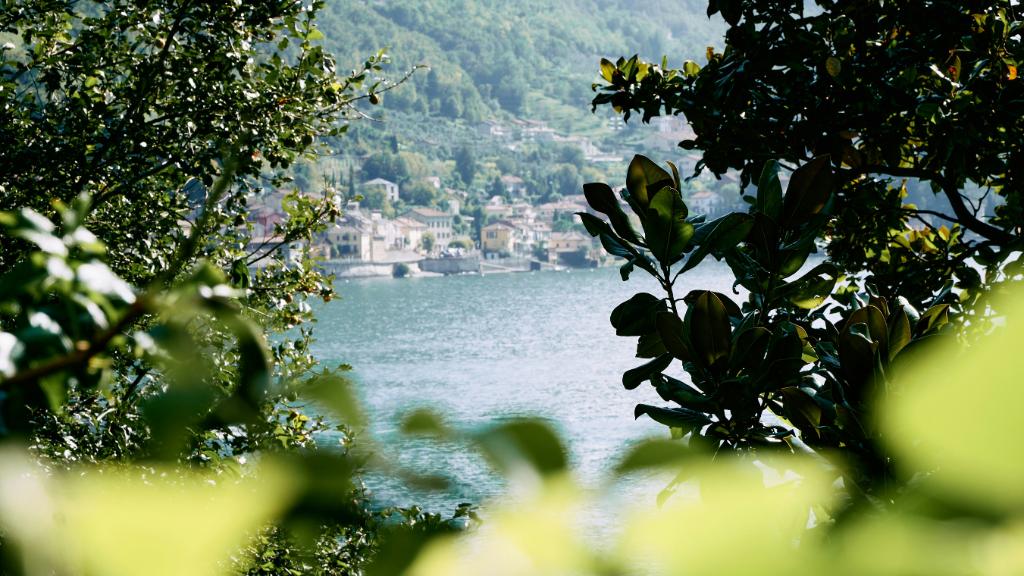}};
\node[right=0.2cm of source_img_matching] (target_img_matching) {\includegraphics[width=2cm,height=2cm,trim=0pt 150pt 0pt 150pt, clip]{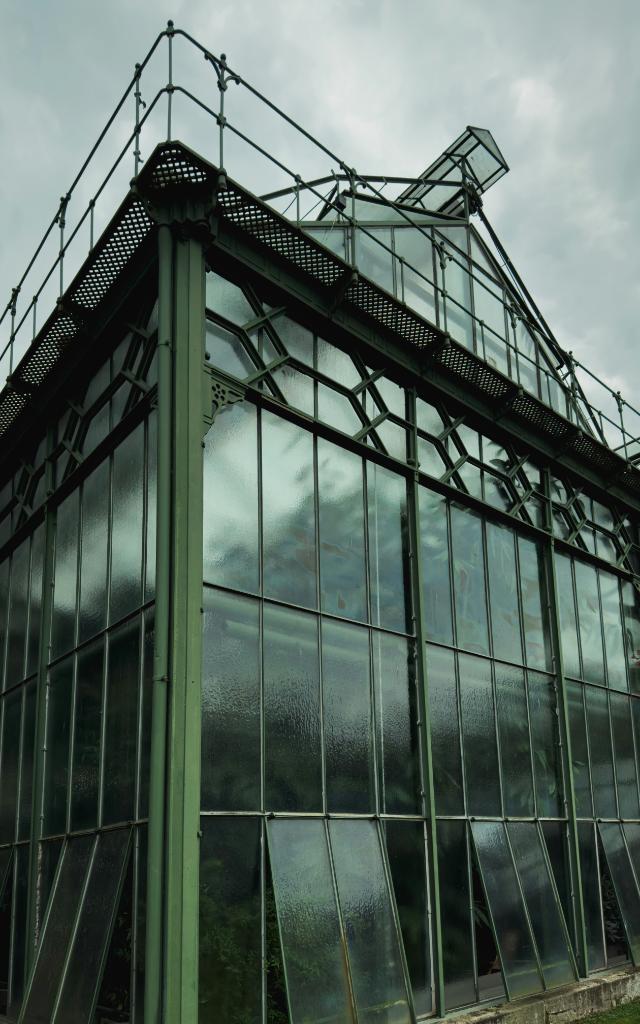}};
\draw[->,shorten >=-1mm, shorten <=-1mm] (source_img_matching) -- (target_img_matching);

\node[below=of $(source_img_matching)!0.5!(target_img_matching)$,fill=orange!30,draw=orange,thick,rounded corners=2pt,inner xsep=12pt,inner ysep=12pt, yshift=-0.4cm] (cdl_matcher) {Color Matching};

\node[below=0.225cm of cdl_matcher.south -| source_img_matching.west,anchor=north west] (result) {\includegraphics[width=2cm,height=2cm,trim=250pt 0pt 250pt 0pt, clip]{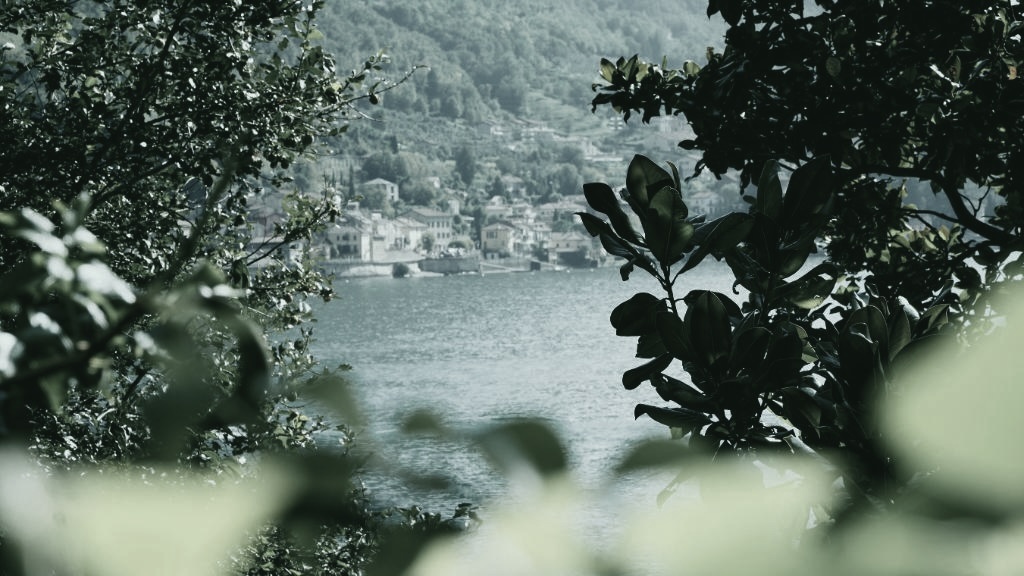}};

\node[right=0.1cm of result, text width=2cm, align=left, font=\fontsize{3.5}{4}\selectfont] (cdl) {
\begin{verbatim}
<SOPNode>
<Slope>
    0.902 0.821 0.892
</Slope>
<Offset>
    -0.006 0.152 0.035
</Offset>
<Power>
    1.587 1.302 1.061
</Power>
</SOPNode>
<SatNode>
<Saturation>0.268</Saturation>
</SatNode>
\end{verbatim}
};

\draw[->] (source_img_matching) -- (cdl_matcher);
\draw[->] (target_img_matching) -- (cdl_matcher);
\draw[->] (cdl_matcher) -- (result);
\draw[->] (cdl_matcher) -- (cdl);

%%% Labels

\draw [decorate, decoration={brace,mirror,raise=6pt}] (image_guidance.south -| general_matching_end.west) -- (image_guidance.south -| general_matching_end.east)
    node[pos=0.5,below=10pt] {Distribution Matching};

\draw [decorate, decoration={brace,mirror,raise=6pt}] (image_guidance.south west) -- (guidance_4.south east)
    node[pos=0.5,below=10pt] {Diffusion Guidance};

\draw [decorate, decoration={brace,mirror,raise=6pt}] (image_guidance.south -| result.west) -- (image_guidance.south -| cdl.east)
node[pos=0.5,below=10pt] {Color Correction};

\end{tikzpicture}
    \caption{\textbf{Applications.}{ SWD can be used in various applications and with \ours we enhance their performances. Here, we show diffusion guidance, color corrections and general distribution matching.}}
    \label{fig:teaser}
\end{center}

\ifarxiv
\renewcommand{\headrulewidth}{0pt}
\fancyhf{}
\fi

\ificlrfinal
\renewcommand{\thefootnote}{\fnsymbol{footnote}}
\footnotetext[2]{Work done during internship at Stability AI.}
\fi

\begin{abstract}
Distribution matching is central to many vision and graphics tasks, where
the widely used Wasserstein distance is too costly to compute for high-dimensional distributions. 
The \emph{Sliced Wasserstein Distance} (SWD) offers a scalable alternative, yet its Monte Carlo estimator suffers from high variance, resulting in noisy gradients and slow convergence. 
We introduce \emph{Reservoir SWD} (\ours), which integrates Weighted Reservoir Sampling into SWD to adaptively retain informative projection directions in optimization steps, resulting in stable gradients while remaining unbiased.
Experiments on synthetic benchmarks and real-world tasks such as color correction and diffusion guidance show that \ours consistently outperforms standard SWD and other variance reduction baselines.

\ificlrfinal
Project page with code: \url{https://ReservoirSWD.github.io}
\fi

\end{abstract}
\section{Introduction}
\label{sec:introduction}

Distribution matching is a central problem in computer vision and graphics: from classical tasks such as histogram matching to more nuanced applications such as color grading~\citep{pitie2005n,Rabin2012wasserstein,Bonneel2015slicedradon, he2024ms-swd}, texture alignment~\citep{elnekave2022gpdm, heitz2021CVPR}, and the guidance of generative models~\cite{lobashev2025color}.
Among the available metrics, the Wasserstein distance has emerged as particularly powerful due to its ability to capture shifts between distributions.
However, the Wasserstein distance suffers from the curse of dimensionality and has a practical convergence rate scaled with the d-dimensional distribution $\mathcal O\bigl(n^{-\frac{1}{d}}\bigr)$, with $n$ data points.
This limits its direct use in many iterative optimization settings.

The \emph{Sliced Wasserstein Distance} (SWD)~\citep{pitie2005n,Bonneel2015slicedradon,heitz2021CVPR} offers a practical alternative by projecting distributions in random directions and averaging their 1-D Wasserstein costs.
This reduces the computational burden to a sequence of 1-D sorting problems but introduces a new challenge: the expectation over projection directions is typically approximated via Monte Carlo (MC) sampling, which suffers from high variance.
Increasing the directions would result in the true expected value, but the complexity of SWD scales with the number of random directions $M$ as $\mathcal O\bigl(M n \log n\bigr)$. 
Hence, the curse of dimensionality is introduced again indirectly with the number of projections, which should increase with the dimensionality of the data.
In optimization scenarios, where such SWD metrics are used as loss functions to optimize neural networks, this variance directly results in noisy gradients and slower convergence.

A variety of variance reduction techniques have been explored for MC estimators, but most remain underutilized in distribution matching objectives.
In this work, we draw inspiration from recent advances in rendering, particularly the ReSTIR resampling framework~\citep{mbitterli2020restir}, and adapt \emph{Weighted Reservoir Sampling} (WRS)~\citep{efraimidis2006weighted,chao1982wrsgeneral} to the SWD setting.
The resulting estimator, which we term \emph{Reservoir SWD} (\ours), continuously reuses and reweighs the most informative projection directions throughout the optimization.
Intuitively, \ours preferentially retains those directions where the two distributions are most dissimilar, thus concentrating computational effort on projections that produce stronger and more stable gradients.
This simple but effective modification significantly reduces stochastic variance while preserving unbiasedness due to the weighting of the WRS, leading to faster and more robust optimization.

We demonstrate the advantages of \ours on two real-world distribution matching problems of color correction and color diffusion guidance, as shown in \fig{teaser}, as well as synthetic general distribution matching problems, showing clear improvements over the standard SWD and existing variance reduction baselines.

\section{Related Work}
\label{sec:related_work}

\inlinesection{Distribution Matching and Optimal Transport.}
Distribution matching is the general task of aligning two distributions. 
A particularly effective way of achieving this is with \emph{Optimal Transport} (OT) by minimizing the cost of moving the probability mass, which is often performed using Wasserstein distances~\citep{villani2008optimal,peyre2019computational}. 
Entropic regularization accelerates OT through Sinkhorn iterations~\citep{cuturi2013sinkhorn}, leading to scalable solvers and Sinkhorn divergences that interpolate between OT and kernel \emph{Maximum Mean Discrepancies} (MMD)~\citep{feydy2019sinkhorn}. 
OT has been widely applied to textures, color transfer, and barycenters~\citep{Rabin2012wasserstein,Bonneel2015slicedradon,pitie2005n}.

To address high-dimensional costs, \emph{Sliced Wasserstein} (SW) replaces couplings with averages of 1-D transports. 
Early works introduced SW barycenters and Radon/Projected variants that are fast and differentiable~\citep{Bonneel2015slicedradon,Rabin2012wasserstein}. 
Extensions include Max-SW, which selects the most contributing directions instead of averaging over them~\citep{deshpande2019max}, and several other techniques to improve efficiency and optimization performance~\citep{kolouri2019generalized,nguyen2021distributional,nguyen2022hierarchical,nguyen2023energy, nguyen2024sliced}. 
SW has also been studied through geometry and flows: gradient flows as generative dynamics with long-time analysis~\citep{cozzi2024longtime,vauthier2025swflows}, intrinsic geometry of SW space~\citep{park2025geometrysw}, extensions to Cartan–Hadamard manifolds~\citep{bonet2025cartanhadamard}, and stereographic spherical SW for curved domains~\citep{tran2024stereographic}.
Recent work further reduces variance via quasi-Monte Carlo and control variates~\citep{nguyen2024quasi,nguyen2024slicedcv}.
In practice, SW is widely used as a loss in graphics, vision, and generative modeling \eg, for textures, patch statistics, and color transfer due to stable gradients and favorable sample complexity~\citep{heitz2021CVPR,elnekave2022gpdm,pitie2005n,wu2023sin3dm}. 

Our work ties into the work selecting the most contributing directions similar to Max-SW but achieves this by building a reservoir of high contributing directions during the training. 
This improves efficiency, as we keep more directions in optimization. 
It also ties in with the work on variance reduction while remaining unbiased, as our technique is inspired by variance reduction techniques in real-time path tracing~\citep{mbitterli2020restir}.

\inlinesection{Color Transfer and Correction.} %
Classical color transfer methods framed recoloring as histogram or distribution matching across images, often using statistical metrics or optimal transport~\citep{pitie2005n,Rabin2012wasserstein,Bonneel2015slicedradon}. 
These techniques established the foundation for perceptual color differences in imaging. 
More recently, multiscale OT and Wasserstein-based metrics have been proposed as perceptually faithful color difference measures~\citep{he2024ms-swd}.

With deep learning, style transfer methods enabled more expressive color and appearance manipulation. 
CNN-based approaches introduced artistic and photorealistic stylization~\citep{gatys2016image,luan2017deep,huang2017arbitrary,li2017universal,li2018closed,yoo2019photorealistic,an2020ultrafast,chiu2022photowct2,hong2021domain}, with adaptive instance normalization and feature transforms widely adopted to control color and tone. 
These methods evolved from hand-crafted global mappings to neural architectures capable of spatially aware and semantically coherent color control.

With the advance of text-guided diffusion models~\citep{rombach2022highresolution,rombach2022ldm}, additional mechanisms to control the style such as LoRA~\citep{hu2021lora}, IP-Adapter~\citep{zhang2023ipadapter}, ControlNets~\citep{zhang2023controlnet} are introduced. 
A recent work by Lobashev \etal~\citep{lobashev2025color} also demonstrated that traditional constraints based on OT can be used during generation to enforce a color distribution.

Our work can be used to enhance traditional color matching works with improved efficiency and to enable more novel color-guided diffusion tasks.

\section{Method}
\label{sec:method}

In our setting, we only ever compare empirical \emph{discrete} distributions, that is, two sets of samples 
$X = \{x_1, \dots, x_{N_X}\} \subset \mathbb{R}^d$ and 
$Y = \{y_1, \dots, y_{N_Y}\} \subset \mathbb{R}^d$, drawn from the underlying distributions
$\mathcal{X}$ and $\mathcal{Y}$. 

\subsection{Preliminaries}

\inlinesection{Wasserstein Distance.} %
The Wasserstein $p$-distance measures differences between two empirical distributions. 
In the 1D case relevant to our setting, given two sets of samples $X$ and $Y$, the Wasserstein distance can be computed simply by sorting both sets and comparing the corresponding order statistics:
\begin{equation}
    \operatorname{W}_p(X,Y) 
    = \left( \tfrac{1}{n} \sum_{i=1}^n |x_{i} - y_{i}|^p \right)^{\sfrac{1}{p}},
\end{equation}
This efficient formulation has complexity $\mathcal{O}(n \log n)$ due to sorting, in contrast to the cubic complexity of the general $d$-dimensional case~\citep{Rabin2012wasserstein, Bonneel2015slicedradon, pitie2005n}.

\inlinesection{Sliced Wasserstein Distance (SWD).} %
The Sliced Wasserstein Distance proposes to stochastically approximate the true Wasserstein distance in multi-dimensional distributions.
It is defined as~\citep{pitie2005n,heitz2021CVPR,elnekave2022gpdm}:
\begin{equation}
    \operatorname{S}_p(X,Y) 
    = \mathbb{E}_{\theta \sim U(S^{d-1})} 
        \Big[ \operatorname{W}_p(\pi_\theta X, \pi_\theta Y) \Big],
\end{equation}
where $\pi_\theta$ indicates the projection onto the unit-normal vector $\theta$ (uniformly sampled from $S^{d-1}$, the $d$-dimensional unit sphere $\mathbb{R}^d$).

In practice, the expectation is often approximated via Monte Carlo (MC) integration:
\begin{equation}
    \operatorname{S}_p(X,Y) \approx w_i \sum_{i=1}^{L} 
        \operatorname{W}_p(\pi_{\theta_i} X, \pi_{\theta_i} Y). \label{eq:expected_value}
\end{equation}
with random uniformly sampled directions $\theta_i$ and $w_i$ being the sampling weights, \ie $w_i=\frac{1}{L}$ with uniform direction sampling. 
The resulting complexity is therefore $\mathcal{O}(L n \log n)$.
The SWD is an unbiased estimate of the true Wasserstein distance~\citep{pitie2005n, heitz2021CVPR}. 
With modern frameworks such as PyTorch or Tensorflow, the entire metric can be trivially implemented fully differentiable, which enables usage in optimization settings.
However, due to the MC integration, SWD can lead to noisy gradients, which we aim to solve with \ours.

\subsection{Reservoir Sliced Wasserstein Distance (ReSWD)}
\label{subsec:wrswd}

\begin{figure}
    \centering
\begin{tikzpicture}[
  >=Latex,
  font=\footnotesize,
  node distance=0.9cm and 1.2cm,
  mbox/.style={draw, rounded corners=2pt, fill=cyan!18, inner xsep=6pt, inner ysep=3pt},
  kbox/.style={draw, rounded corners=2pt, fill=orange!20, inner xsep=6pt, inner ysep=3pt},
  db/.style={draw, shape=cylinder, shape border rotate=90, aspect=0.3,
             minimum height=1.2cm, minimum width=1.1cm, align=center, fill=orange!10},
  ell/.style={draw, ellipse, inner xsep=8pt, inner ysep=5pt},
  dot/.style={circle, draw, inner sep=1.1pt},
  lbl/.style={inner sep=1pt}
]

% --- Left stack: New candidate Directions (M boxes) ---
\node[lbl, align=center, anchor=west] (mTitle) at (0,1.9) {New candidate\\Directions};
\node[mbox, below=0.25cm of mTitle.south west, anchor=north west] (m1) {$M_1$};
\node[mbox, right=0.18cm of m1] (m2) {$M_2$};
\node[mbox, right=0.18cm of m2] (m3) {$M_3$};

% --- K row ---
\node[kbox] (k1) at (0,-1.3) {$K_1$};
\node[kbox, right=0.18cm of k1] (k2) {$K_2$};
\node[kbox, right=0.18cm of k2] (k3) {$K_3$};
\node[kbox, right=0.18cm of k3] (k4) {$K_4$};

% --- Reservoir (parked above K's; no arrows to K or Projection) ---
\node[db, above=0.45cm of $(k1)!0.5!(k4)$] (res) {Reservoir};

% --- Projection (ellipse) ---
\node[ell, right=1.4cm of res] (proj) {Projection};

% --- Only M3 and K4 connect to Projection ---
\draw[->] (m3) to[out=0,in=120] (proj);
\draw[->] (k4) to[out=0,in=230] (proj);

% --- Distributions (compact) ---
\def\dy{0.6}
\coordinate (base) at ($(proj.east)+(0.9,1.75)$);
\node[dot] (t0) at ($(base)+(0,\dy)$)  {};
\node[dot] (b0) at ($(base)+(0.3,-\dy)$)  {};
\node[dot] (t1) at ($(base)+(1.2,\dy)$) {};
\node[dot] (b1) at ($(base)+(1.0,-\dy)$) {};
\node[dot] (t2) at ($(base)+(1.8,\dy)$) {};
\node[dot] (b2) at ($(base)+(2.2,-\dy)$) {};
\node[dot] (t3) at ($(base)+(2.5,\dy)$) {};
\node[dot] (b3) at ($(base)+(2.5,-\dy)$) {};

% Horizontal guide lines
\begin{scope}[on background layer]
\draw[teal!40,line width=1.5pt] ($(b0)+(-0.55,0)$) -- ($(b3)+(0.25,0)$);
\draw[purple!40,line width=1.5pt] ($(t0)+(-0.25,0)$) -- ($(t3)+(0.25,0)$);
\end{scope}

% Zig-zag connections
\foreach \i in {0,1,2,3}{
  \draw (t\i) -- (b\i);
}

% Labels
\node[lbl, above=3pt of $(t0)!0.5!(t3)$] {Distribution 1};
\node[lbl, below=3pt of $(b0)!0.5!(b3)$] {Distribution 2};

\draw[->] (proj) to[out=0,in=180] (base);

% --- weight and L (kept close; short arrows) ---
\node (weight) at ($(base)+(4.0,0.0)$) {Weight $w$};
\draw[->] ($(base)+(2.6,0.0)$) -- (weight);
\draw[->, dashed] (weight) to[out=220,in=350] node[lbl,pos=0.15,below right](update_lbl){Update Reservoir} (k4);

\node[kbox, below=0.1cm of update_lbl.south west, font=\tiny, inner xsep=3pt, inner ysep=1.5pt] (k1_upd) {$K_1$};
\node[mbox, right=0.18cm of k1_upd, font=\tiny, inner xsep=3pt, inner ysep=1.5pt] (k2_upd) {$M_3$};
\node[kbox, right=0.18cm of k2_upd, font=\tiny, inner xsep=3pt, inner ysep=1.5pt] (k3_upd) {$K_3$};
\node[kbox, right=0.18cm of k3_upd, font=\tiny, inner xsep=3pt, inner ysep=1.5pt] (k4_upd) {$K_4$};

\node[below=1cm of update_lbl] (loss) {Calculate Loss $W_p^p$};
\draw[->] ($(k1_upd.south)!0.5!(k4_upd)$) -- (loss);

\end{tikzpicture}
    \titlecaption{Overview}{During the optimization we keep a reservoir of highly influential directions. We use the Sliced Wasserstein Distance (SWD) as a proxy metric for the reservoir update and the final optimization loss. Note that only directions in the reservoir influence the optimization to remain unbiased.}
    \label{fig:explainer}
\end{figure}

Although SWD is highly efficient in the calculation at high dimensions, it still suffers from high variance due to the MC integration from random directions.
Especially, when SWD is used as a loss during an optimization, this leads to noisy gradients.
In computer graphics, variance is often also an issue, especially in real-time path tracing where MC integration is also usually employed.
Inspired by ReSTIR’s \emph{resampled-importance-sampling}~\citep{mbitterli2020restir}, we incorporate the Weighted Reservoir Sampling (WRS) mechanism into SWD.
This allows the reuse of information between optimization steps, and hence faster convergence overall with minimal performance penalties. 

Weighted Reservoir Sampling is a family of algorithms that draw a fixed subset from a weighted stream of candidates, \ie directions, in a single pass~\citep{chao1982wrsgeneral, efraimidis2006weighted}.
Suppose that we wish to maintain a reservoir of $K$ candidates while processing a candidate pool of $K+M$ items, where $M$ denotes the number of newly drawn candidates.
Each candidate $\theta_j$ is associated with a non-negative weight $w_j$, and the goal is to select exactly $K$ survivors such that the marginal inclusion probability of each element is proportional to $w_j$.
This is achieved by assigning to every candidate a random key~\citep{efraimidis2006weighted}
\begin{equation}
    k_j = u_j^{\sfrac{1}{w_j}}, \qquad u_j \sim \mathcal U(0,1),
\end{equation}
retaining the $K$ elements with the smallest keys.
The resulting selection is unbiased: the expected inclusion probability of each element is proportional to its weight, while the reservoir size remains fixed~\citep{efraimidis2006weighted}.
In our context, WRS allows efficient reallocation of computational effort toward projections with higher contribution to the optimization loss, while preserving Monte Carlo unbiasedness.

In \fig{explainer}, we present an overview of our proposed method.
Let $\mathcal P$ be a pool of $K+M$ 1-D projection directions $\theta\subset S^{d-1}$.
For every $\theta\in\mathcal P$ we compute the 1-D $p$-\emph{power} Wasserstein cost
\begin{equation}
  D(\theta)=\operatorname{W}_p\!\bigl(\pi_\theta\mu,\pi_\theta\nu\bigr)
\end{equation}

Then each distance calculation performs the following steps.

\inlinesection{Step 0: Time-decay reweighting.} %
As SWD is most often used in optimization scenarios, we introduce $\tau$ as the time decay constant to place more emphasis on newer random directions to account for the shifting optimization field.
Before drawing new candidates, we \emph{age} the stored reservoir weights:
\begin{equation}
\label{eq:decay}
   \tilde w_i \;\leftarrow\;
   w_i \,\exp\!\bigl[-(t-t_i)\!/\tau\bigr],\qquad
   \tilde k_i \;\leftarrow\;
   k_i \,\exp\!\bigl[-(t-t_i)\!/\tau\bigr],
\end{equation}
where $t_i$ is the step when $\theta_i$ entered the reservoir.
This exponential decay (enabled when $\tau\!>\!0$) steadily forgets stale projections so the sampler adapts to \emph{non-stationary} optimization trajectories.

\inlinesection{Step 1: Reservoir construction.}
At optimization step $t$ we maintain a persistent reservoir $\mathcal R_{t-1}$ of $K$ directions from the previous optimization step $t-1$ and draw $M$ new directions $\mathcal N_t$.
The candidate set is $\mathcal P_t=\mathcal R_{t-1}\cup\mathcal N_t$.

\inlinesection{Step 2: Weighted reservoir sampling.} %
Following Efraimidis \& Spirakis \citep{efraimidis2006weighted} we assign each $\theta\in\mathcal P_t$ a key $k(\theta)=u^{1/D(\theta)}$, $u\sim\mathcal U(0,1)$ and keep the $K$ smallest keys.
This selects each $\theta$ with probability $q(\theta)=\frac{D(\theta)}{\sum_{\theta'\in\mathcal P_t}D(\theta')}$, \ie proportionally to its contribution to the loss.

\inlinesection{Step 3: Self-normalized weights.} %
For the survivors $\{\theta_i\}_{i=1}^{K}$ we calculate the loss with weights as:
\begin{equation}
\label{eq:wrswd_est}
   \widehat{S}_{p}(\mu,\nu)
   =\sum_{i=1}^{K} \underbrace{
        \frac{1/q(\theta_i)}
             {\sum_{j} 1/q(\theta_j)}
      }_{w_i}\,D(\theta_i),
\end{equation}
This loss can then be used in optimizations with a detached gradient calculation for importance weights $w_i$.
This remains an \emph{unbiased} MC estimate of $\mathbb E_{\theta}[W_p]$ while focusing computation on projections which highlight larger differences.

\inlinesection{Step 4: ESS-based reservoir reset.} %
We monitor the effective sample size~\citep{mbitterli2020restir} $\mathrm{ESS}= (\!\sum_i w_i)^2 / \sum_i w_i^{\,2}$ and flush the reservoir whenever $\mathrm{ESS}<\alpha K$ ($\alpha\!=\!0.5$ in all experiments), preventing weight collapse.

\inlinesection{Handling unequal sample counts.} %
If $N_\mu\neq N_\nu$ we repeat the shorter projection so that both 1-D arrays have length $n=\max(N_\mu,N_\nu)$ before sorting. Repetitions are picked uniformly with replacement, following \citet{elnekave2022gpdm}.

\inlinesection{Complexity.} %
The dominant cost is sorting $n$ scalars for each of $(K\!+\!M)$ directions: $\mathcal O\bigl((K\!+\!M)\,n\log n\bigr)$.
Key generation, ESS evaluation, and gradient accumulation are $\mathcal O(K\!+\!M+d)$.

\inlinesection{Algorithmic summary.} %
The overview of the algorithm is shown in Algorithm \ref{alg:wrswd}.

\begin{algorithm}[htb]
\caption{\ours estimator (per optimisation step)}
\label{alg:wrswd}
\small
\begin{algorithmic}[1]
\Require batches $\{x_n\}_{n=1}^{N_\mu}$, $\{y_m\}_{m=1}^{N_\nu}$, reservoir $\mathcal R_{t-1}$, hyper-params $K,M,p,\alpha,\tau$  
\For{$(\theta_i,w_i,k_i,t_i)\in\mathcal R_{t-1}$} \Comment{Step 0: Time-decay}
      \State $w_i\!\leftarrow w_i\exp\!\bigl[-(t{-}t_i)/\tau\bigr]$
      \State $k_i\!\leftarrow k_i\exp\!\bigl[-(t{-}t_i)/\tau\bigr]$
\EndFor
\State $\mathcal N_t \gets$ \Call{DrawDirections}{$M,d$} \Comment{Step 1: Reservoir Construction}
\State $\mathcal P_t \gets \mathcal R_{t-1}\cup\mathcal N_t$
\For{$\theta\in\mathcal P_t$} \Comment{Step 2: Reservoir Sampling}
      \State $D(\theta)\gets W_{p}(\pi_\theta\mu,\pi_\theta\nu)$
      \State $k(\theta)\gets u^{1/D(\theta)},\;u\!\sim\!\mathcal U(0,1)$
\EndFor
\State $\mathcal R_t \gets$ $K$ directions with smallest $k(\theta)$
\State compute $q(\theta)$, weights $w_i$, and
       $\widehat{W}_{p}$ via Eq.\,(\ref{eq:wrswd_est}) \Comment{Step 3: Self normalizing weights}
\If{ESS $<\alpha K$}\;\Return $\widehat{W}_{p},\;\varnothing$ \Comment{Step 4: ESS-Reset}
\Else\;\Return $\widehat{W}_{p},\;\mathcal R_t$
\EndIf
\end{algorithmic}
\end{algorithm}

\subsection{Applications}
\label{sec:applications}

We showcase our SWD modifications on two real-world applications.

\inlinesection{Color Correction.} %
Often in movie production two shots have to be matched in terms of their color appearance. 
We tackle this approach by matching a source image using a Color Decision List (CDL)~\citep{asc_cdl} with a reference image.
The source image $x$ is first passed through a differentiable CDL implementation, applying the \textit{slope} $s$, \textit{offset} $o$, \textit{power} $p$ and $\lambda$ \textit{saturation} adjustments according to $x^\prime = \operatorname{S} \bigl((s \times x + o)^p; \lambda \bigr)$. 
The saturation is defined as $\operatorname{S}(x;\lambda)=\operatorname{L}(x) + \lambda \bigl(x - \operatorname{L}(x)\bigr)$ and $\operatorname{L}(x) = 0.2126 x_r + 0.7152 x_g + 0.0722 x_b$, where $r,g,b$ define the color channels.
The source and reference are then converted to the CIELAB space. 
We found that this space improves the color accuracy, which is consistent with a recent paper proposing a perceptual color metric~\citep{he2024ms-swd}.
As we perform the matching on a pixel level, we find that lower resolutions result in faster matching speeds with similar performance. Thus, we leverage a resolution of $128$ in the maximum dimension.
Within 150 steps, the CDL parameters are optimized and can be applied to the full-res image or even a video clip.

\inlinesection{Diffusion Guidance.} %
\citet{lobashev2025color} proposed to incorporate SWD guidance in diffusion models to shift the final generation towards a certain color distribution. 
The main concept is to optimize a small offset on top of the current latents with $i$ SWD steps in each diffusion step. 
Here, we use the predicted $x_0$ in each step and decode it using the VAE decoder. 
This prediction is then matched with the reference image.
We apply this approach to the more recent flow matching model SD3.5~\citep{stablediffusion35github}. 
This required several modifications. 
\citet{lobashev2025color} calculated the full gradient from the input, through the VAE and the U-net. 
With the increased complexity of larger transformer models, we opted to employ a gradient stop after the backbone similar to SDS~\citep{poole2022dreamfusion}.
Hence, we only backpropagate through the VAE decoder.
Similarly to our color-correction approach, we also opted to perform the matching in the CIELAB space.
In addition, we replace the simple gradient descent of \citet{lobashev2025color} with an Adam optimizer. 
With these changes, we use a learning rate of $\expnumber{3}{-3}$ and a total of 6 steps for $95\%$ of the total denoising steps.
We do not reset the reservoir after each denoising step, but twice during the generation, as we only perform 6 SWD steps, which would not suffice to build a reservoir.
This general method can then be applied to medium, large, and large-turbo SD3.5 models using the recommended CFG and step counts.

\section{Results}
\label{sec:results}

We perform synthetic as well as real-world tests with \ours and study the influence of our main hyperparameter, the number of fresh candidates in each optimization step.

\begin{figure}[ht]
\begin{minipage}[t]{.56\textwidth}
    \centering
\titlecaptionof{table}{Comparison on 1D distribution matching}{Mean-W$_1$ over 1000 distribution matches for various methods alongside the respective running time. Here, we can see that ours provides the best performance with a comparatively low run-time cost.
}
\label{tab:comparison_1d}
\resizebox{0.95\linewidth}{!}{ %
\begin{tabular}{lcc}
Method & Mean-W$_1$ [$10^3$] $\downarrow$ & Time per step [ms] $\downarrow$ \\
\midrule
SWD &  0.733 & \best{1.03} \\
\cite{nguyen2024slicedcv} (LCV) & 0.735 & 1.81 \\
\cite{nguyen2024slicedcv} (UCV) & 0.726 & 2.19\\
\cite{nguyen2024quasi} (QMC) & \secondbest{0.670} & \secondbest{1.51} \\
\midrule
\ours & \best{0.622} & 1.92 \\
\end{tabular}
}
\end{minipage}
\hfill
\begin{minipage}[t]{.4\textwidth}
    \titlecaptionof{table}{Influence of fresh candidates}{With a fixed budget of 64 projections, we analyze the influence of the fresh candidates.
}
\label{tab:candidate_1d}
\centering
\resizebox{0.95\linewidth}{!}{ %
\begin{tabular}{lcc}
\# Candidates & Mean-W$_1$ [$10^3$] $\downarrow$ & Time per step [ms] $\downarrow$ \\
\midrule
2 & 0.721 & 1.99 \\
4 & \secondbest{0.673} & 1.96  \\
\textbf{8} & \best{0.622} & \secondbest{1.92} \\
16 & 0.746 & \best{1.91} \\
32 & 1.192 & 1.98 \\
48 & 2.122 & 1.93 \\
56 & 3.811 & 1.85 \\
\end{tabular}
}

\end{minipage}
\\[1em]
\begin{minipage}[t]{.48\textwidth}
    \centering
\begin{tikzpicture}
\begin{axis}[
    width=\linewidth,
    height=0.75\linewidth,
    xlabel={Step},
    ylabel={Mean W$_1$ Distance},
    ymode=log,
    grid=major,
    legend style={at={(0.5,-0.3)}, anchor=north, legend columns=2},
    thick,
]

% -- Plot definitions: {filename}{legend label}
% \addplot table [x=step, y=mean_w1, col sep=comma] {metrics_vs_step_methodA.csv};
% \addlegendentry{Method A}

% \addplot table [x=step, y=mean_w1, col sep=comma] {metrics_vs_step_methodB.csv};
% \addlegendentry{Method B (Improved)}

\addplot+[no marks] table [x=step, y=mean_w1, col sep=comma] {\detokenize{figures/general_matching/metrics_vs_step_nc-0_dist-l1_mode-gaussian_ucv-False_lcv-False.csv}};
\addlegendentry{SWD}

\addplot+[no marks] table [x=step, y=mean_w1, col sep=comma] {\detokenize{figures/general_matching/metrics_vs_step_nc-8_dist-l1_mode-gaussian_ucv-False_lcv-False.csv}};
\addlegendentry{\ours}

\end{axis}
\end{tikzpicture}
\titlecaptionof{figure}{True Wasserstein metric over steps}{
%When comparing with the true Wasserstein distance at each optimization, the effect of the reservoir warmup is evident. Initially, our method performs slightly worse due to lower final projections in the loss, but our method can outperform SWD in the end.
The effect of reservoir warmup is clear when comparing \ours with the true Wasserstein distance during optimization. Initially, \ours performs slightly worse due to lower projections in the loss, but can outperform SWD in the end.
}
\label{fig:w1_vs_steps}
\end{minipage}
\hfill
\begin{minipage}[t]{.48\textwidth}
    \centering
\begin{tikzpicture}
\begin{axis}[
    width=\linewidth,
    height=0.75\linewidth,
    xlabel={Step},
    ylabel={Correlation $\rho$(SWD, W$_1$)},
    % ymode=log,
    grid=major,
    legend style={at={(0.5,-0.3)}, anchor=north, legend columns=2},
    thick,
]

% -- Plot definitions: {filename}{legend label}
% \addplot table [x=step, y=mean_w1, col sep=comma] {metrics_vs_step_methodA.csv};
% \addlegendentry{Method A}

% \addplot table [x=step, y=mean_w1, col sep=comma] {metrics_vs_step_methodB.csv};
% \addlegendentry{Method B (Improved)}

\addplot+[no marks] table [x=step, y=pearson_corr, col sep=comma] {\detokenize{figures/general_matching/metrics_vs_step_nc-0_dist-l1_mode-gaussian_ucv-False_lcv-False.csv}};
\addlegendentry{SWD}

% \addplot+[no marks] table [x=step, y=mean_w1, col sep=comma] {\detokenize{figures/general_matching/metrics_vs_step_nc-0_dist-l1_mode-gaussian_ucv-True_lcv-False.csv}};
% \addlegendentry{SWD+UCV}

% \addplot+[no marks] table [x=step, y=mean_w1, col sep=comma] {\detokenize{figures/general_matching/metrics_vs_step_nc-0_dist-l1_mode-gaussian_ucv-False_lcv-True.csv}};
% \addlegendentry{SWD+LCV}

% \addplot+[no marks] table [x=step, y=mean_w1, col sep=comma] {\detokenize{figures/general_matching/metrics_vs_step_nc-0_dist-l1_mode-qmc_ucv-False_lcv-False.csv}};
% \addlegendentry{SWD+QMC}

\addplot+[no marks] table [x=step, y=pearson_corr, col sep=comma] {\detokenize{figures/general_matching/metrics_vs_step_nc-8_dist-l1_mode-gaussian_ucv-False_lcv-False.csv}};
\addlegendentry{\ours}

% \addplot+[no marks] table [x=step, y=mean_w1, col sep=comma] {\detokenize{figures/general_matching/metrics_vs_step_nc-8_dist-l1_mode-qmc_ucv-False_lcv-False.csv}};
% \addlegendentry{Ours+QMC}

\end{axis}
\end{tikzpicture}
\titlecaptionof{figure}{Pearson correlation with the true Wasserstein Distance}{Our method achieves a high correlation with the true Wasserstein loss, while improving upon pure SWD (See \fig{w1_vs_steps}). This indicates the unbiased nature of our proposed method.}
\label{fig:corr_vs_steps}
\end{minipage}
\end{figure}

\inlinesection{General Distribution Matching.} %
For a general test of our proposed method we create 1000 $d=3$ distribution pairs (normal, uniform, bimodal normal) and align them based on 1024 samples in 300 steps. All methods leverage 64 projections. 
In \Table{comparison_1d}, our method clearly outperforms previous SWD techniques with a slight additional performance cost.
To evaluate the optimization behavior of our method, we calculate the mean W$_1$ true Wasserstein score for each dimension in each step. 
This allows us to plot the convergence behavior of our method in \fig{w1_vs_steps}. 
Here, it is evident that pure SWD and our ReSTIR-based modification have a similar trajectory, but building the reservoir initially results in slightly slower trajectory, which produces better results after roughly 140 steps. 
This also allows us to investigate the correlation between our proposed loss and the true Wasserstein distance. 
In \fig{corr_vs_steps}, our method achieves a high correlation with the true loss, indicating the unbiased nature of our method.

\begin{figure*}[htb] 
\centering
\setlength{\tabcolsep}{3pt}
\renewcommand{\arraystretch}{0.8} 
\begin{tabular}{@{}p{0.05cm}*{4}{m{3.02cm}}@{}}
\rotatebox[origin=c]{90}{\fontsize{4pt}{4.2pt}\selectfont Source} & 
\includegraphics[width=3.0cm]{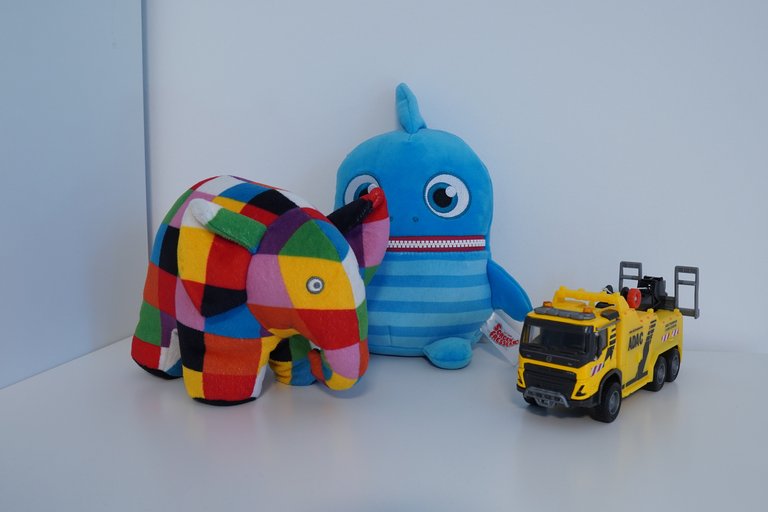} &
\includegraphics[width=3.0cm]{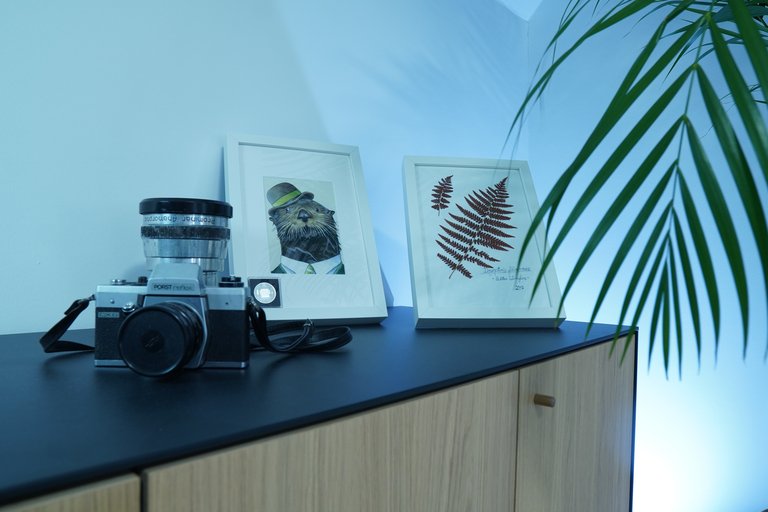} &
\includegraphics[width=3.0cm]{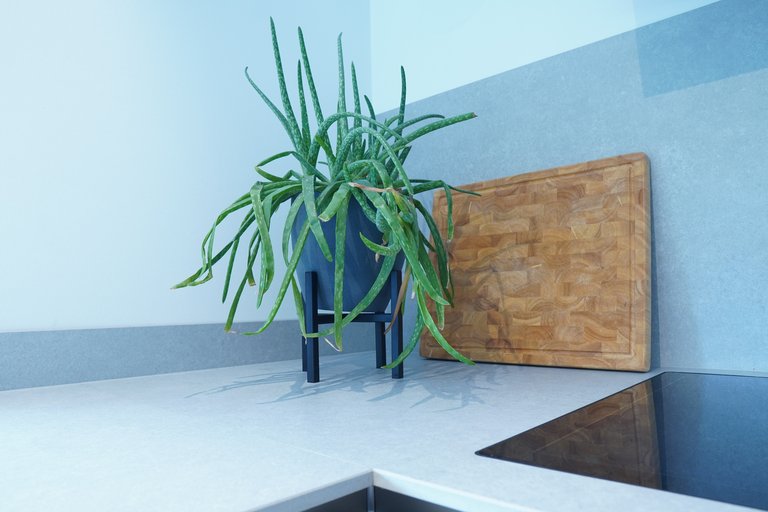} &
\includegraphics[width=3.0cm]{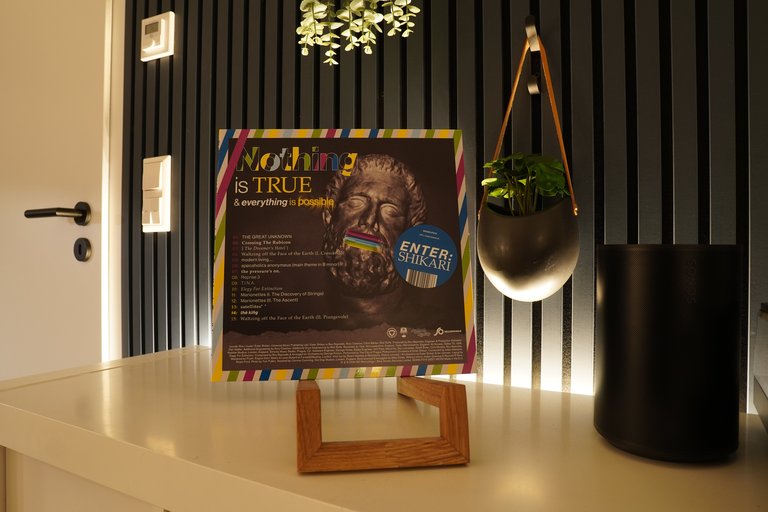} \\

\rotatebox[origin=c]{90}{\fontsize{4pt}{4.2pt}\selectfont Target} & %
\begin{tikzpicture}[spy using outlines={circle, magnification=2, connect spies}, inner sep=0.0cm]
    \node {\includegraphics[width=3.0cm]{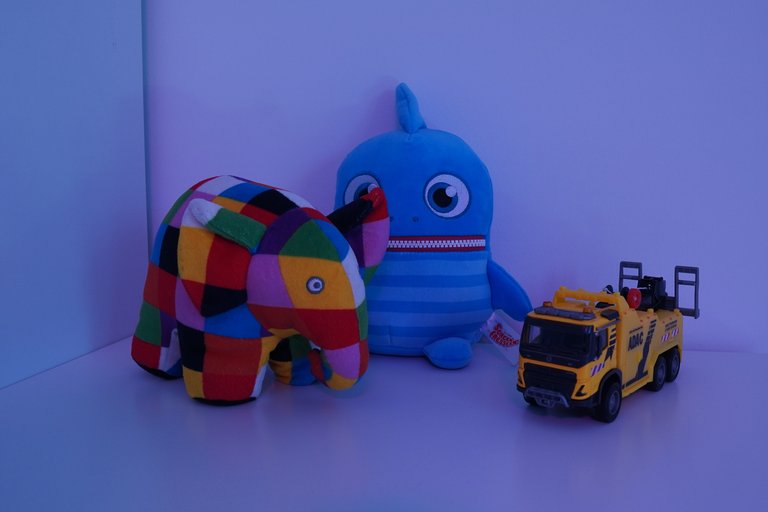}};
    \spy[size=0.75cm] on (-0.5, -0.45) in node at (-1.1, 0.6);
    \spy[size=0.75cm] on (0.95, -0.225) in node at (1.1, 0.6);
\end{tikzpicture} &
\begin{tikzpicture}[spy using outlines={circle, magnification=2, connect spies}, inner sep=0.0cm]
    \node {\includegraphics[width=3.0cm]{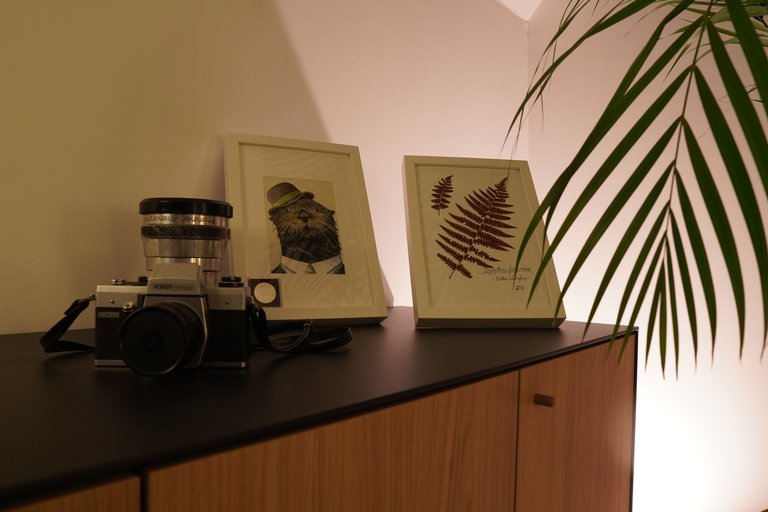}};
    \spy[size=0.75cm] on (-0.4, -0.75) in node at (-1.1, 0.6);
    \spy[size=0.75cm] on (-0.55, 0.5) in node at (1.1, 0.6);
\end{tikzpicture} &
\begin{tikzpicture}[spy using outlines={circle, magnification=2, connect spies}, inner sep=0.0cm]
    \node {\includegraphics[width=3.0cm]{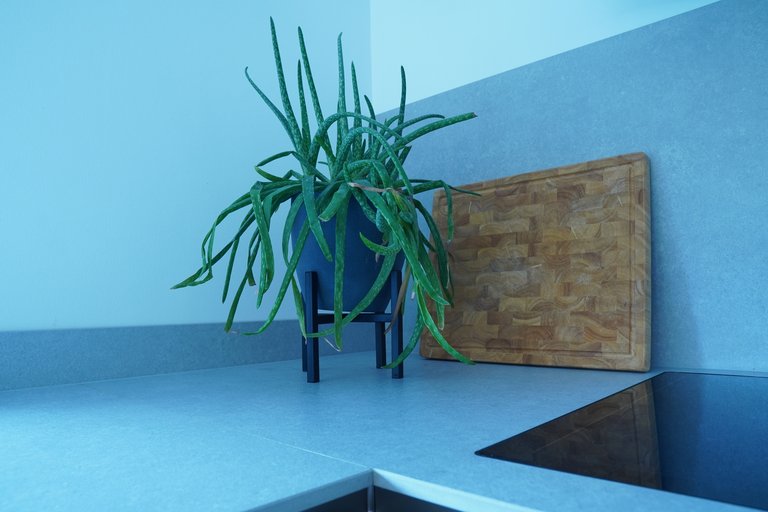}};
    \spy[size=0.75cm] on (-0.25, 0.2) in node at (-1.1, 0.6);
    \spy[size=0.75cm] on (1.0, -0.45) in node at (1.1, 0.6);
\end{tikzpicture} %
&
\begin{tikzpicture}[spy using outlines={circle, magnification=2, connect spies}, inner sep=0.0cm]
    \node {\includegraphics[width=3.0cm]{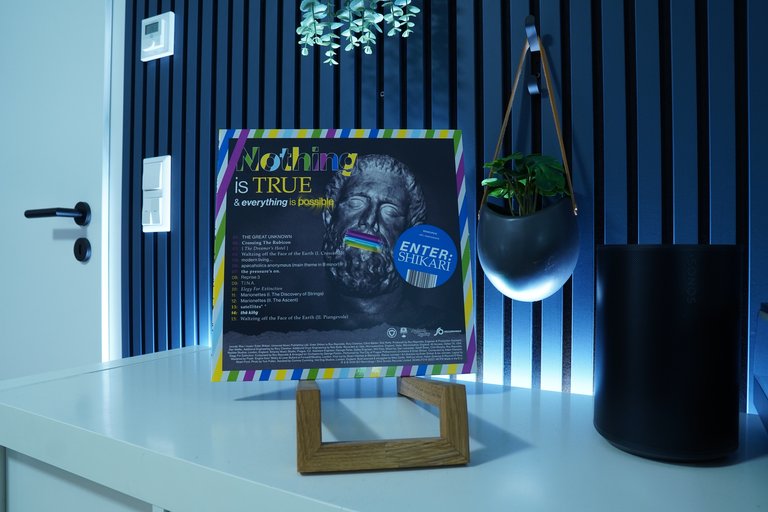}};
    \spy[size=0.75cm] on (-0.65, 0.4) in node at (-1.1, 0.6);
    \spy[size=0.75cm] on (-0.35, -0.5) in node at (1.1, 0.6);
\end{tikzpicture} \\
\midrule

%%%
%Reinhard
%%%

\rotatebox[origin=c]{90}{\fontsize{4pt}{4.2pt}\selectfont\citet{reinhard2001color}} & 
\begin{tikzpicture}[spy using outlines={circle, magnification=2, connect spies}, inner sep=0.0cm]
    \node {\includegraphics[width=3.0cm]{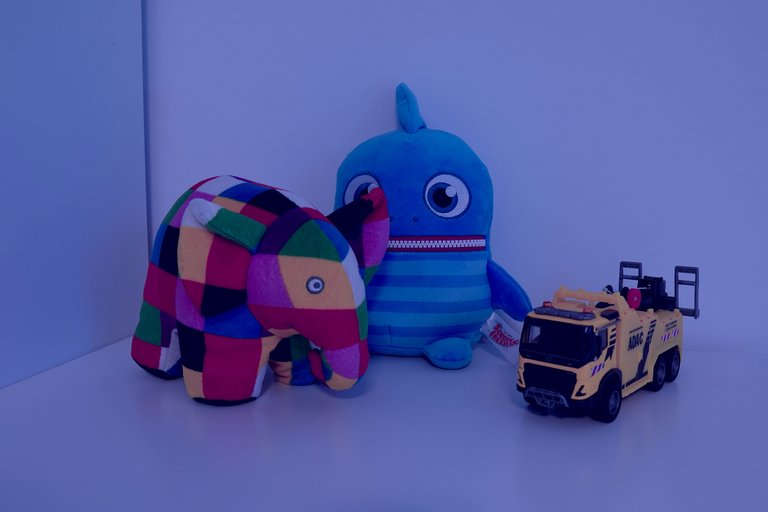}};
    \spy[size=0.75cm] on (-0.5, -0.45) in node at (-1.1, 0.6);
    \spy[size=0.75cm] on (0.95, -0.225) in node at (1.1, 0.6);
\end{tikzpicture} &
\begin{tikzpicture}[spy using outlines={circle, magnification=2, connect spies}, inner sep=0.0cm]
    \node {\includegraphics[width=3.0cm]{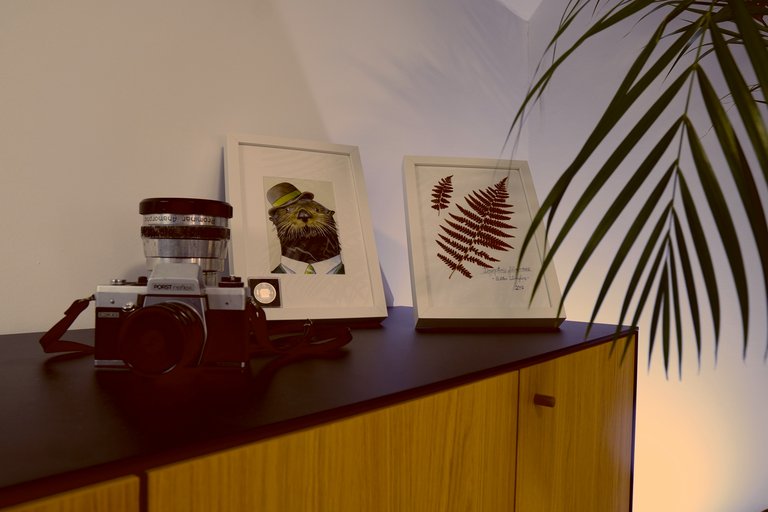}};
    \spy[size=0.75cm] on (-0.4, -0.75) in node at (-1.1, 0.6);
    \spy[size=0.75cm] on (-0.55, 0.5) in node at (1.1, 0.6);
\end{tikzpicture} &
\begin{tikzpicture}[spy using outlines={circle, magnification=2, connect spies}, inner sep=0.0cm]
    \node {\includegraphics[width=3.0cm]{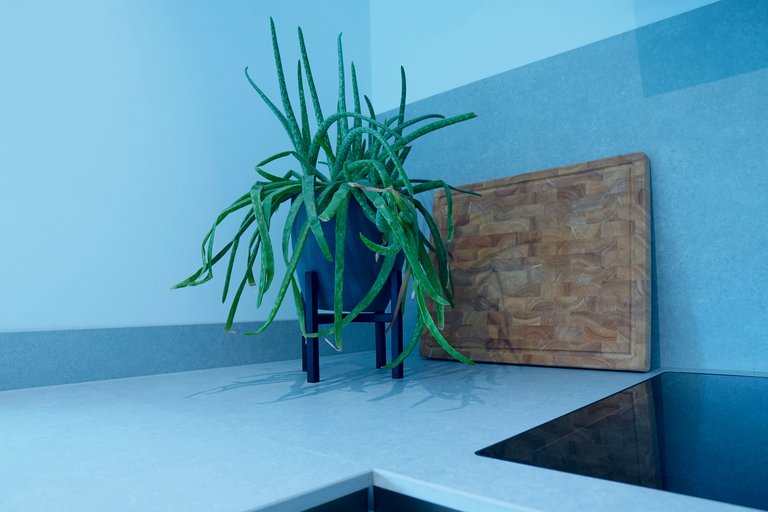}};
    \spy[size=0.75cm] on (-0.25, 0.2) in node at (-1.1, 0.6);
    \spy[size=0.75cm] on (1.0, -0.45) in node at (1.1, 0.6);
\end{tikzpicture} %
&
\begin{tikzpicture}[spy using outlines={circle, magnification=2, connect spies}, inner sep=0.0cm]
    \node {\includegraphics[width=3.0cm]{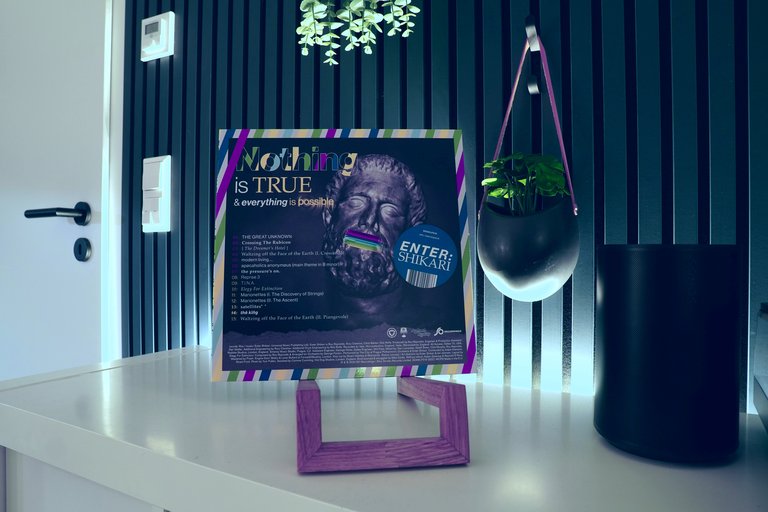}};
    \spy[size=0.75cm] on (-0.65, 0.4) in node at (-1.1, 0.6);
    \spy[size=0.75cm] on (-0.35, -0.5) in node at (1.1, 0.6);
\end{tikzpicture} \\

%%%%%
% Nguyen
%%%%%
\rotatebox[origin=c]{90}{\fontsize{4pt}{4.2pt}\selectfont\citet{nguyenIlluminantAwareGamutBased2014}} & 
\begin{tikzpicture}[spy using outlines={circle, magnification=2, connect spies}, inner sep=0.0cm]
    \node {\includegraphics[width=3.0cm]{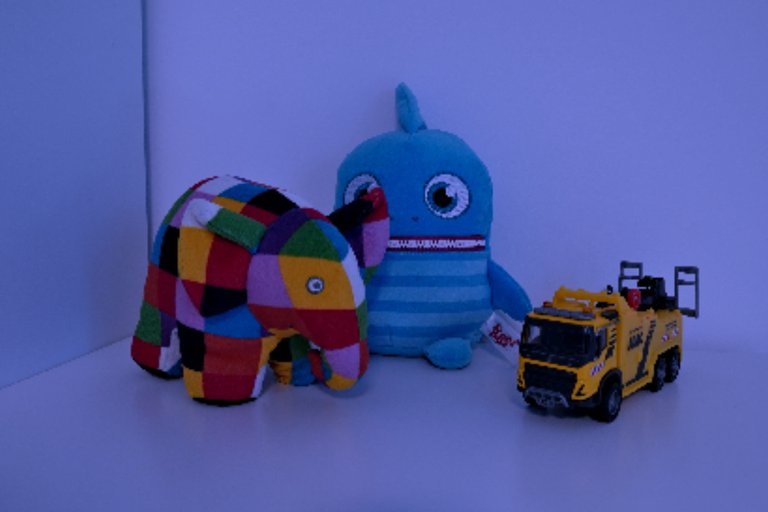}};
    \spy[size=0.75cm] on (-0.5, -0.45) in node at (-1.1, 0.6);
    \spy[size=0.75cm] on (0.95, -0.225) in node at (1.1, 0.6);
\end{tikzpicture} &
\begin{tikzpicture}[spy using outlines={circle, magnification=2, connect spies}, inner sep=0.0cm]
    \node {\includegraphics[width=3.0cm]{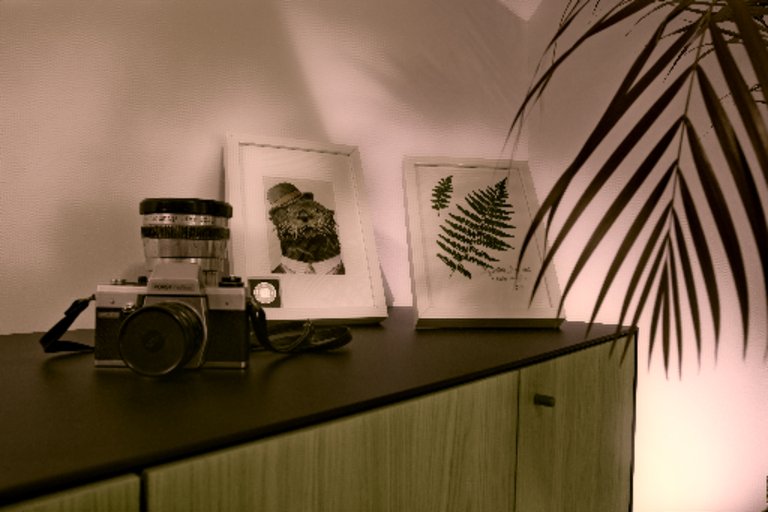}};
    \spy[size=0.75cm] on (-0.4, -0.75) in node at (-1.1, 0.6);
    \spy[size=0.75cm] on (-0.55, 0.5) in node at (1.1, 0.6);
\end{tikzpicture} &
\begin{tikzpicture}[spy using outlines={circle, magnification=2, connect spies}, inner sep=0.0cm]
    \node {\includegraphics[width=3.0cm]{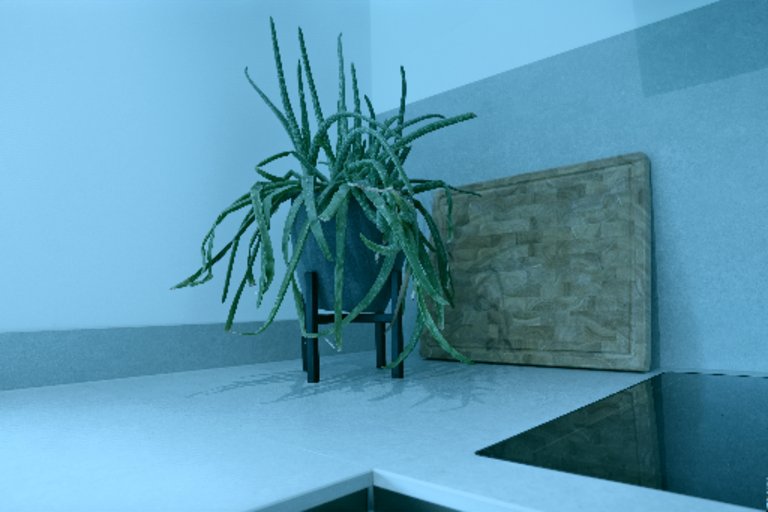}};
    \spy[size=0.75cm] on (-0.25, 0.2) in node at (-1.1, 0.6);
    \spy[size=0.75cm] on (1.0, -0.45) in node at (1.1, 0.6);
\end{tikzpicture} %
&
\begin{tikzpicture}[spy using outlines={circle, magnification=2, connect spies}, inner sep=0.0cm]
    \node {\includegraphics[width=3.0cm]{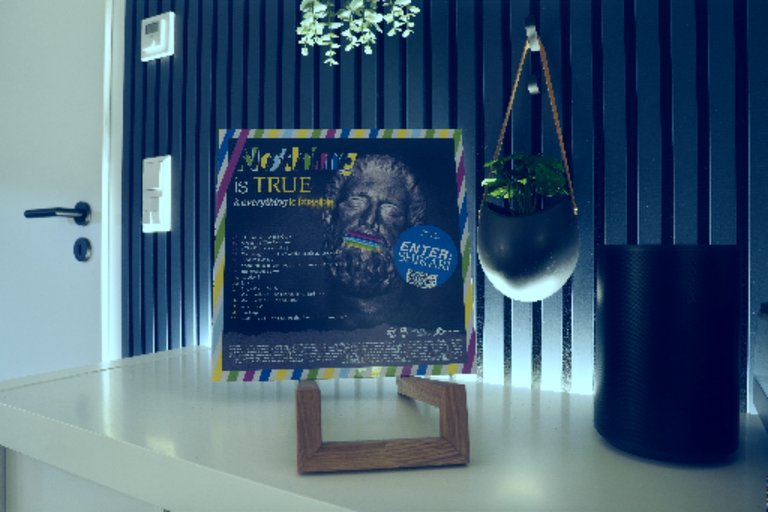}};
    \spy[size=0.75cm] on (-0.65, 0.4) in node at (-1.1, 0.6);
    \spy[size=0.75cm] on (-0.35, -0.5) in node at (1.1, 0.6);
\end{tikzpicture} \\

%%%%%%
% Larchenko
%%%%%%

\rotatebox[origin=c]{90}{\fontsize{4pt}{4.2pt}\selectfont\citet{larchenko_color_2025}} & 
\begin{tikzpicture}[spy using outlines={circle, magnification=2, connect spies}, inner sep=0.0cm]
    \node {\includegraphics[width=3.0cm]{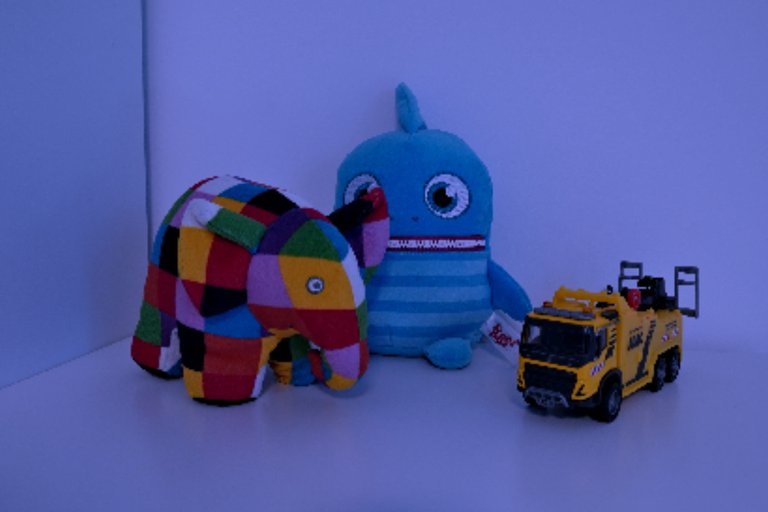}};
    \spy[size=0.75cm] on (-0.5, -0.45) in node at (-1.1, 0.6);
    \spy[size=0.75cm] on (0.95, -0.225) in node at (1.1, 0.6);
\end{tikzpicture} &
\begin{tikzpicture}[spy using outlines={circle, magnification=2, connect spies}, inner sep=0.0cm]
    \node {\includegraphics[width=3.0cm]{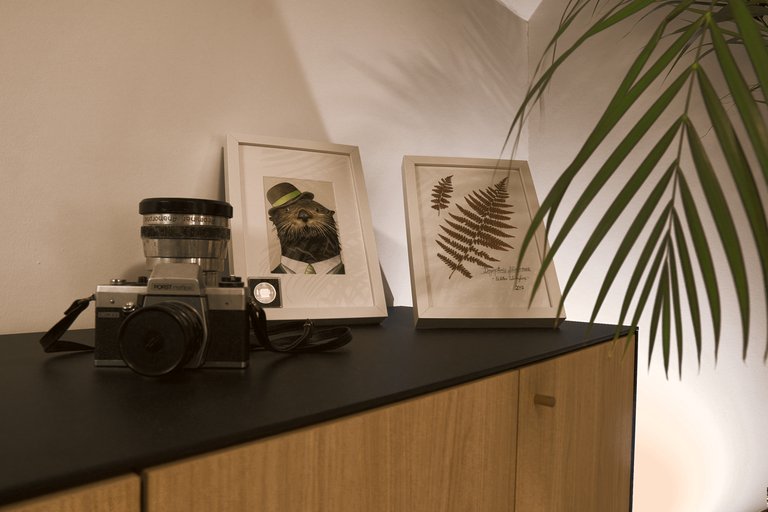}};
    \spy[size=0.75cm] on (-0.4, -0.75) in node at (-1.1, 0.6);
    \spy[size=0.75cm] on (-0.55, 0.5) in node at (1.1, 0.6);
\end{tikzpicture} &
\begin{tikzpicture}[spy using outlines={circle, magnification=2, connect spies}, inner sep=0.0cm]
    \node {\includegraphics[width=3.0cm]{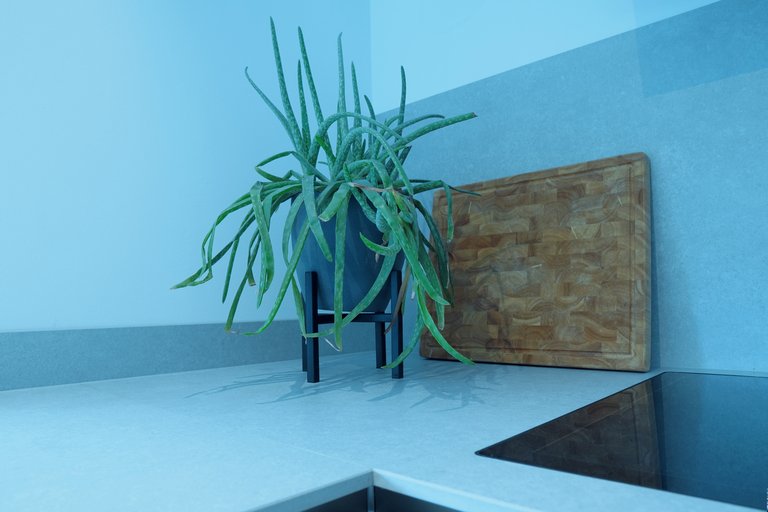}};
    \spy[size=0.75cm] on (-0.25, 0.2) in node at (-1.1, 0.6);
    \spy[size=0.75cm] on (1.0, -0.45) in node at (1.1, 0.6);
\end{tikzpicture} %
&
\begin{tikzpicture}[spy using outlines={circle, magnification=2, connect spies}, inner sep=0.0cm]
    \node {\includegraphics[width=3.0cm]{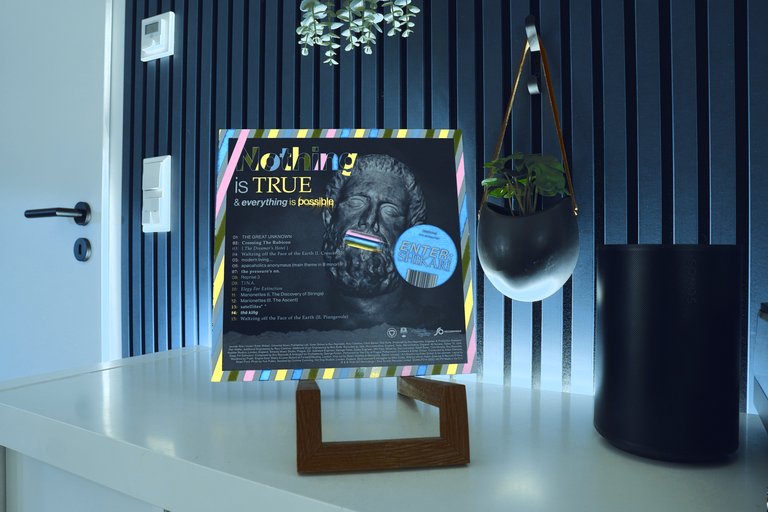}};
    \spy[size=0.75cm] on (-0.65, 0.4) in node at (-1.1, 0.6);
    \spy[size=0.75cm] on (-0.35, -0.5) in node at (1.1, 0.6);
\end{tikzpicture} \\

%%%%%%
% Ours W/o SWD
%%%%%

% \rotatebox[origin=c]{90}{\fontsize{4pt}{4.2pt}\selectfont Ours SWD} & 
% \begin{tikzpicture}[spy using outlines={circle, magnification=3, connect spies}, inner sep=0.0cm]
%     \node {\includegraphics[width=3.0cm]{figures/color_matching/images/dual_illum/toys_converted/vector-swd_no-wrs/Toys1-Illum2_adjusted.jpg}};
%     \spy[size=0.75cm] on (-0.5, -0.25) in node at (-1.1, 0.6);
%     \spy[size=0.75cm] on (0.95, -0.225) in node at (1.1, 0.6);
% \end{tikzpicture} %
% &
% % \includegraphics[width=3.0cm]{figures/color_matching/images/dual_illum/brunsviga_converted/vector-swd/Brunsviga-Illum1_adjusted.jpg} &
% \includegraphics[width=3.0cm]{figures/color_matching/images/dual_illum/frame_converted/vector-swd_no-wrs/Frames-Illum2_adjusted.jpg} &
% \includegraphics[width=3.0cm]{figures/color_matching/images/dual_illum/plant_converted/vector-swd_no-wrs/Plant-Illum1_adjusted.jpg} &
% % \includegraphics[width=3.0cm]{figures/color_matching/images/dual_illum/toys3_converted/vector-swd_no-wrs/Toys3-Illum2_adjusted.jpg} &
% \includegraphics[width=3.0cm]{figures/color_matching/images/dual_illum/vinyl_converted/vector-swd_no-wrs/Record-Illum2_adjusted.jpg} \\

%%%%%
% Ours
%%%%%

\rotatebox[origin=c]{90}{\fontsize{4pt}{4.2pt}\selectfont\textbf{\ours}} & 
\begin{tikzpicture}[spy using outlines={circle, magnification=2, connect spies}, inner sep=0.0cm]
    \node {\includegraphics[width=3.0cm]{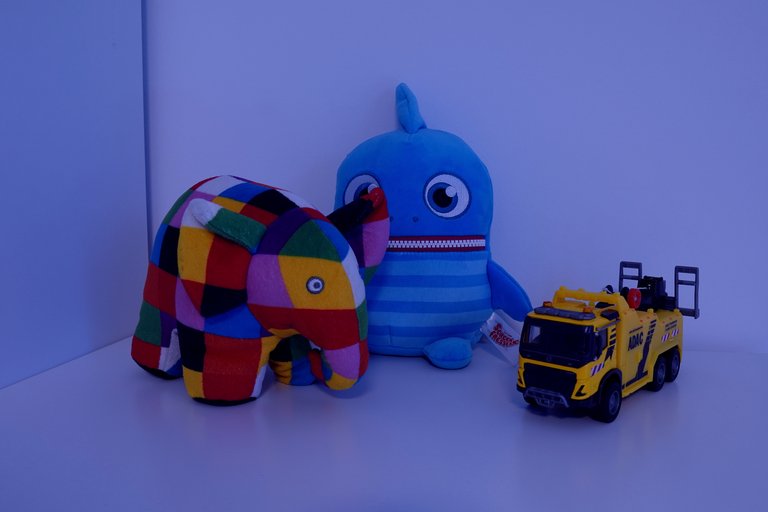}};
    \spy[size=0.75cm] on (-0.5, -0.45) in node at (-1.1, 0.6);
    \spy[size=0.75cm] on (0.95, -0.25) in node at (1.1, 0.6);
\end{tikzpicture} &
\begin{tikzpicture}[spy using outlines={circle, magnification=2, connect spies}, inner sep=0.0cm]
    \node {\includegraphics[width=3.0cm]{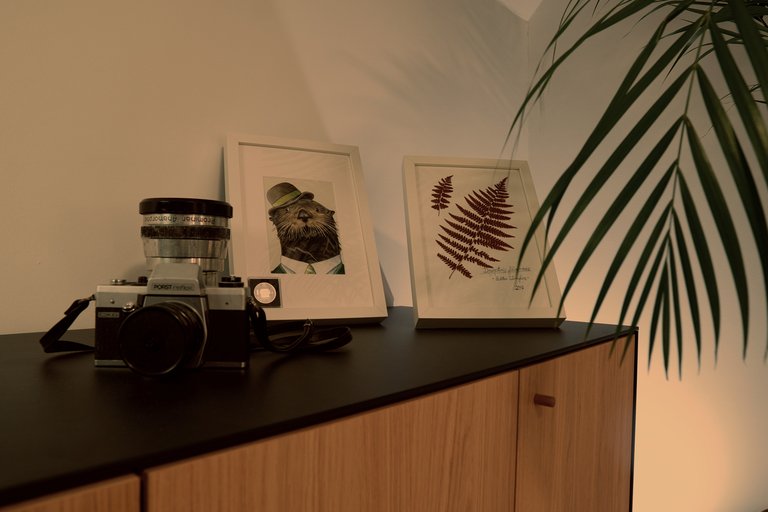}};
    \spy[size=0.75cm] on (-0.4, -0.75) in node at (-1.1, 0.6);
    \spy[size=0.75cm] on (-0.55, 0.5) in node at (1.1, 0.6);
\end{tikzpicture} &
\begin{tikzpicture}[spy using outlines={circle, magnification=2, connect spies}, inner sep=0.0cm]
    \node {\includegraphics[width=3.0cm]{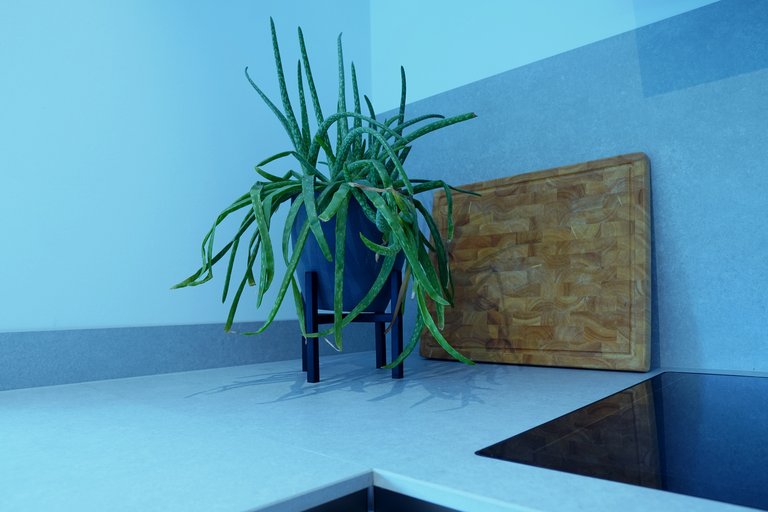}};
    \spy[size=0.75cm] on (-0.25, 0.2) in node at (-1.1, 0.6);
    \spy[size=0.75cm] on (1.0, -0.45) in node at (1.1, 0.6);
\end{tikzpicture} %
&
\begin{tikzpicture}[spy using outlines={circle, magnification=2, connect spies}, inner sep=0.0cm]
    \node {\includegraphics[width=3.0cm]{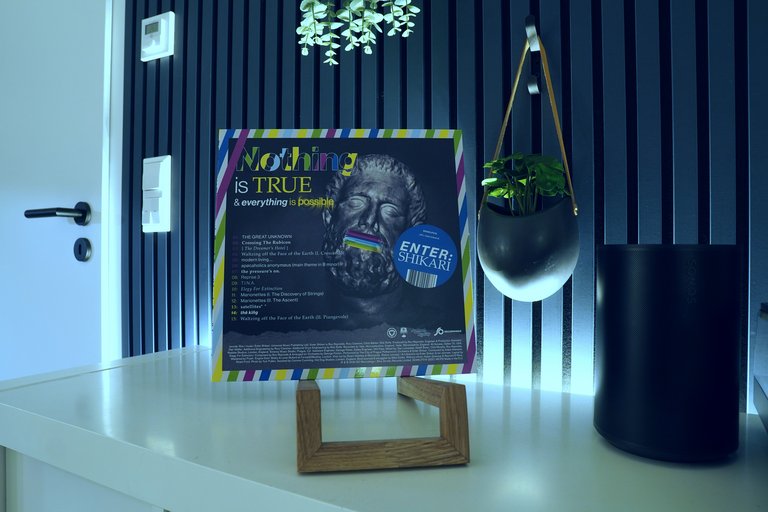}};
    \spy[size=0.75cm] on (-0.65, 0.4) in node at (-1.1, 0.6);
    \spy[size=0.75cm] on (-0.35, -0.5) in node at (1.1, 0.6);
\end{tikzpicture} \\

\end{tabular}
\titlecaption{Color Matching}{Notice the consistent images our method produces with accurate shot matching. \ours consistently produces good results which match the final color distribution and without introducing any artifacts. We highlighted challenging areas which our method handled well. Note that the reference implementation or \citet{larchenko_color_2025} and \citet{nguyenIlluminantAwareGamutBased2014} do not support high resolution images.} 
\label{fig:color_match} 
\end{figure*}

\inlinesection{Color Matching.}
To evaluate the color matching based on our differentiable color correction pipeline, we created a dataset of 10 scenes with two different illumination settings each. For each setting, we also take a photo using a calibration color chart. We match the illumination pairs and compute the following metrics from the color checker data extracted from the additional image pairs: Color data PSNR and RMSE (between adjusted and ground truth color checker images), transform error, i.e. deviation from an identity transform between target and adjusted colors as RMSE as well as an adapted CTQM metric~\citep{CTQMpanettaNovelMulticolorTransfer2016} describing the overall color transfer quality.
We select \citet{reinhard2001color}, \citet{nguyenIlluminantAwareGamutBased2014}, and \citet{larchenko_color_2025} as our baseline comparisons, as well as our proposed method but without the addition of \ours.
Table~\ref{tab:color_matching_dual_illum} shows that our approach achieves the lowest transform error at a competitive runtime.
In \fig{color_match}, we present the visual comparison between the methods. It is also evident that our method outperforms the existing baseline in qualitative results as well.
Additionally, it is worth pointing out that our results using a parametric model also offer real-world advantages because of better pipeline integration and further manual edits being possible.

\begin{table}[t]
\titlecaptionof{table}{Comparison on color matching}{Here, we compare \ours with a baseline SWD method and prior color matching works by ~\citet{reinhard2001color}, \citet{nguyenIlluminantAwareGamutBased2014} and \citet{larchenko_color_2025}. We report the errors between the color charts of adjusted and ground truth images as PSNR, transform and CDL error and the image quality as CTQM. Additionally, we compare the runtimes of the methods.
}
\label{tab:color_matching_dual_illum}
\centering
\resizebox{0.85\linewidth}{!}{ %
\begin{tabular}{lccccc}
Method & {Color PSNR $\uparrow$} & {Transform err. (RMSE) $\downarrow$} & {CDL err. (RMSE) $\downarrow$} & {CTQM $\uparrow$} & {Time per match [s] $\downarrow$}\\
\midrule
\cite{reinhard2001color} &
21.94 & \best{0.31} & 0.14 & 5.12 & \best{1}\\
\cite{nguyenIlluminantAwareGamutBased2014} &
18.76 & 1.27 & 0.33 & \secondbest{5.16} & \secondbest{3} \\
\cite{larchenko_color_2025} &
14.80 & 0.47 & 0.20 & 5.05 & 24 \\
\midrule
% Ours SWD (patches) &
% 24.62(6.27) & \best{0.29(0.48)} & \secondbest{0.10(0.08)} & 5.16(1.82) & 7  \\
% %
% Ours SWD $+$ ReSTIR  (patches)&
% \best{24.70(6.22)} & \secondbest{0.30(0.51)} & \secondbest{0.10(0.08)} & \secondbest{5.17(1.82)} & 7 \\
%
Ours (SWD) &
\secondbest{24.30} & \secondbest{0.34} & \secondbest{0.11} & 5.15 & 5 \\
Ours (\textbf{\ours}) &
\best{24.64} & \best{0.31} & \best{0.10} & \best{5.17} & 5 \\

\end{tabular}
}
\end{table}

\begin{figure*}[htb] 
\centering
\setlength{\tabcolsep}{3pt}
\renewcommand{\arraystretch}{0.8} 
\begin{tabular}{@{}p{0.05cm}*{7}{m{1.72cm}}@{}}
\rotatebox[origin=c]{90}{\tiny{Reference}} & 
\includegraphics[width=1.7cm]{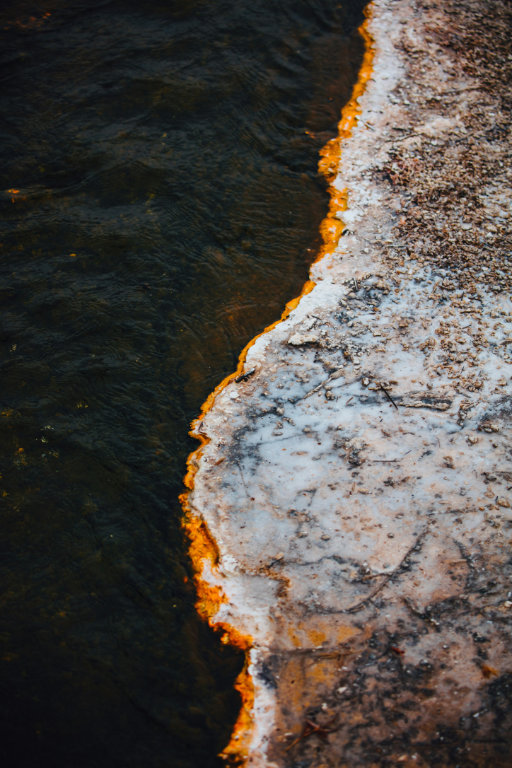} & \includegraphics[width=1.7cm]{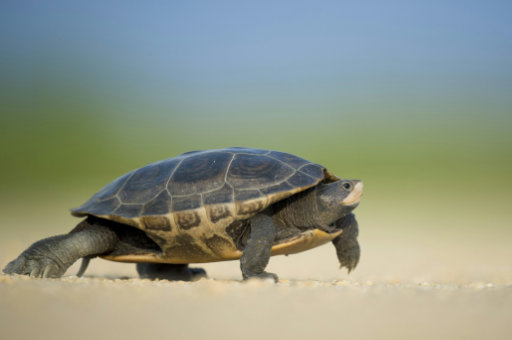} & \includegraphics[width=1.7cm]{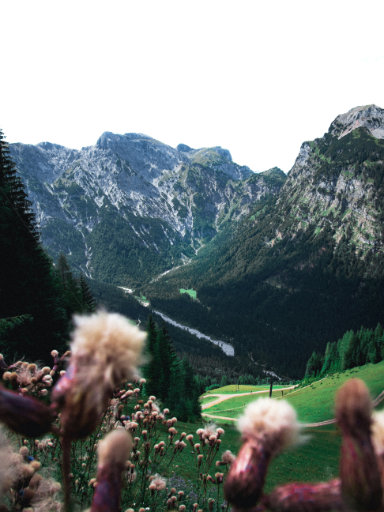} & \includegraphics[width=1.7cm]{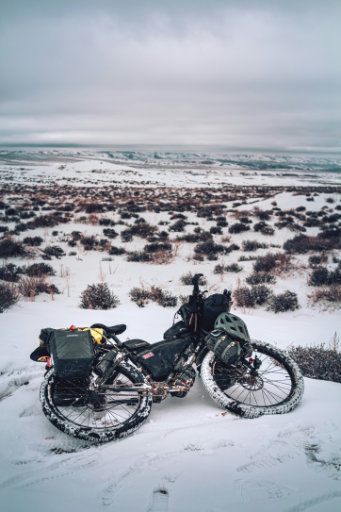} & \includegraphics[width=1.7cm]{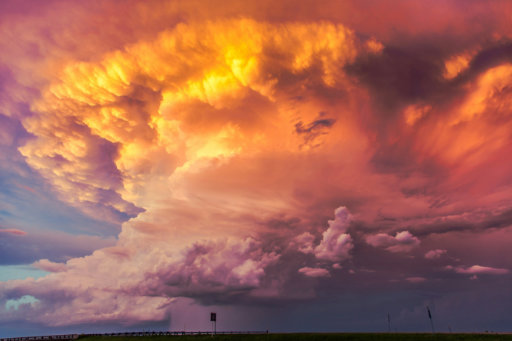} & \includegraphics[width=1.7cm]{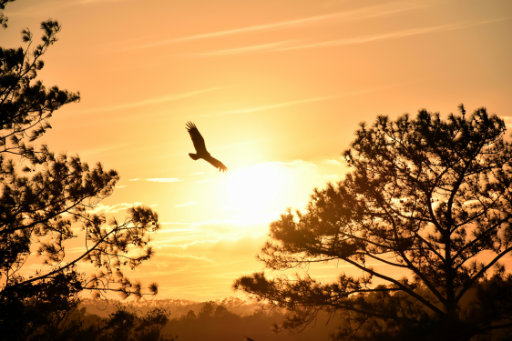} & \includegraphics[width=1.7cm]{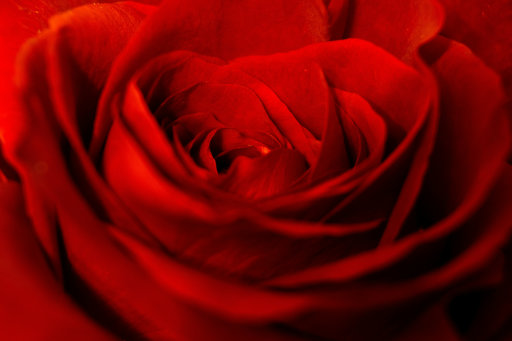} \\ % 
\rotatebox[origin=c]{90}{\tiny{\cite{lobashev2025color}}} & \includegraphics[width=1.7cm]{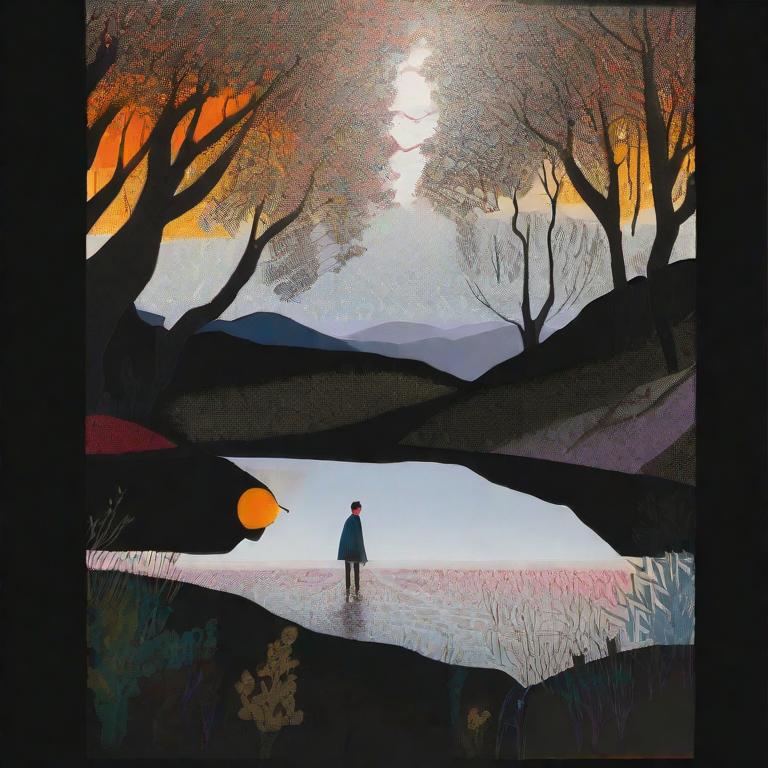} & \includegraphics[width=1.7cm]{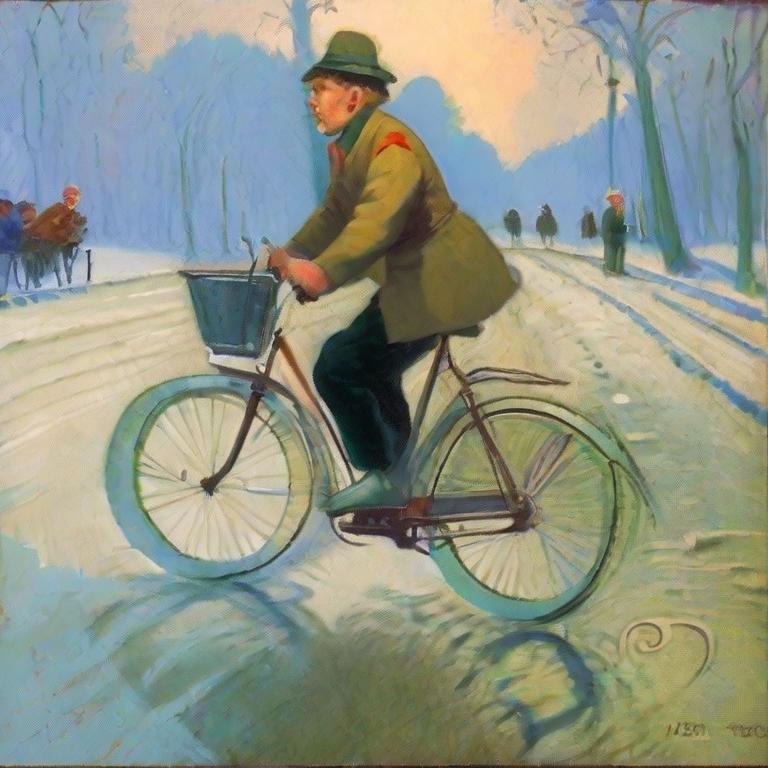} & \includegraphics[width=1.7cm]{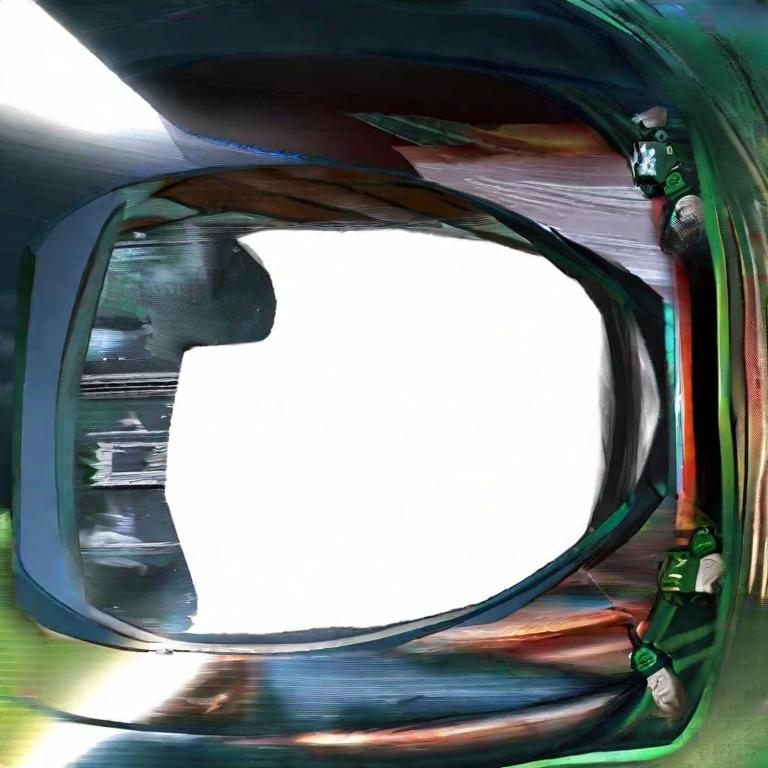} & \includegraphics[width=1.7cm]{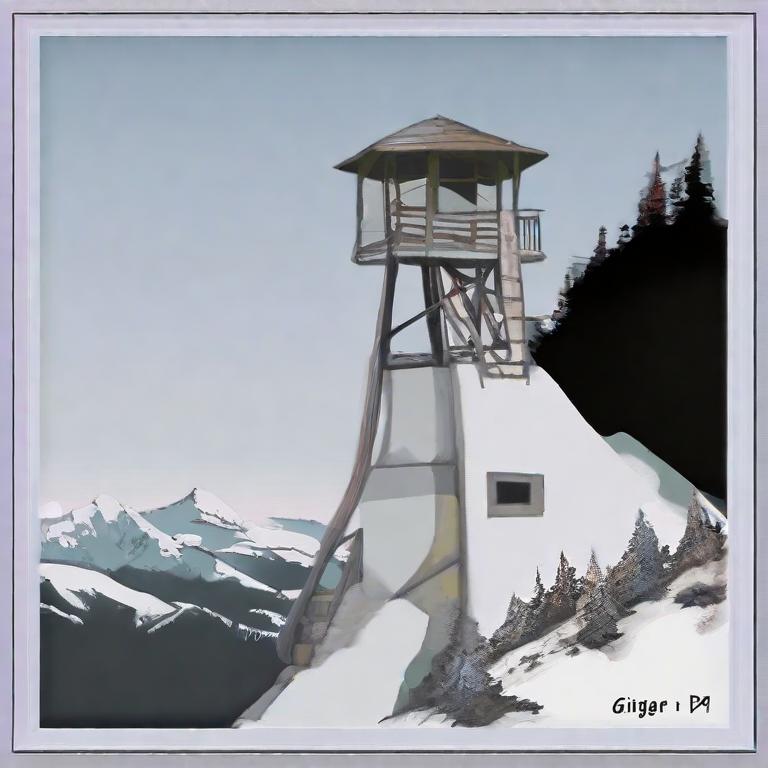} & \includegraphics[width=1.7cm]{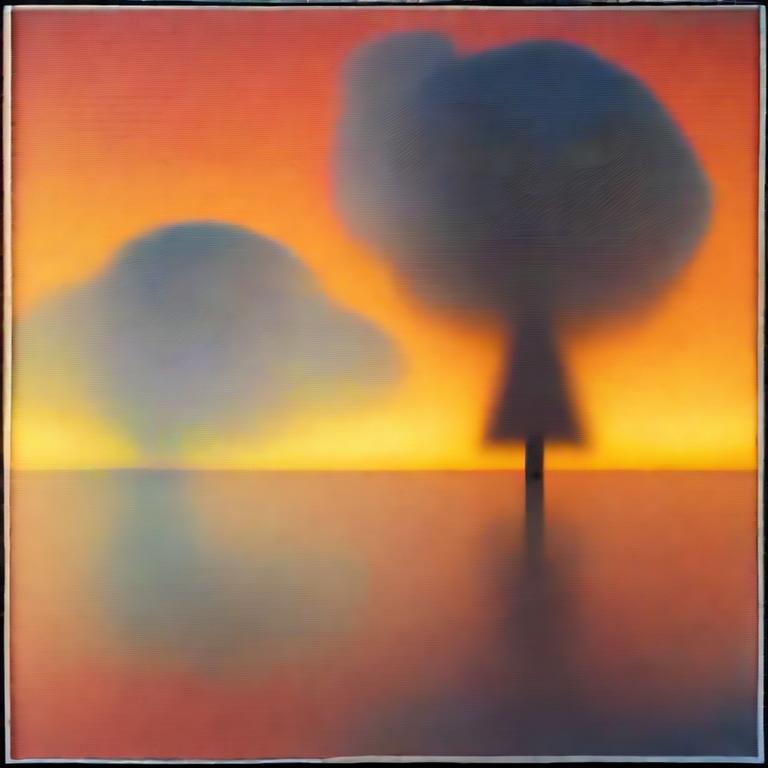} & \includegraphics[width=1.7cm]{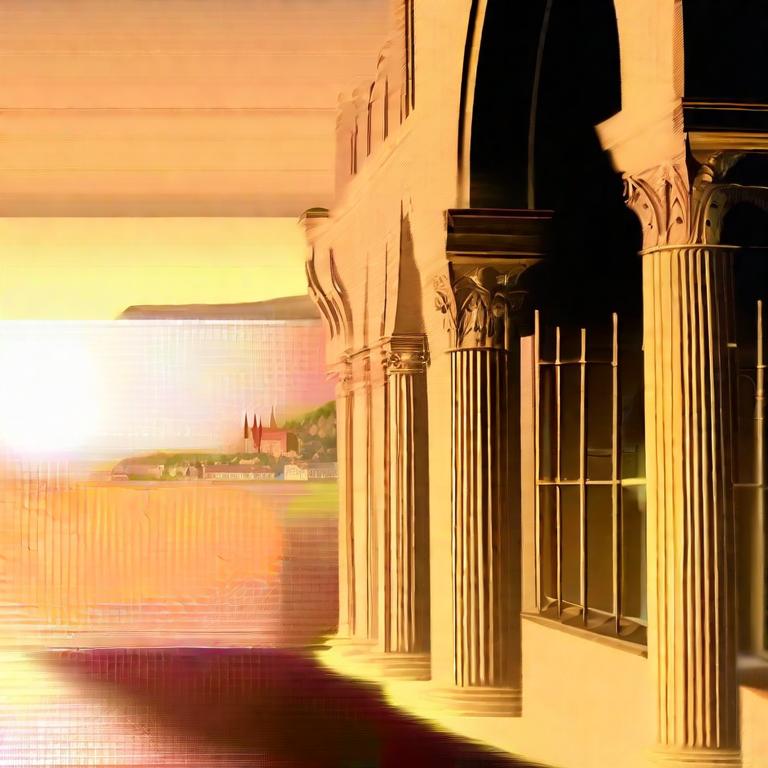} & \includegraphics[width=1.7cm]{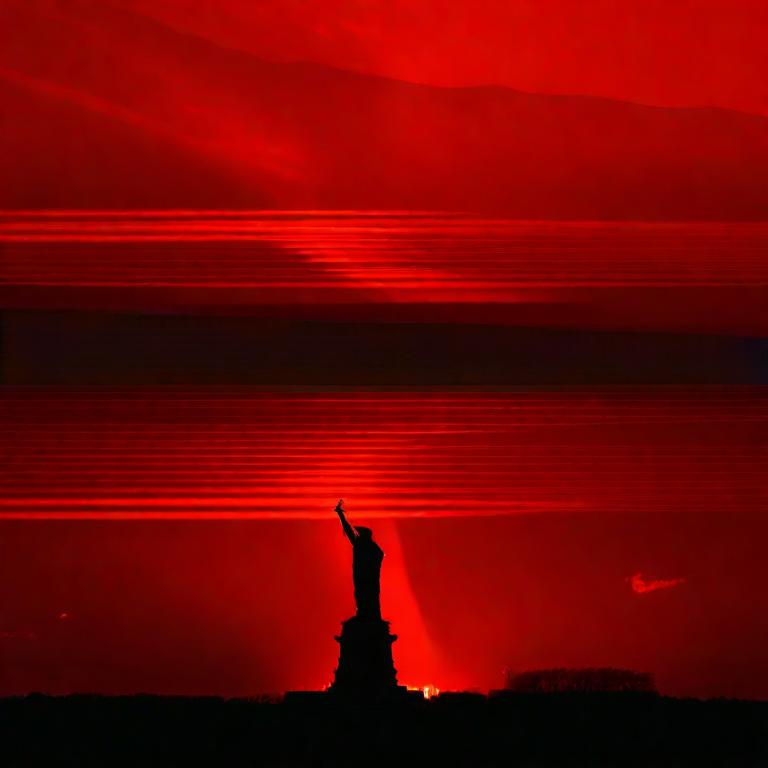} \\ %
\rotatebox[origin=c]{90}{\tiny{\textbf{\ours}}} & \includegraphics[width=1.7cm]{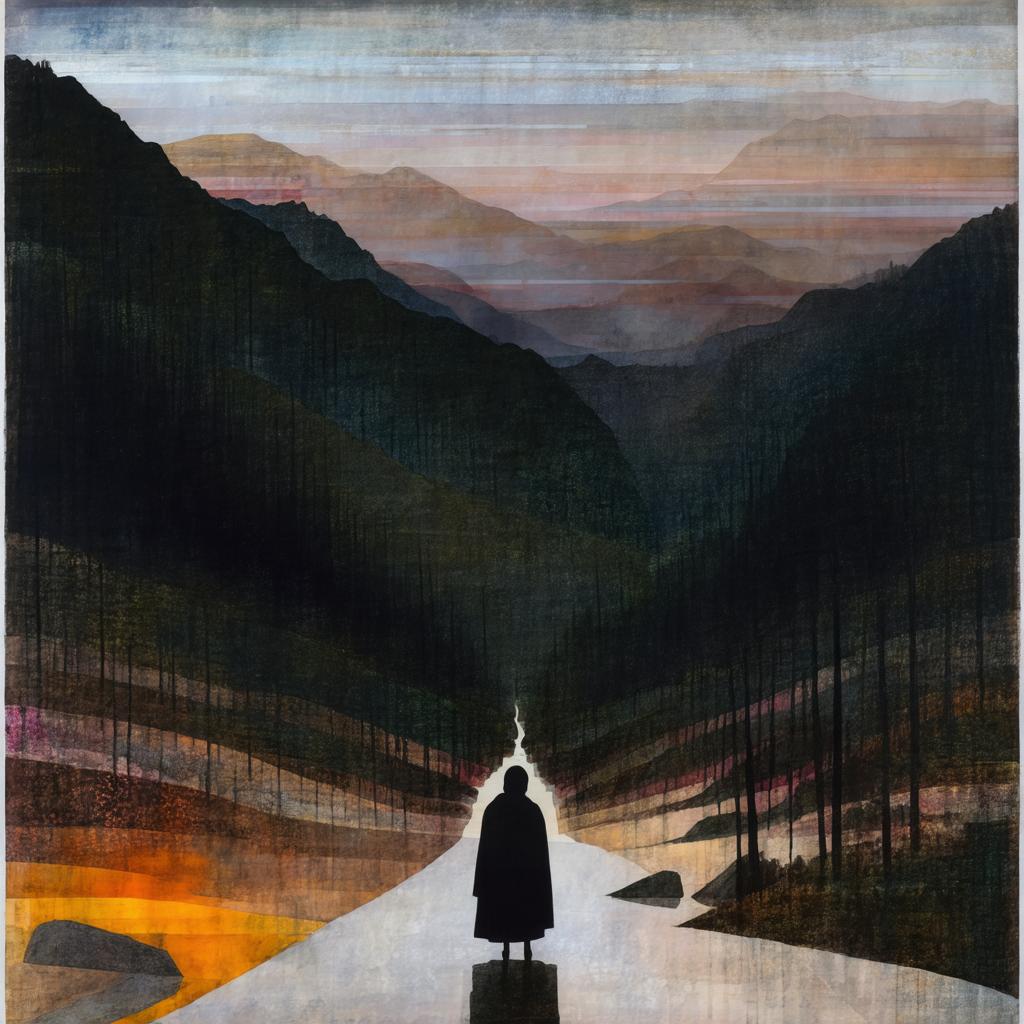} & \includegraphics[width=1.7cm]{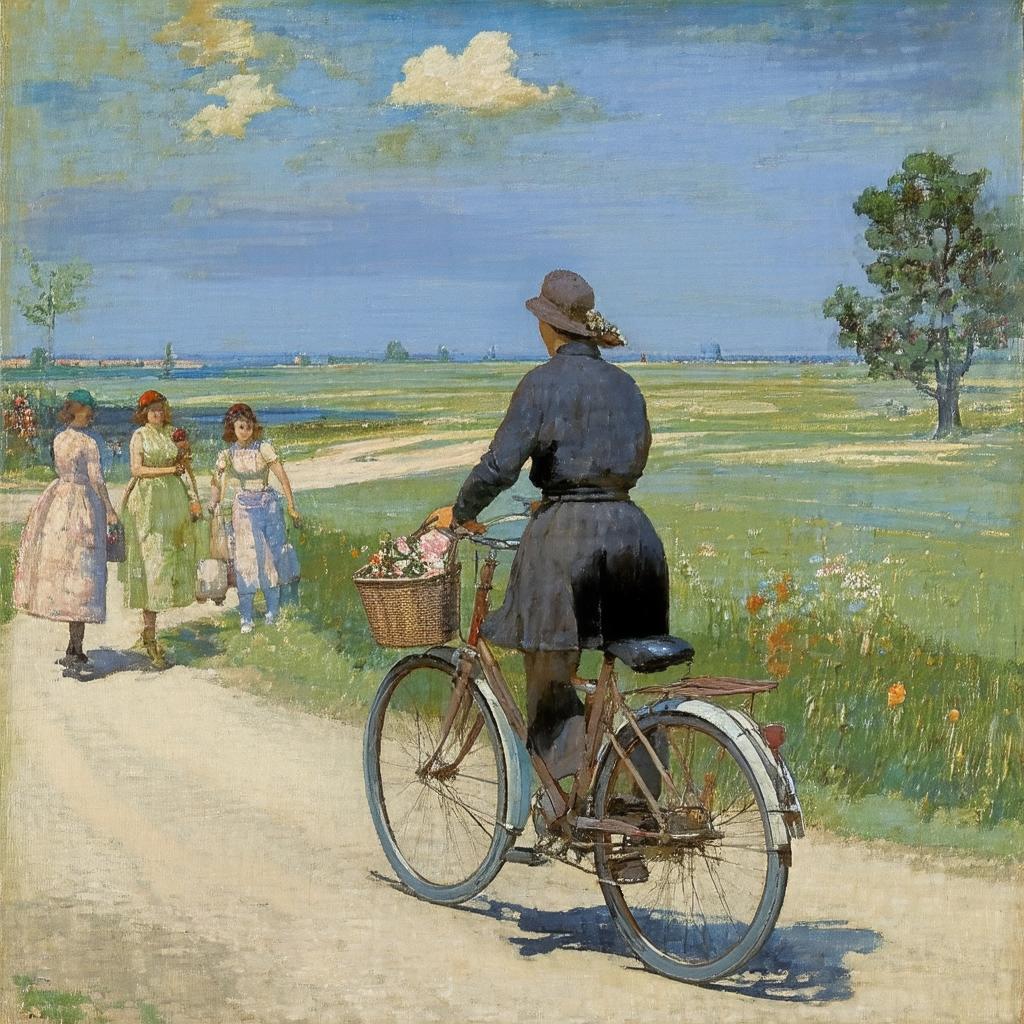} & \includegraphics[width=1.7cm]{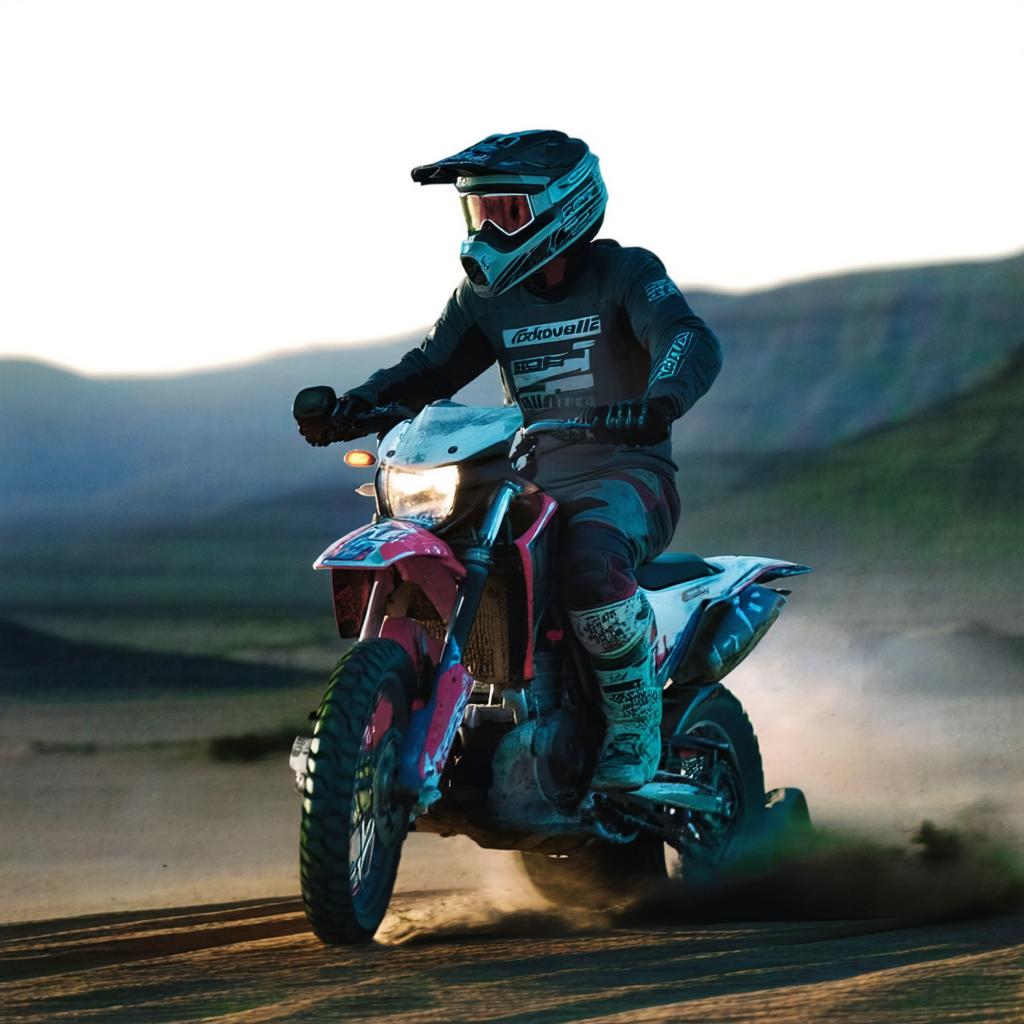} & \includegraphics[width=1.7cm]{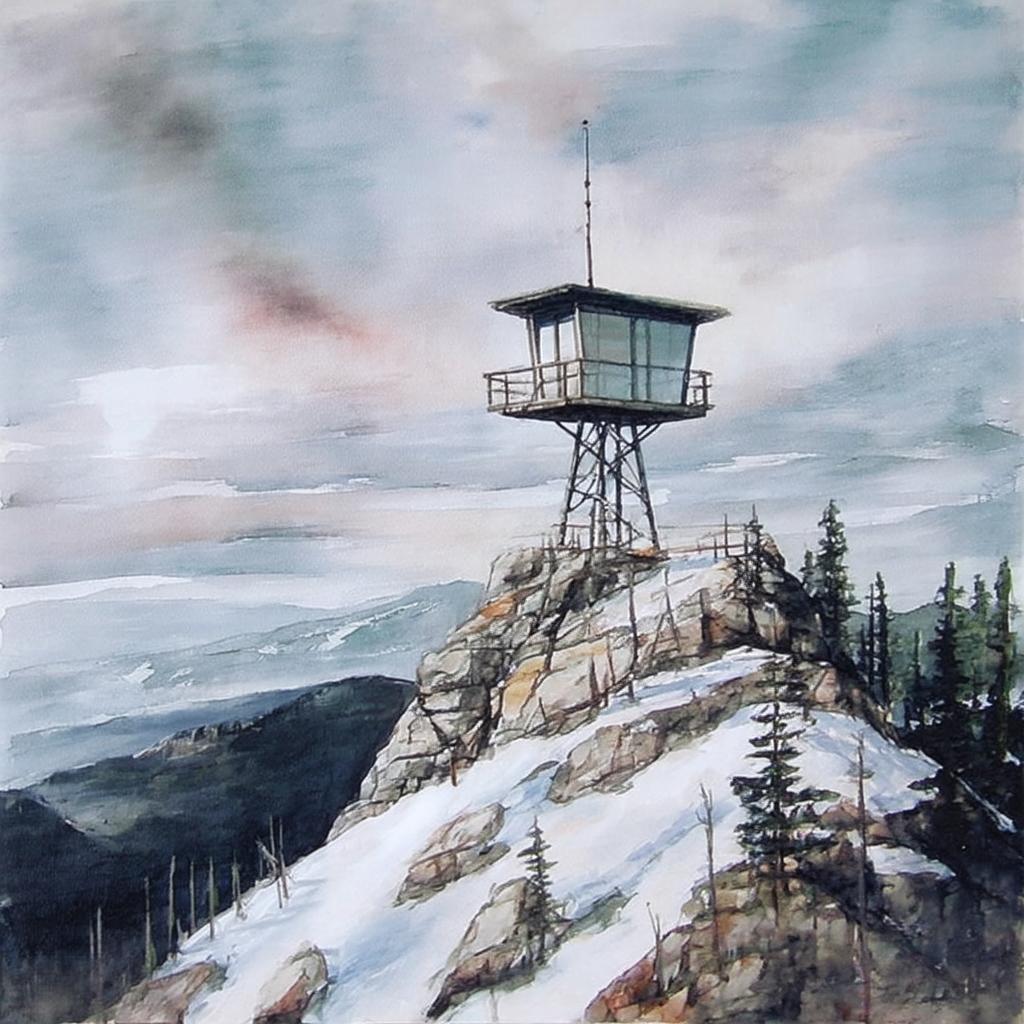} & \includegraphics[width=1.7cm]{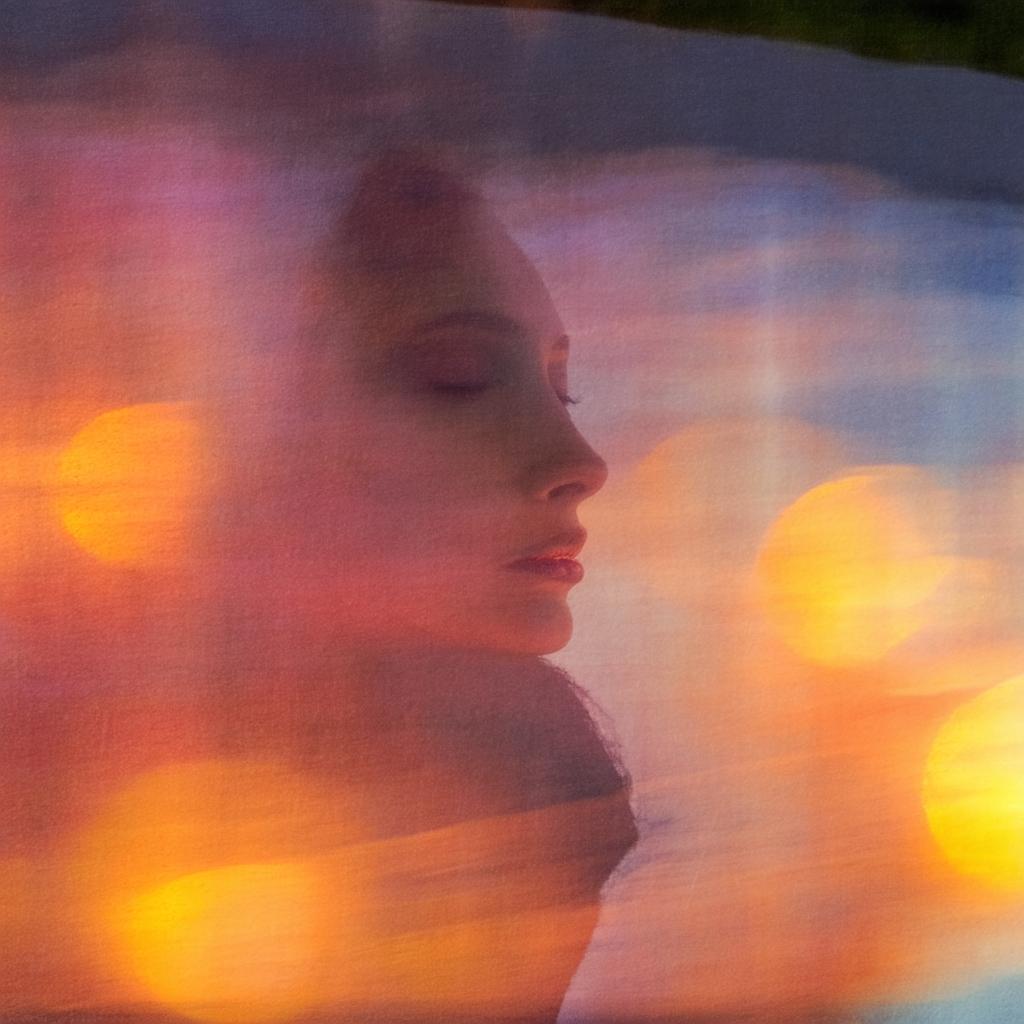} & \includegraphics[width=1.7cm]{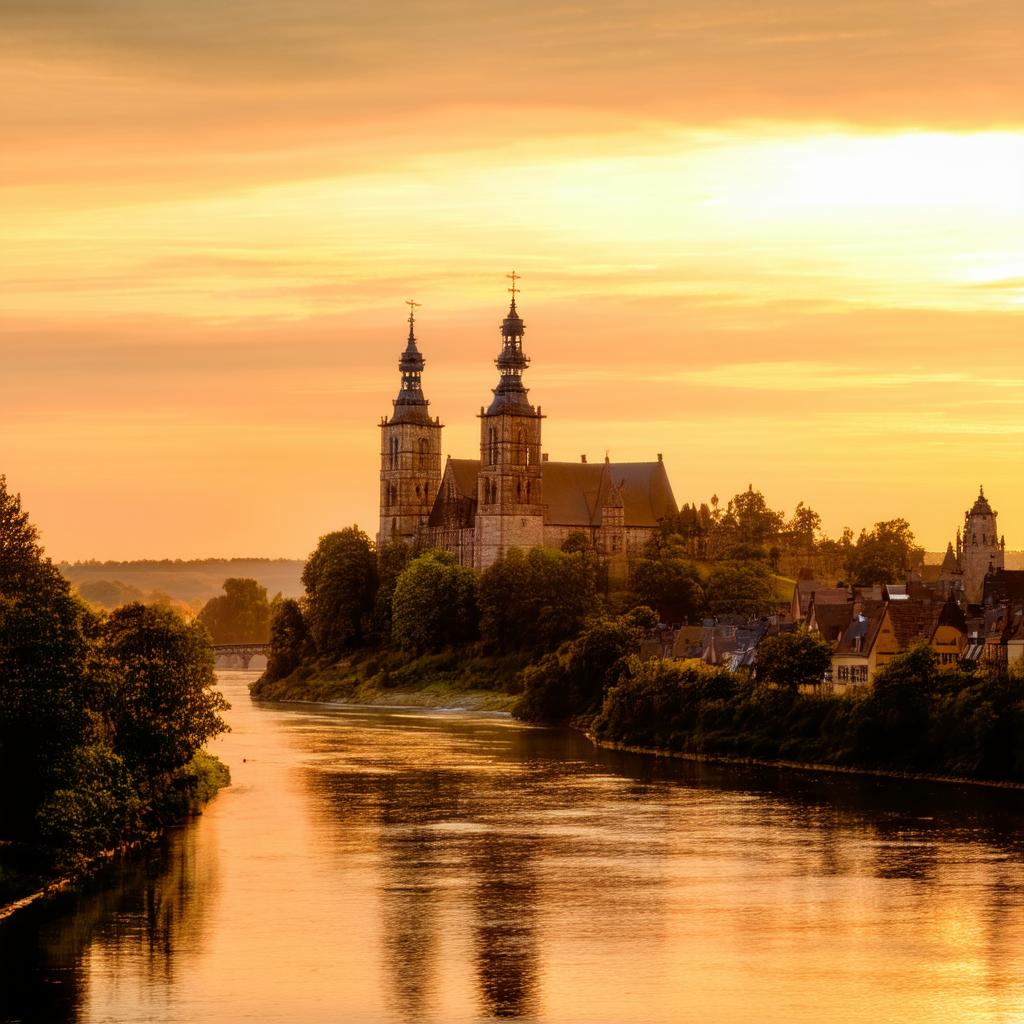} & \includegraphics[width=1.7cm]{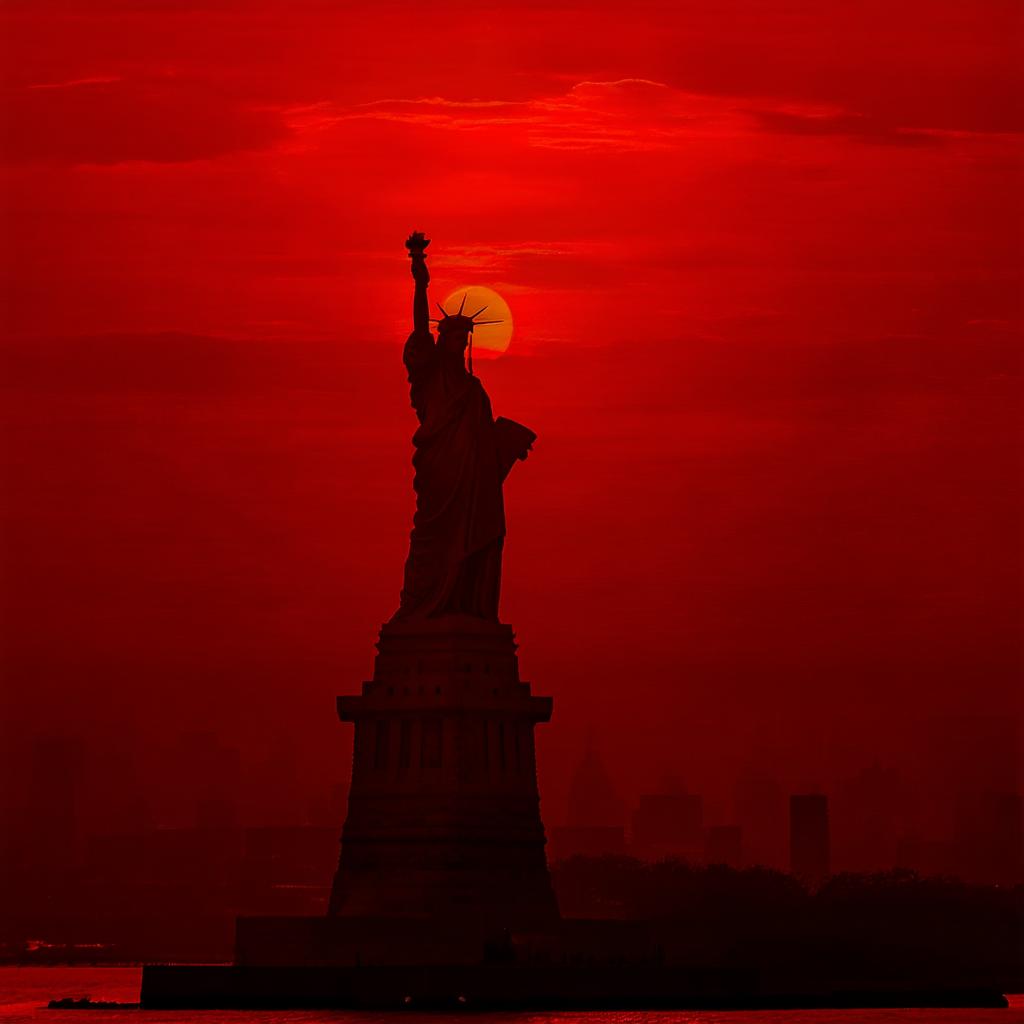} \\ % 
\midrule % 
\rotatebox[origin=c]{90}{\tiny{Reference}} & 
\includegraphics[width=1.7cm]{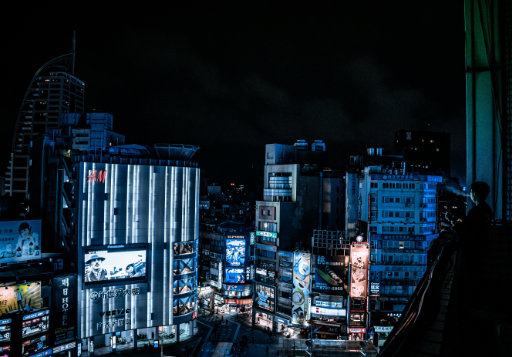} & \includegraphics[width=1.7cm]{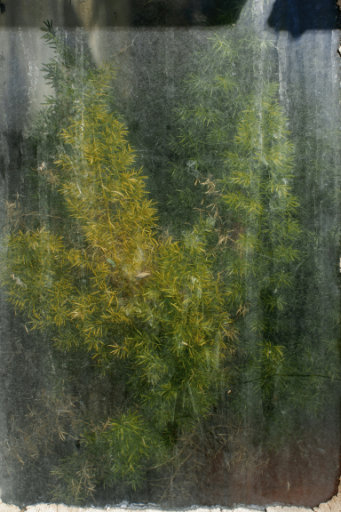} & \includegraphics[width=1.7cm]{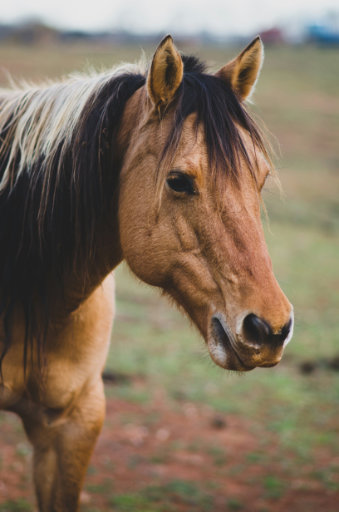} & \includegraphics[width=1.7cm]{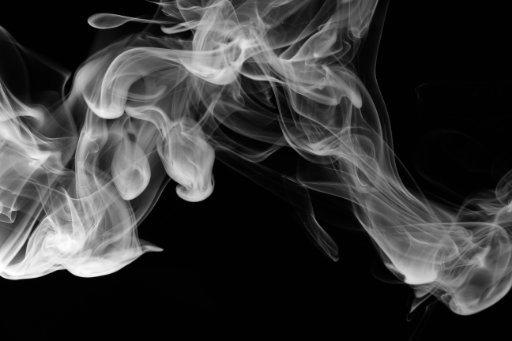} & \includegraphics[width=1.7cm]{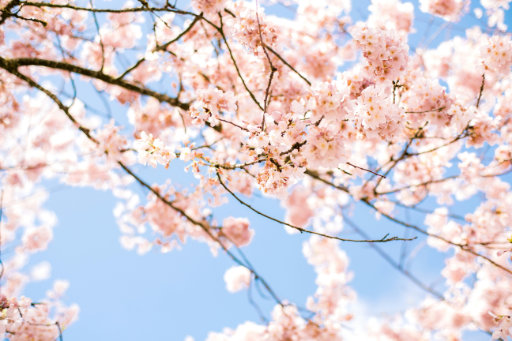} & \includegraphics[width=1.7cm]{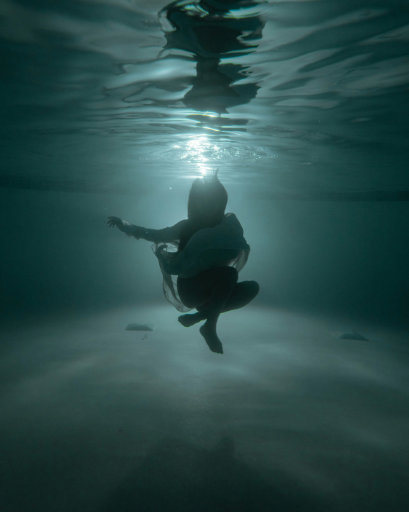} & \includegraphics[width=1.7cm]{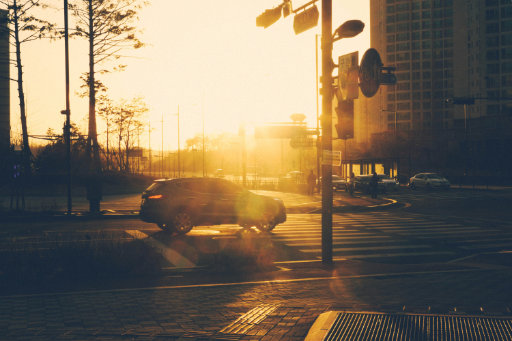} \\ % 
\rotatebox[origin=c]{90}{\tiny{\cite{lobashev2025color}}} & \includegraphics[width=1.7cm]{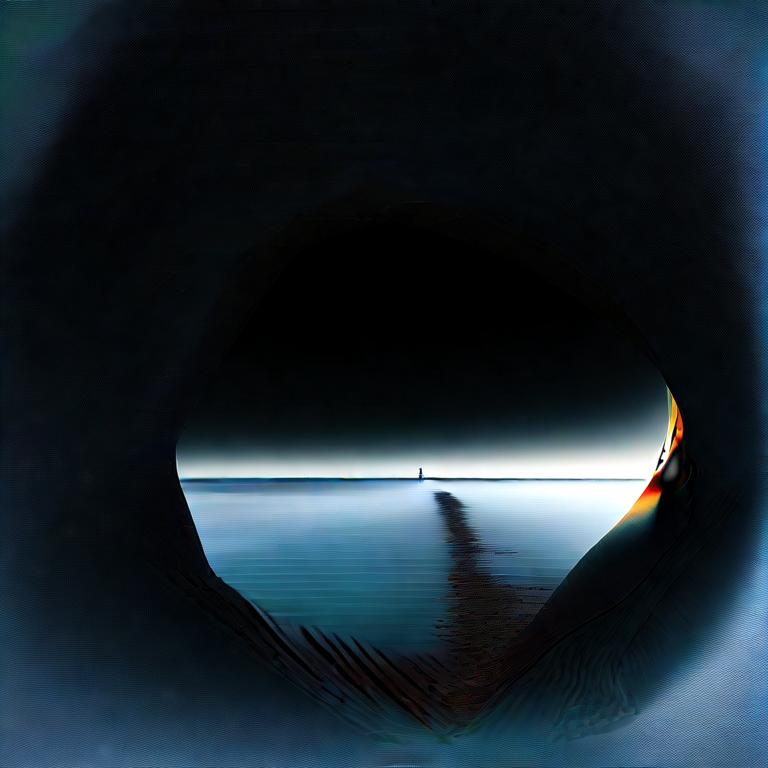} & \includegraphics[width=1.7cm]{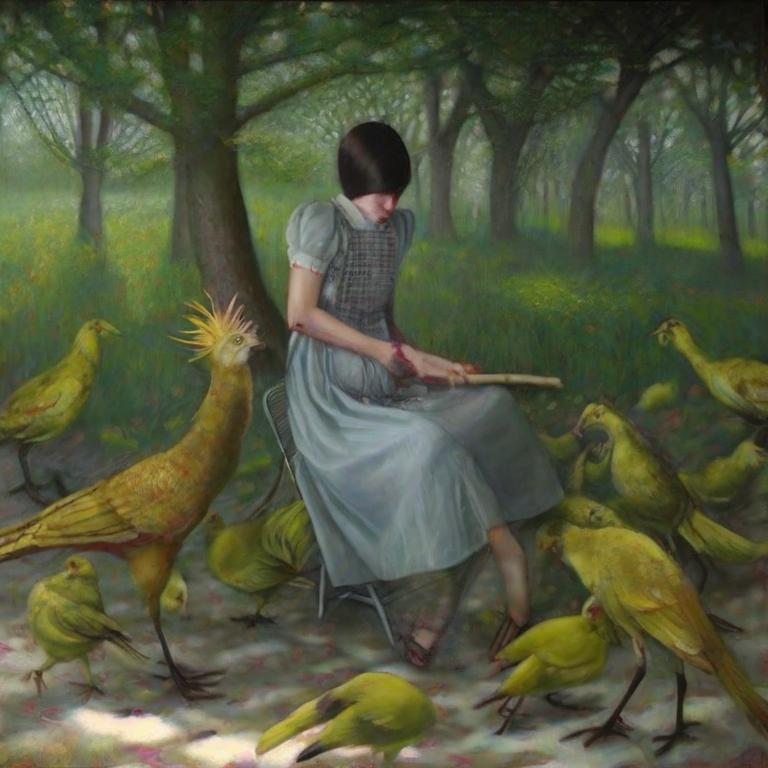} & \includegraphics[width=1.7cm]{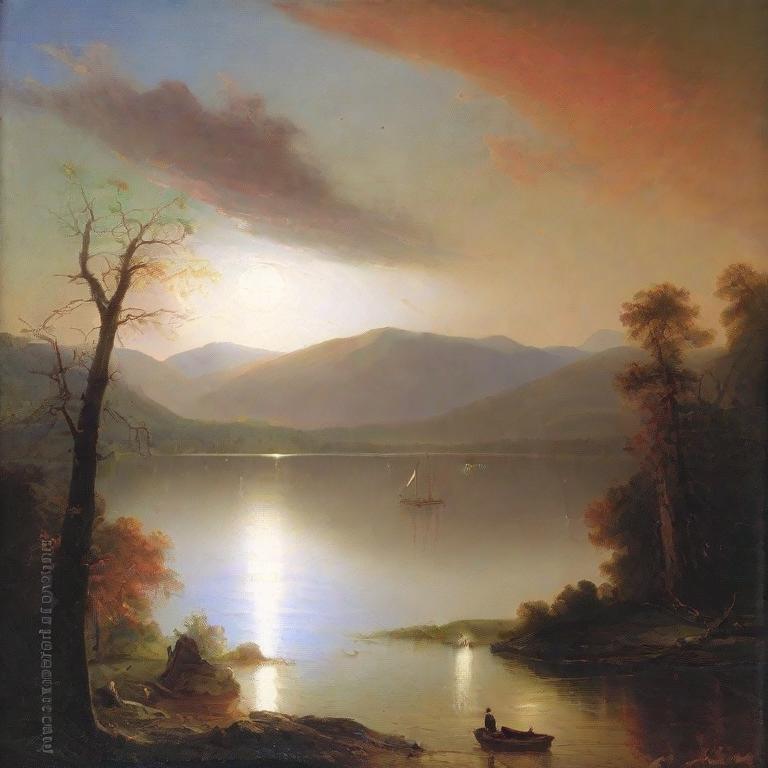} & \includegraphics[width=1.7cm]{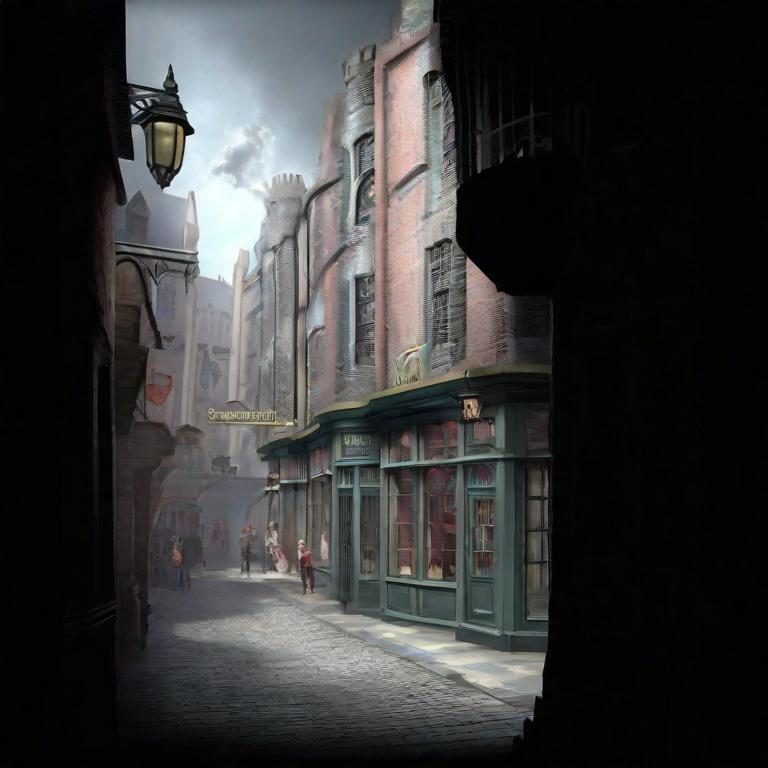} & \includegraphics[width=1.7cm]{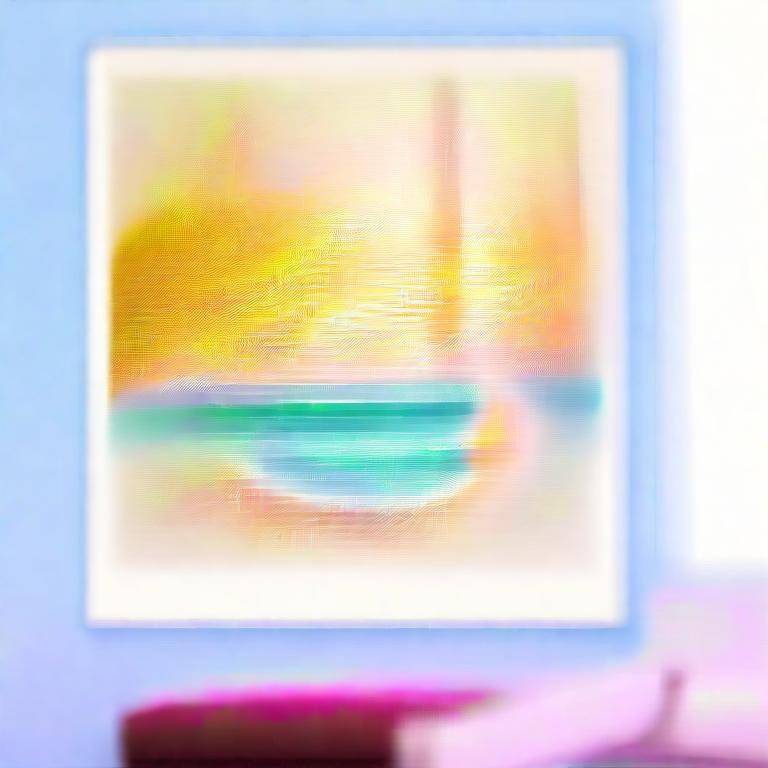} & \includegraphics[width=1.7cm]{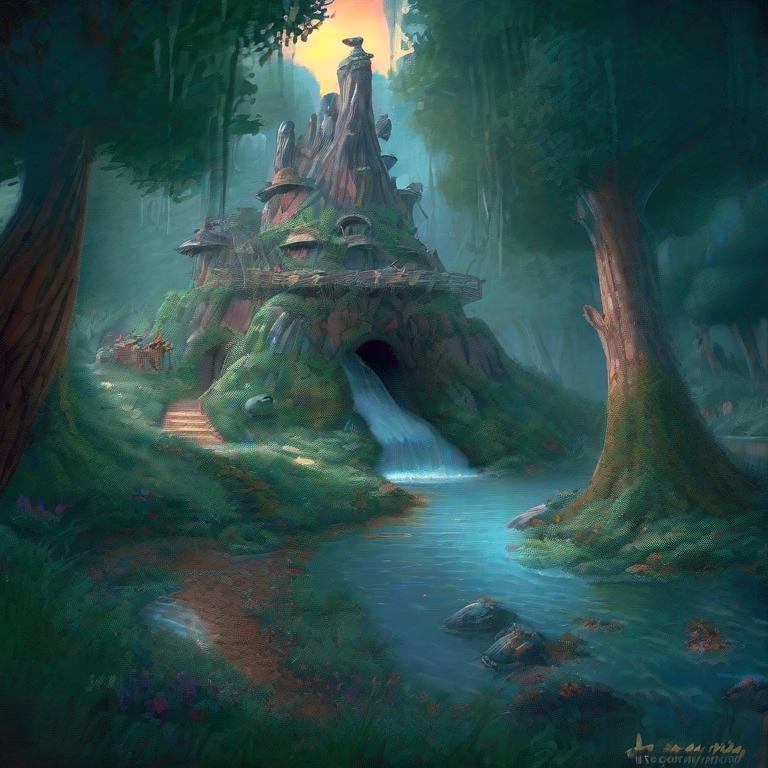} & \includegraphics[width=1.7cm]{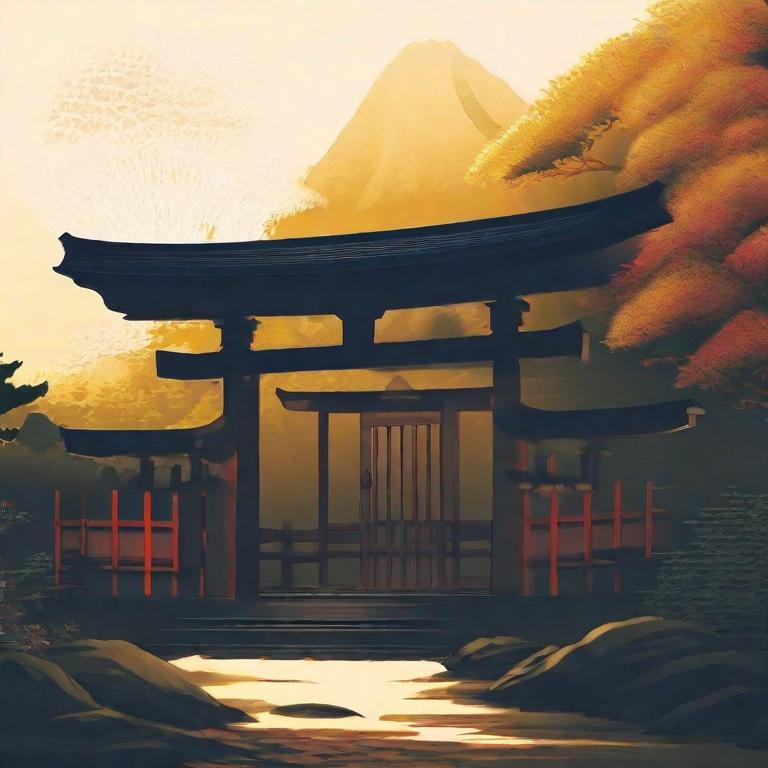}\\ % 
\rotatebox[origin=c]{90}{\tiny{\textbf{\ours}}} & \includegraphics[width=1.7cm]{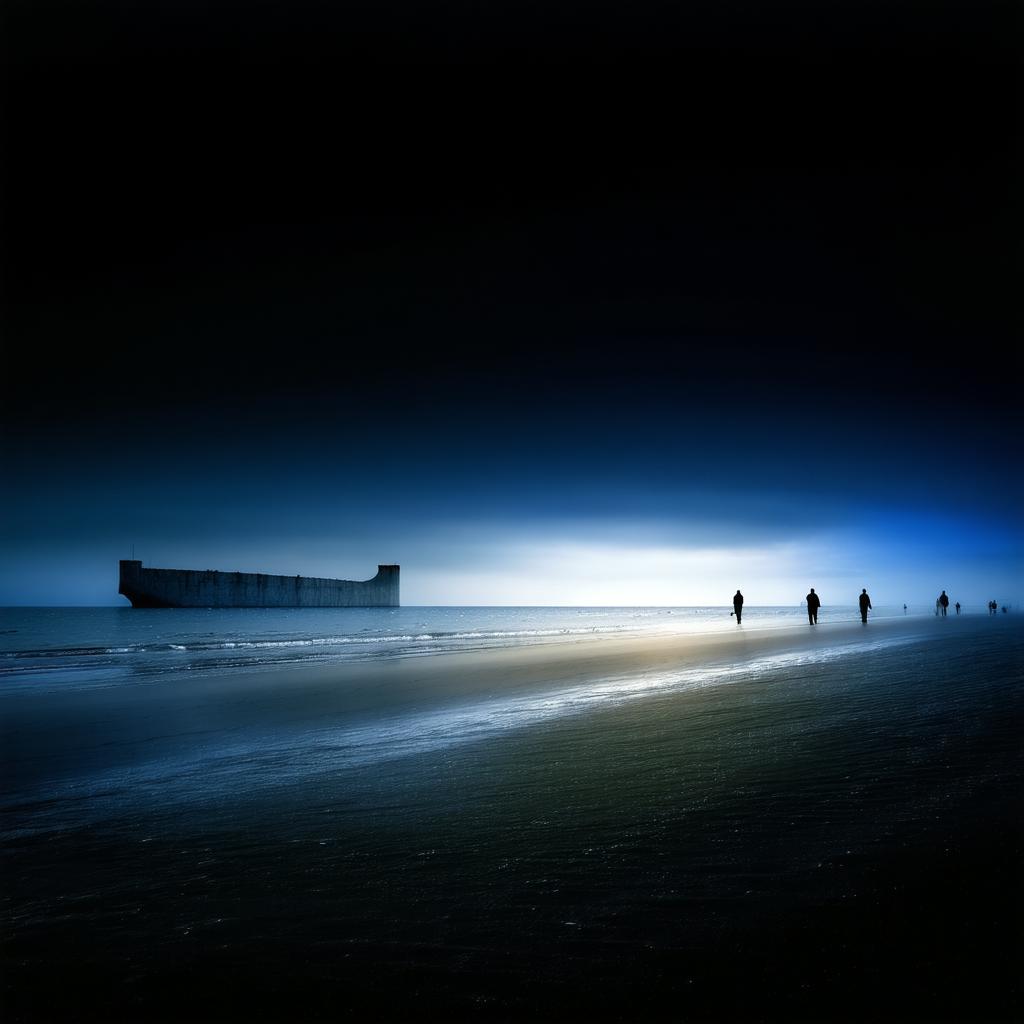} & \includegraphics[width=1.7cm]{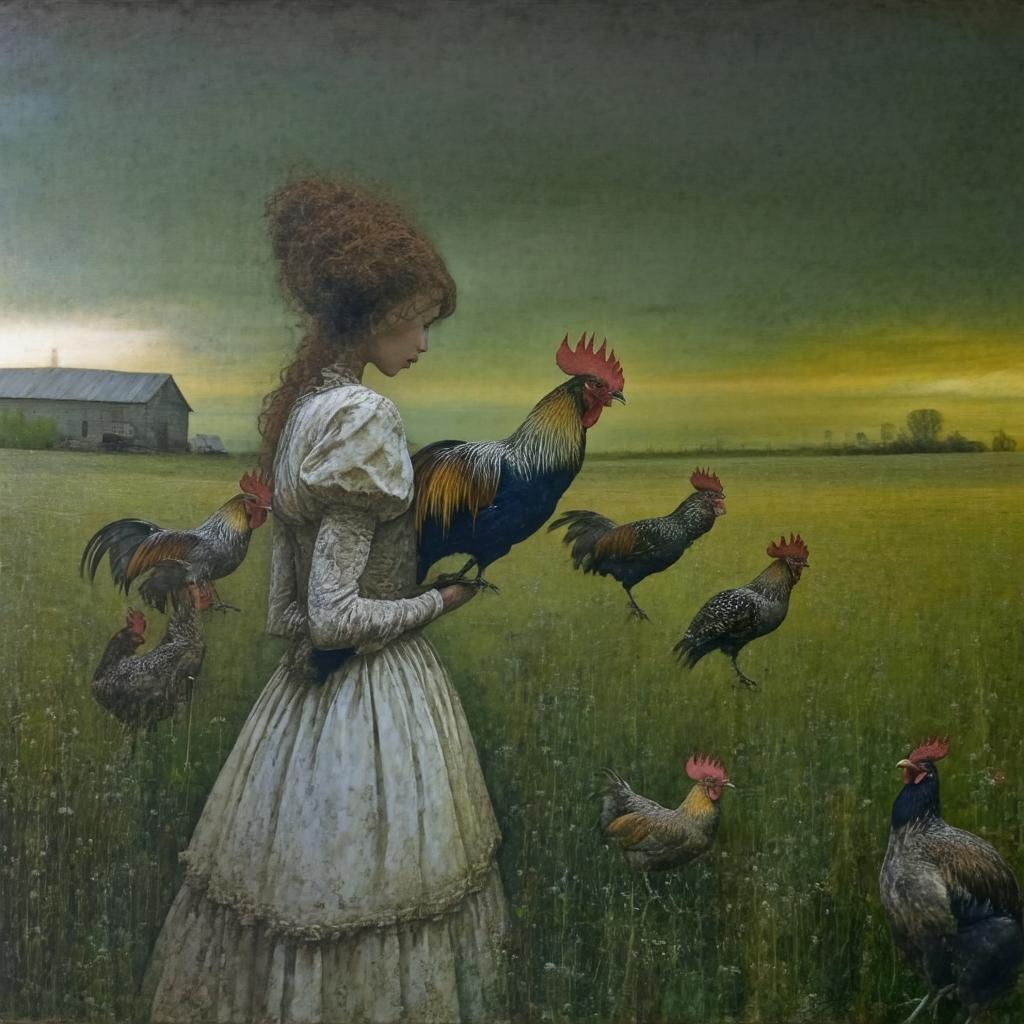} & \includegraphics[width=1.7cm]{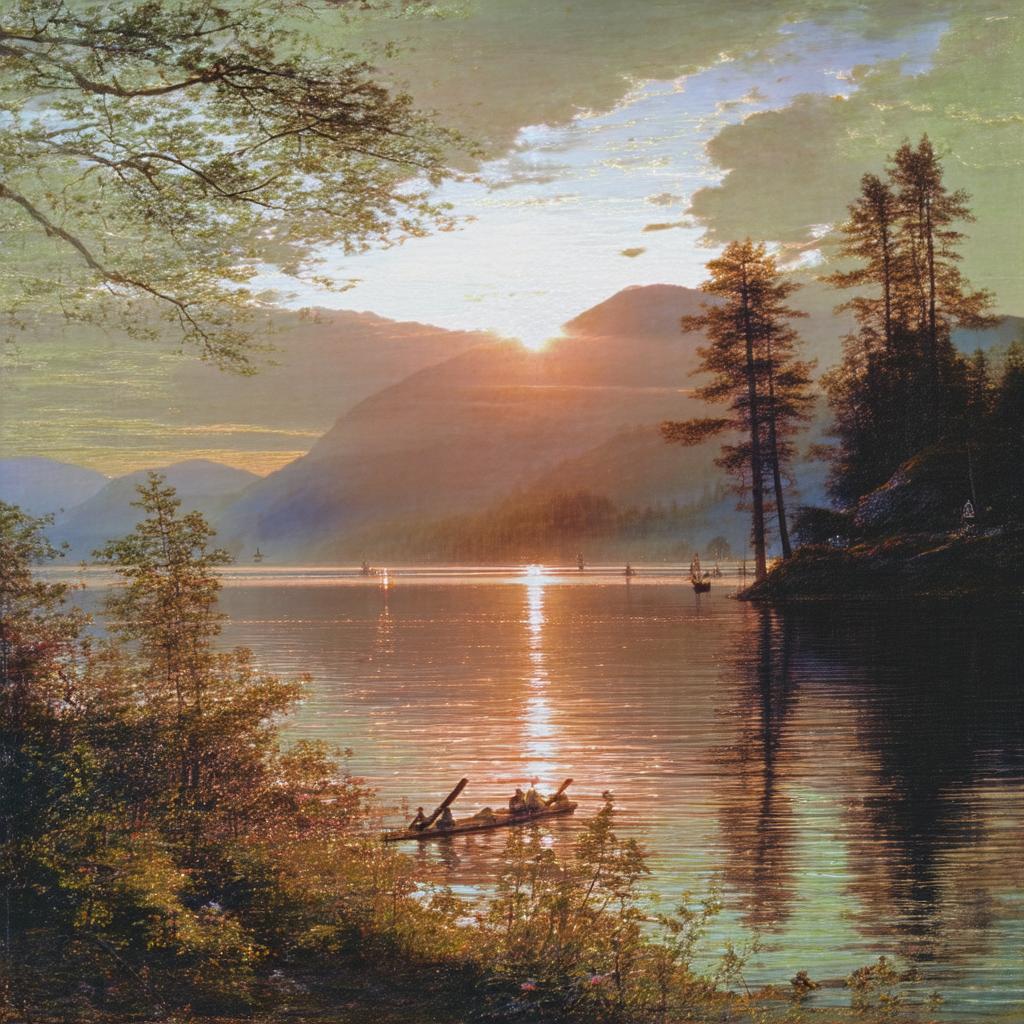} & \includegraphics[width=1.7cm]{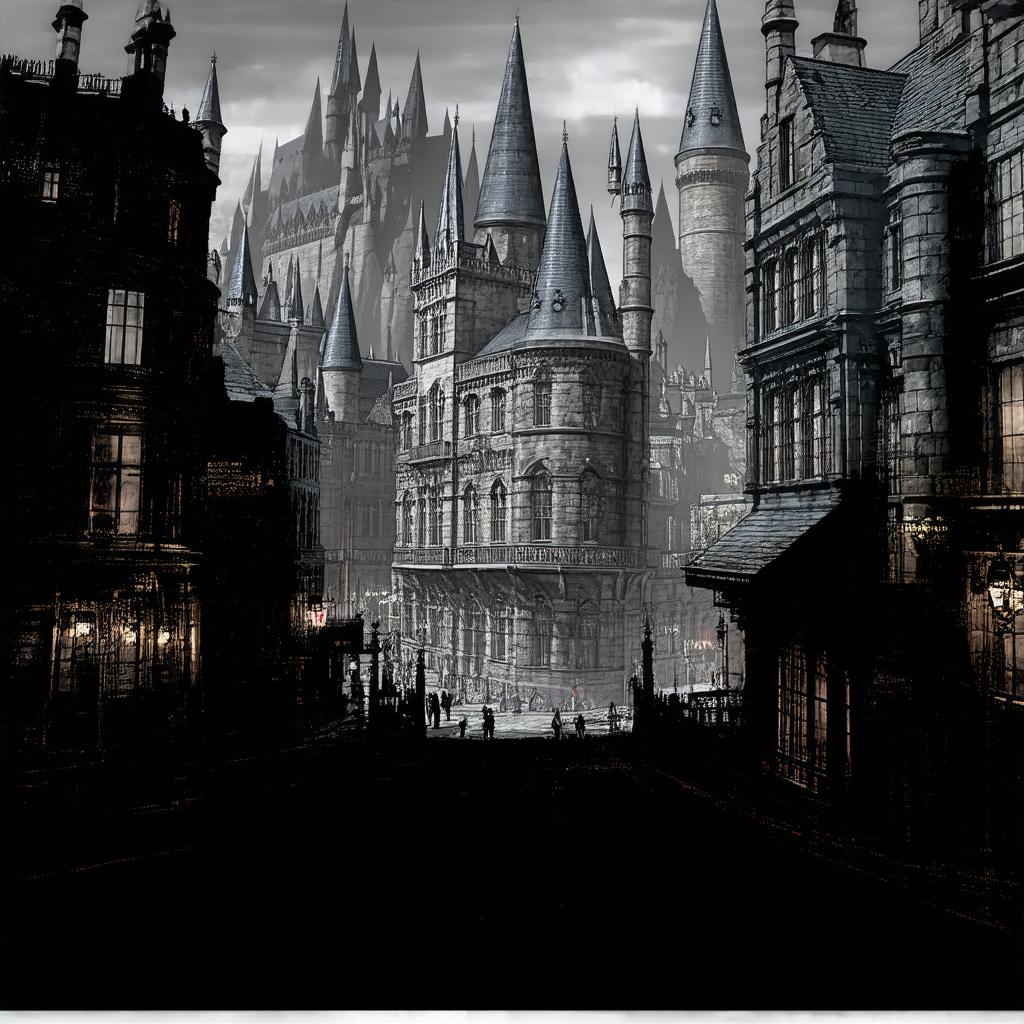} & \includegraphics[width=1.7cm]{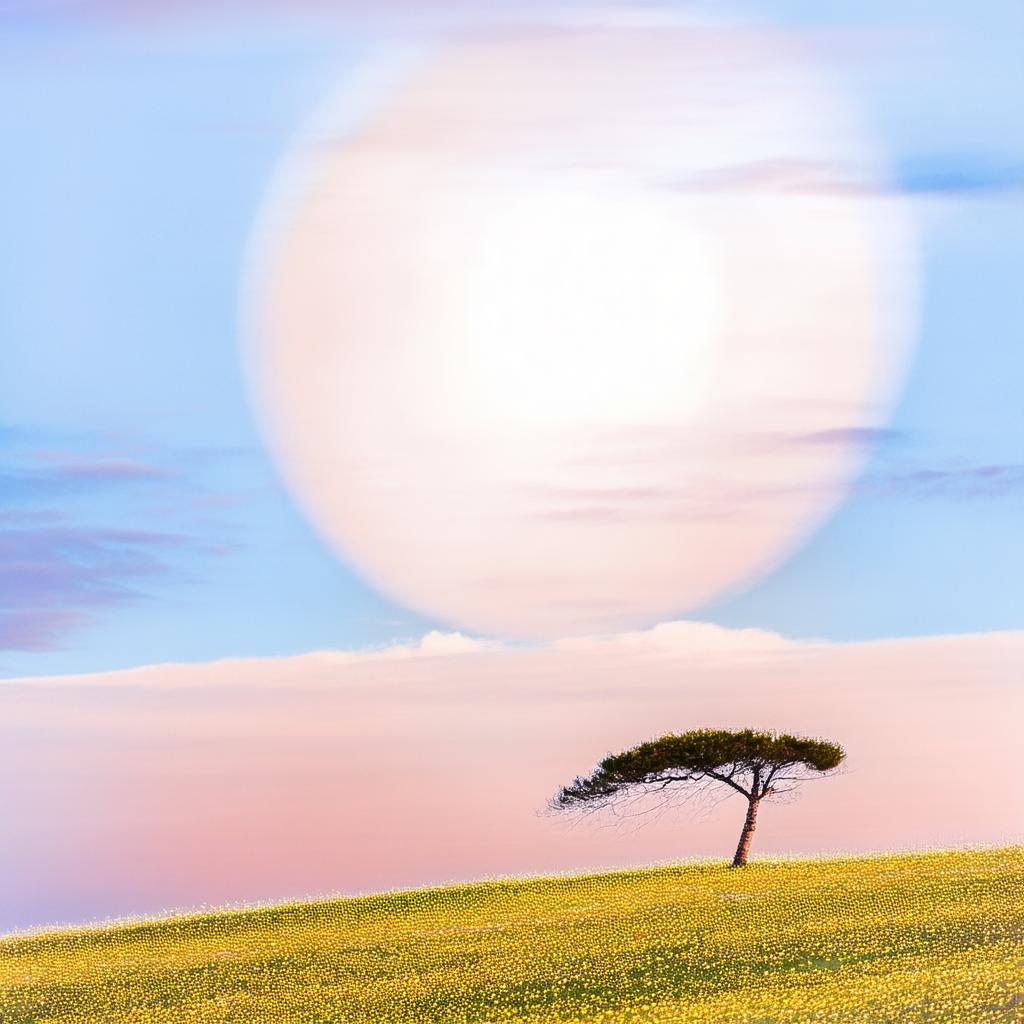} & \includegraphics[width=1.7cm]{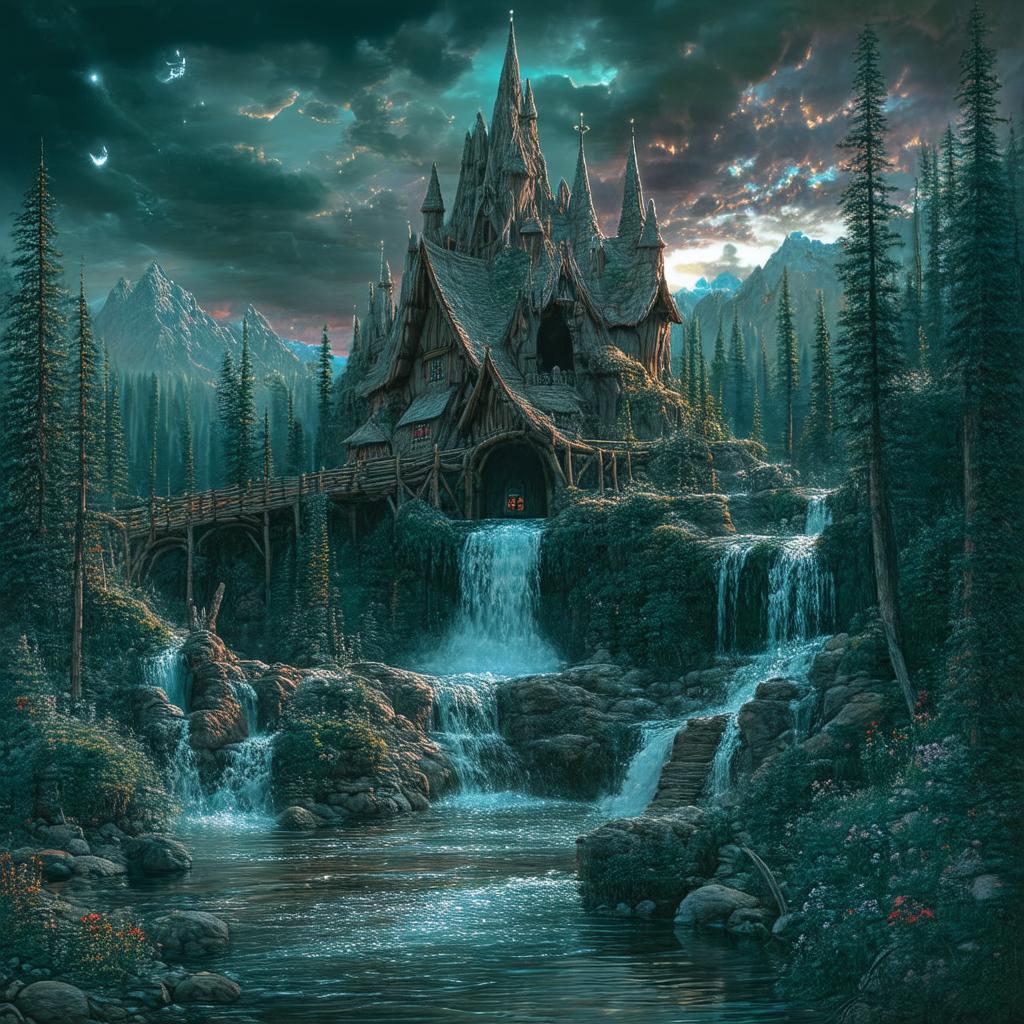} & \includegraphics[width=1.7cm]{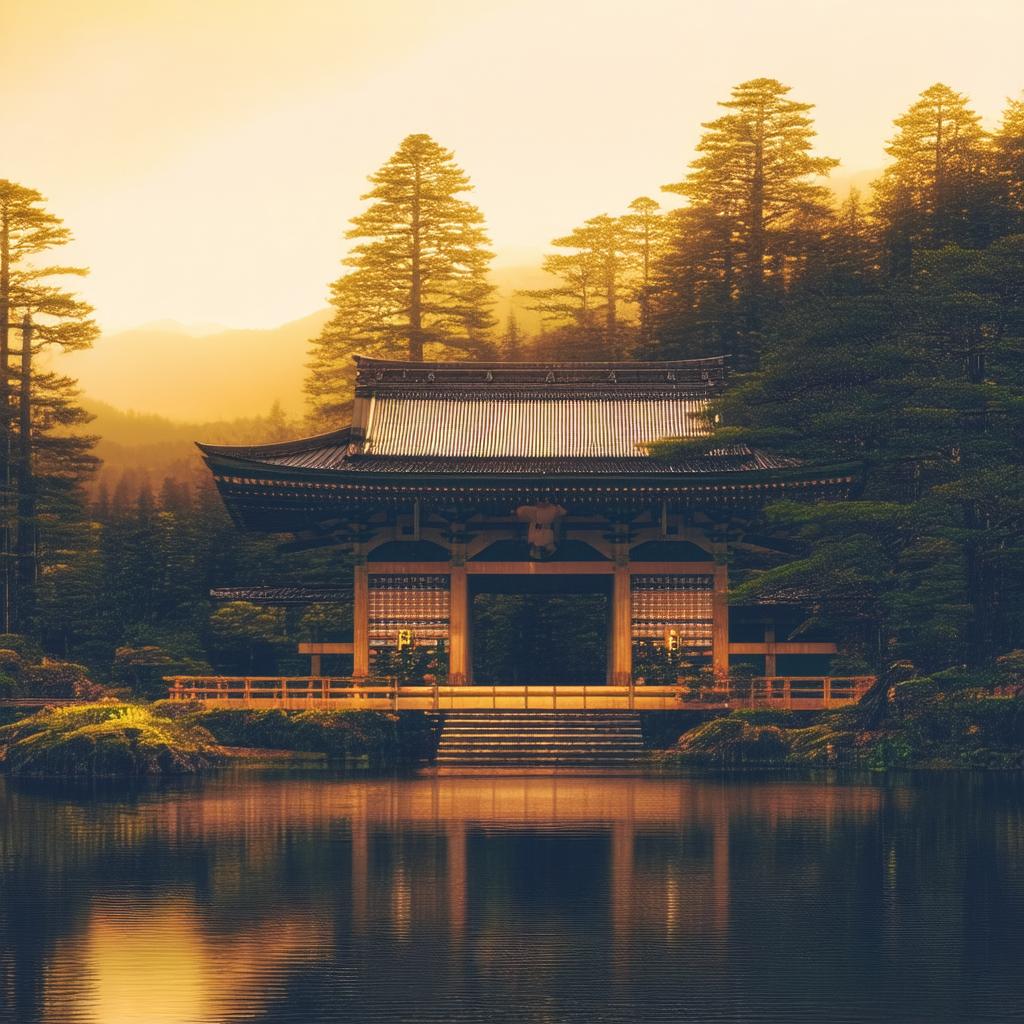}\\ % 
\end{tabular}

\titlecaption{Diffusion Guidance}{Visualization of our color guidance with SD3.5-Large. Notice, how closely the colors match the reference and the detailed generation quality of our method. Also we noticed that \citet{lobashev2025color} sometimes degenerates (Third example top, First example bottom) whereas \ours handles these challenging guidance signal well.} 
\label{fig:guidance} 
\end{figure*}

\begin{table}[t]
\titlecaptionof{table}{Comparison on diffusion guidance}{Here, we compare \ours with \citet{lobashev2025color} on color guidance in image generation. As our main proposed method uses the more recent SD3.5, we also compare with our modifications on SDXL. Even on the same base model our modifications improve upon the base technique.
}
\label{tab:diffusion_guidance}
\centering
\resizebox{0.85\linewidth}{!}{ %
\begin{tabular}{lcccc}
Method & CLIP-IQA $\uparrow$ & CLIP-T $\uparrow$ & Mean-W$_2$ [$10^2$] $\downarrow$ & Time per generation [s] $\downarrow$ \\
\midrule
\cite{lobashev2025color} &
0.671 & 14.861 & 1.941 & 124\\
\ours $+$ SDXL &
0.696 & 14.870 & 1.213 & 34 \\
\midrule
\ours $+$ SD3.5-medium &
0.786 & 14.783 & 0.815 & \secondbest{32} \\
\ours $+$ SD3.5-turbo &
\best{0.800} & \best{14.882} & \secondbest{0.675} & \best{4} \\
\ours $+$ SD3.5-large &
\secondbest{0.793} & \secondbest{14.817} & \best{0.55} & 69 \\
\end{tabular}
}
\end{table}

\inlinesection{Diffusion Guidance.}
We select \citet{lobashev2025color} as our main comparison. Furthermore, we also implemented our modifications to the matching in the SDXL model to compare our alterations with the baseline, which is based on SDXL.
This delineates the influence of the improved SD3.5 model from our matching and alterations. 
For evaluations we follow Lobashev~\etal and select 1000 prompts from the ContraStyles dataset\footnote{\url{https://huggingface.co/datasets/tomg-group-umd/ContraStyles}} and 1000 images from Unsplash Lite\footnote{\url{https://unsplash.com/data}}.
We provide the W$_2$ Wasserstein distance, CLIP-IQA~\citep{wang2023clipiqa}, and CLIP-T~\citep{radford2021clip} to present the color matching performance, highlight the quality of the generation, and prompt adherence, respectively.

In \Table{diffusion_guidance} our modifications clearly outperform the prior state-of-the-art technique of Labashev \etal. 
Even with the same base model (SDXL), our method is faster and provides better results.
The main speed difference comes from our gradient stopping implementation which does not have any influence on the guidance.
With the upgraded base model, \ours clearly outperforms the previous method. 
Here, it is interesting that the distilled Turbo variants provides better quality, which we attribute to the straighter trajectories. 

We show example generations of this dataset in \fig{guidance} along with the reference and the prompts. 
As evident, the improved generation capacity of SD3.5 along with our necessary changes resulted in drastically better generation which are faithful to the color distribution defined by the input image.

\inlinesection{Ablation.}
In \Table{candidate_1d}, we assess the influence of the number of new candidates in our general distribution matching task. 
Too few new candidates starve the reservoir of new directions, resulting in static directions during optimization. 
Too many new candidates reduce the reservoir size, which drives the optimization solely, resulting in too few directions supporting the optimization. 
With 8 candidates, we achieved the best overall results in all applications given 64 total projections.

\inlinesection{Limitations.}
Our current technique is limited to simple matrix projections, and an extension to learned convolution kernels similar to \citet{elnekave2022gpdm} did not provide satisfying results, as the search space for kernels is too large to randomly obtain drastically better kernels. 
This degraded the results similar to limiting the projection directions and reducing the redrawing of new directions to every few optimization steps.

\section{Conclusion}

Our novel combination of real-time rendering-influenced variance reduction techniques in SWD optimization offers a more efficient and unbiased solution compared to other recent variance reduction techniques. This results in reduced overhead while maintaining optimal performance. Our technique has demonstrated superior performance in various real-world and synthetic applications, achieving state-of-the-art results.

{
    \small
    \bibliographystyle{iclr2026_conference}
    \bibliography{references}
}
\appendix
\section*{Supplements}

In the supplements, we present more results from our applications.

\section{Diffusion Guidance}
\begin{figure*}[htb] 
\centering
\setlength{\tabcolsep}{3pt}
\renewcommand{\arraystretch}{0.8} 
\begin{tabular}{@{}p{0.05cm}*{4}{m{3.02cm}}@{}}
\rotatebox[origin=c]{90}{\tiny{Reference}} & 
\includegraphics[width=3cm]{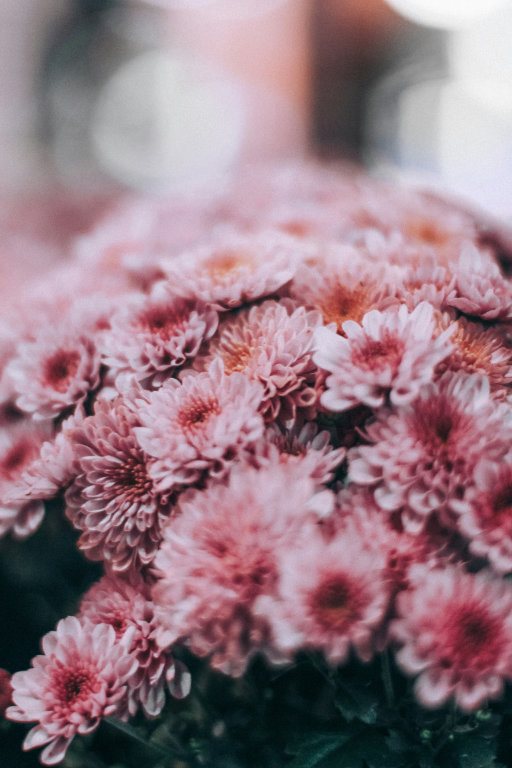} & 
\includegraphics[width=3cm]{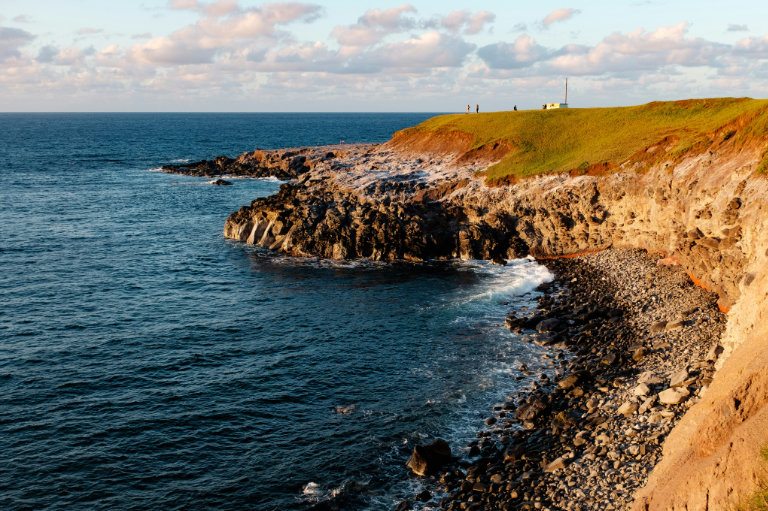} & 
\includegraphics[width=3cm]{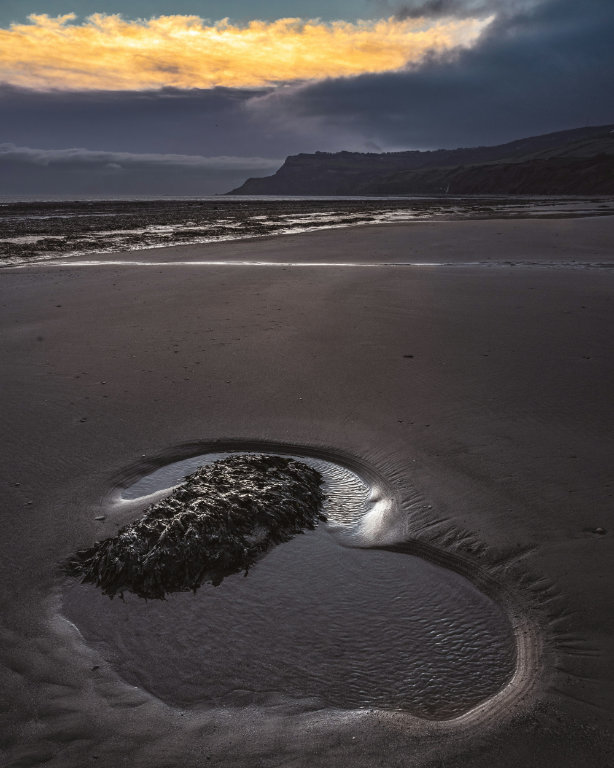} & 
\includegraphics[width=3cm]{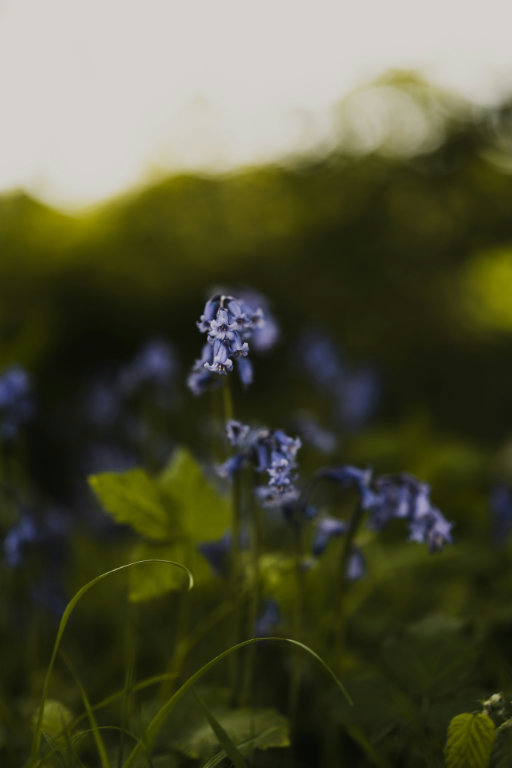}  \\ % 
\rotatebox[origin=c]{90}{\tiny{\cite{lobashev2025color}}} & 
\includegraphics[width=3cm]{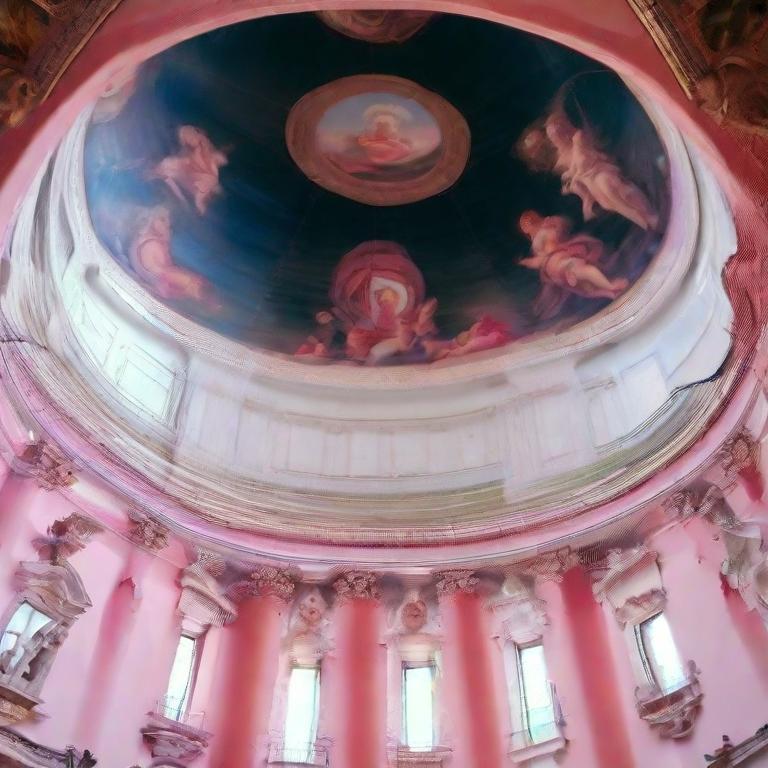} & 
\includegraphics[width=3cm]{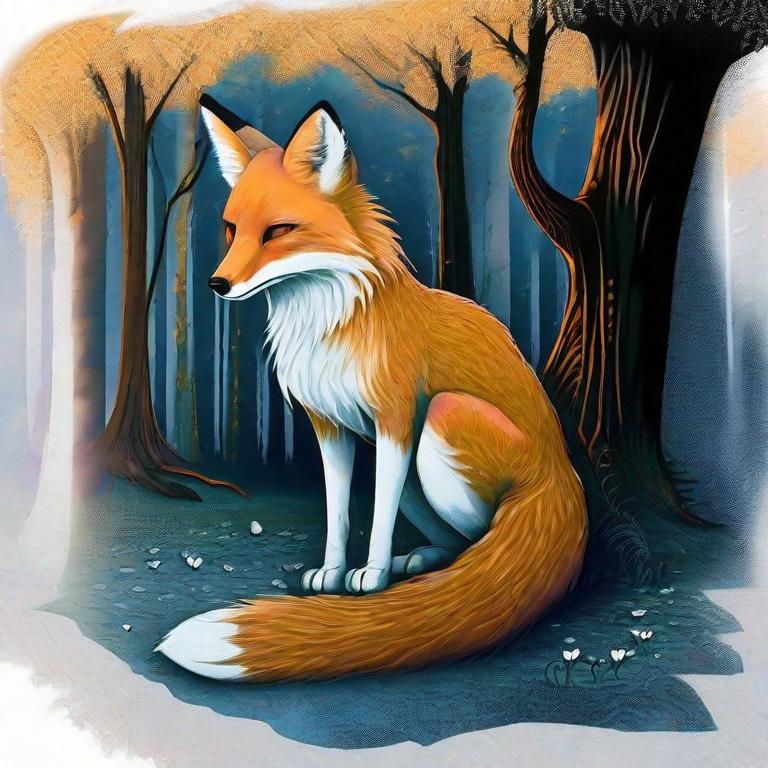} & 
\includegraphics[width=3cm]{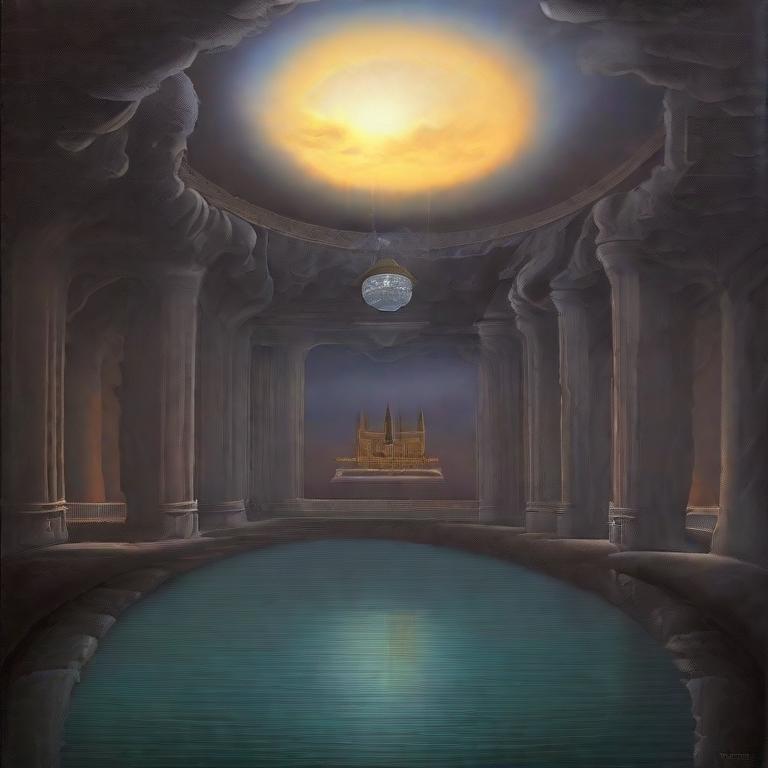} & 
\includegraphics[width=3cm]{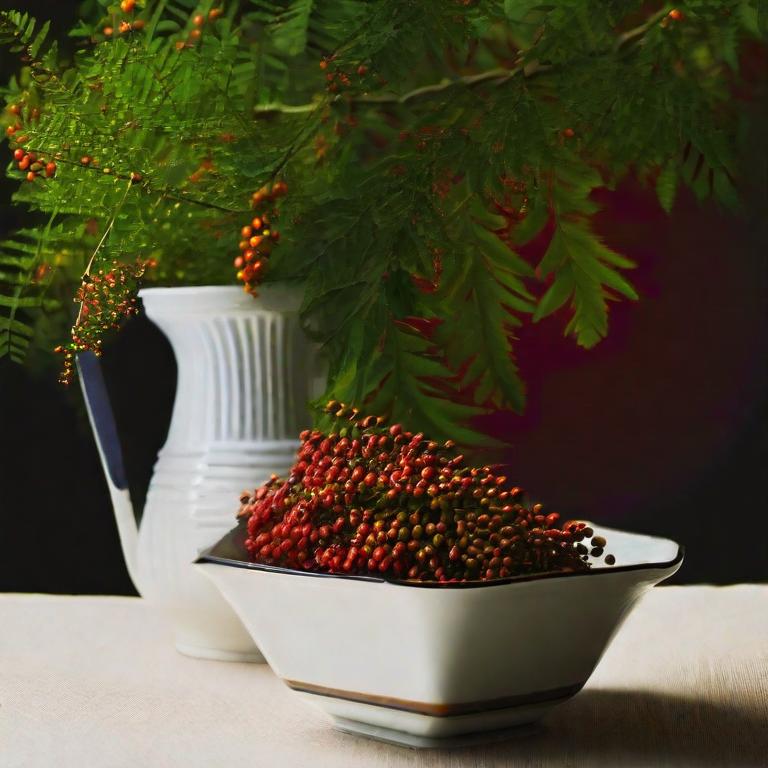}  \\ %
\rotatebox[origin=c]{90}{\tiny{\textbf{\ours}}} & 
\includegraphics[width=3cm]{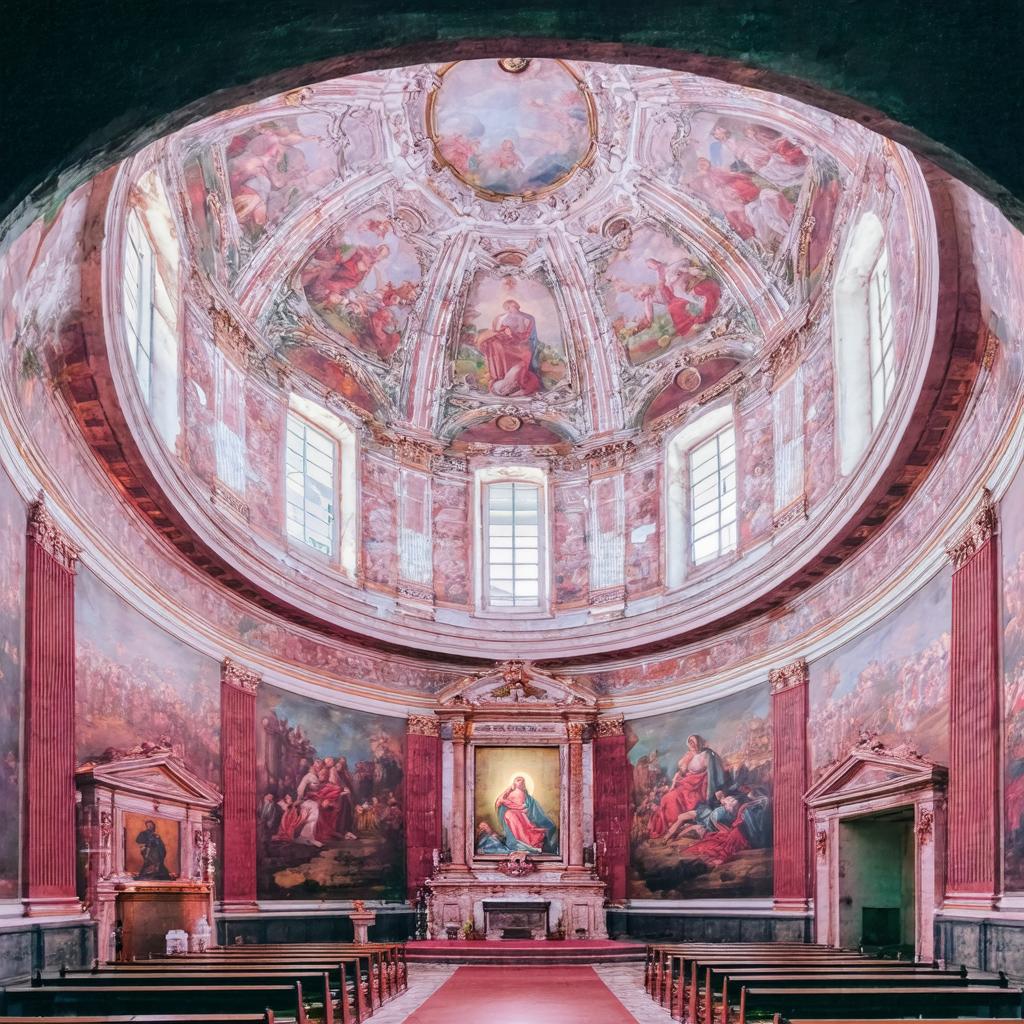} & 
\includegraphics[width=3cm]{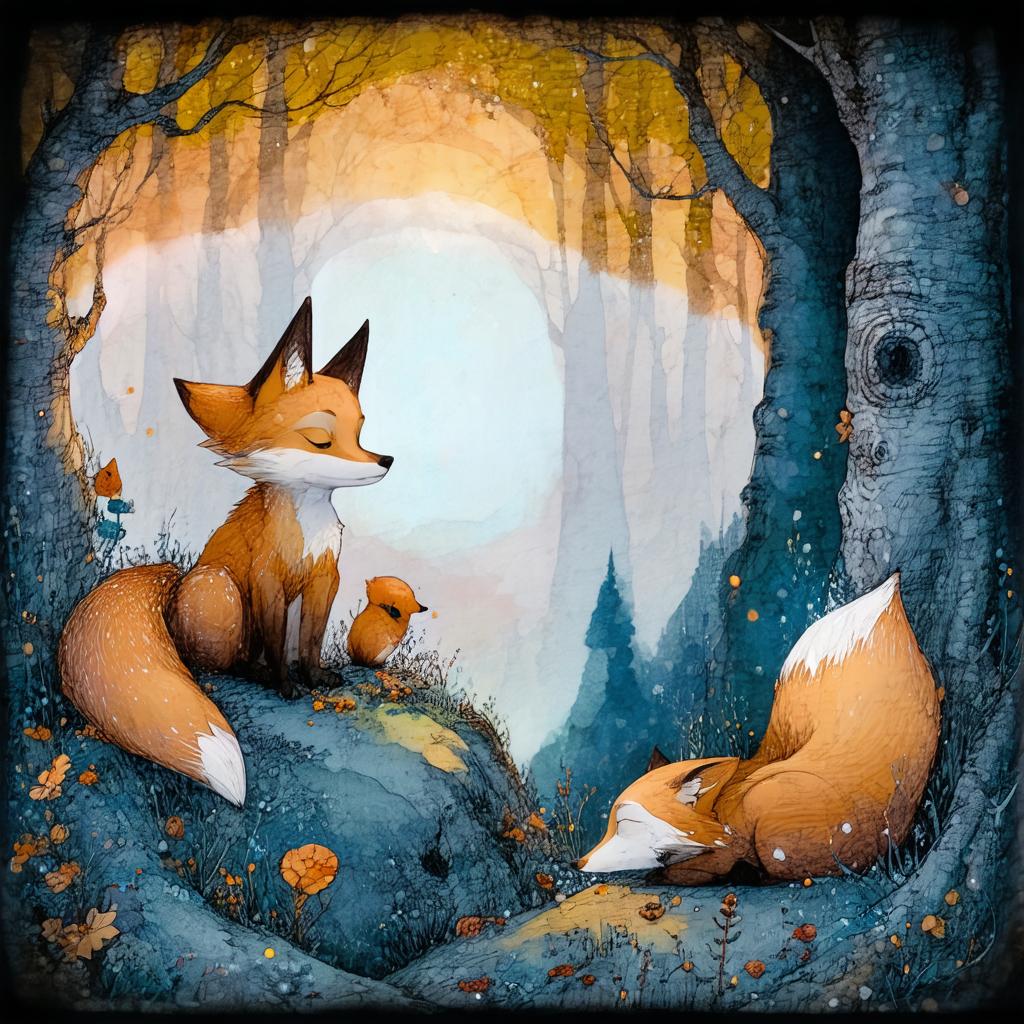} & 
\includegraphics[width=3cm]{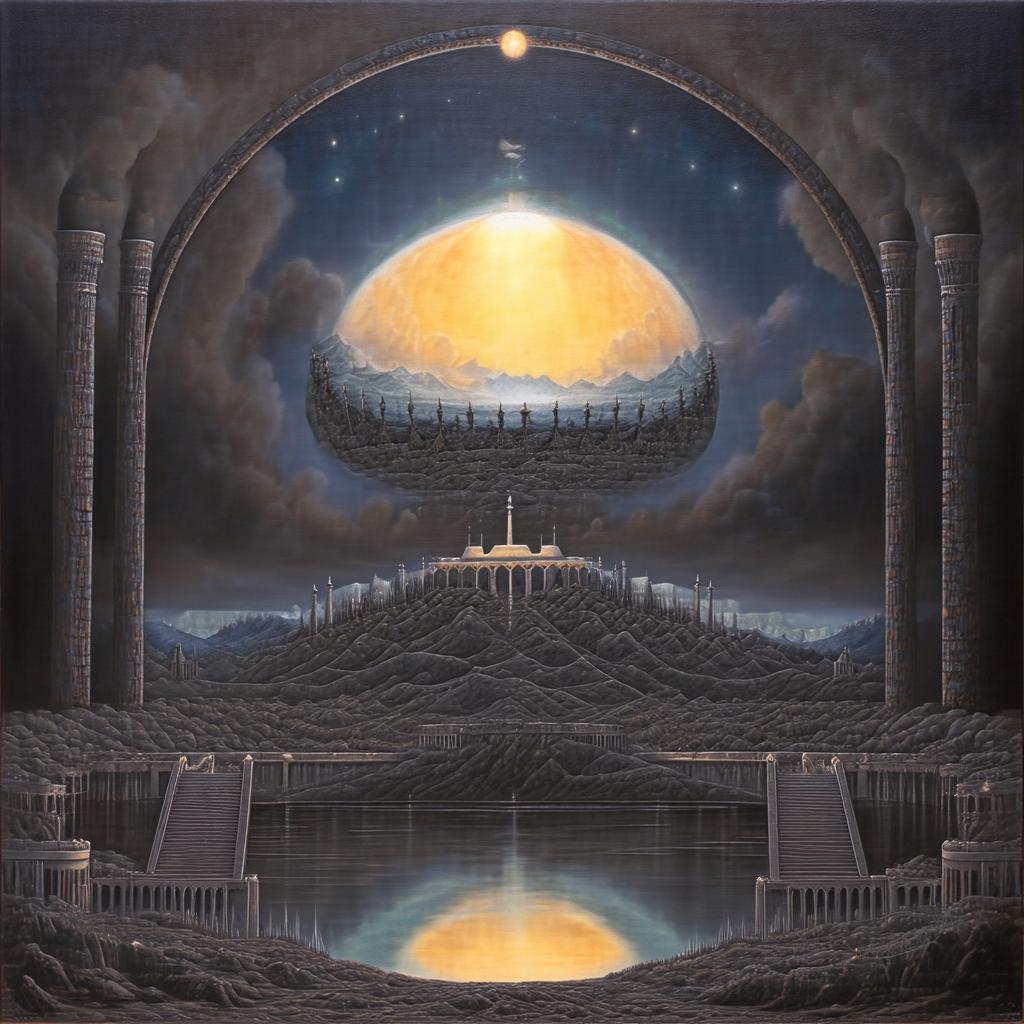} & 
\includegraphics[width=3cm]{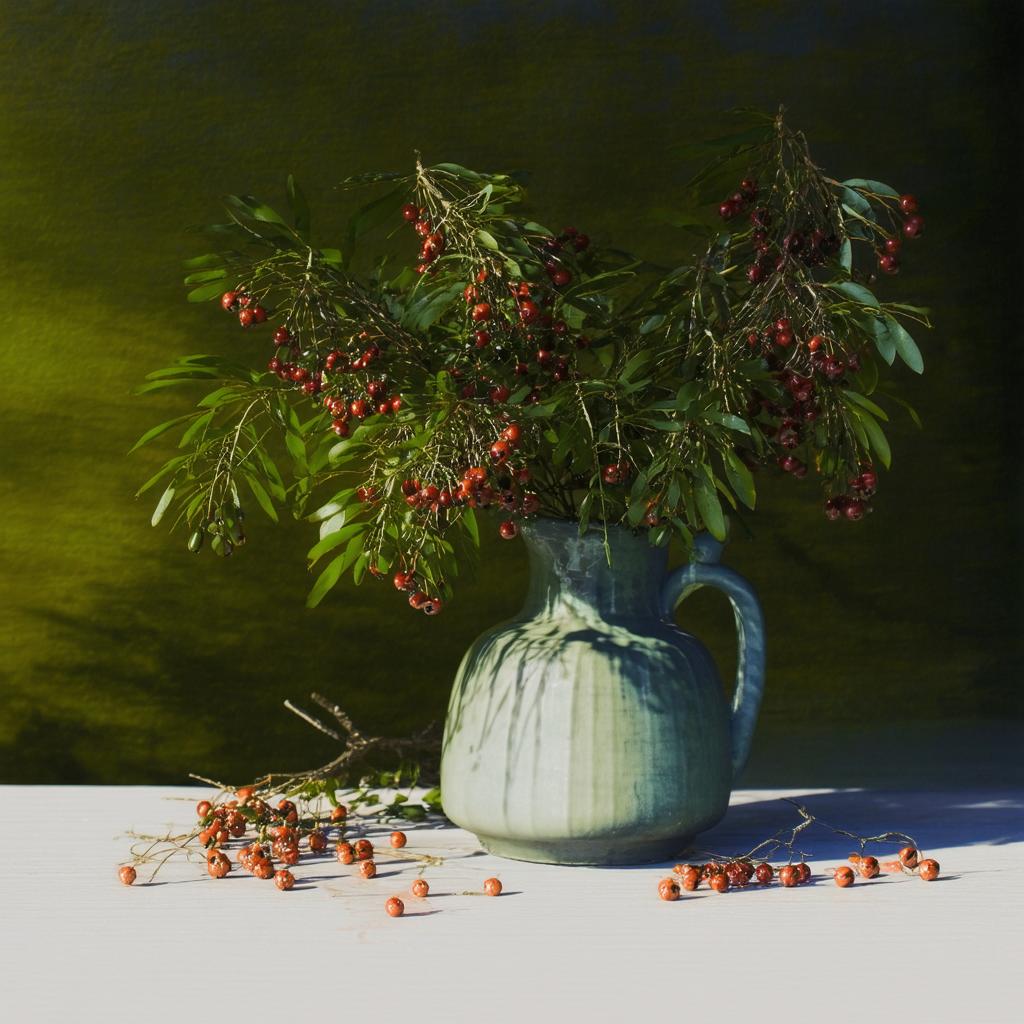}  \\ % 
\midrule % 
\rotatebox[origin=c]{90}{\tiny{Reference}} & 
\includegraphics[width=3cm]{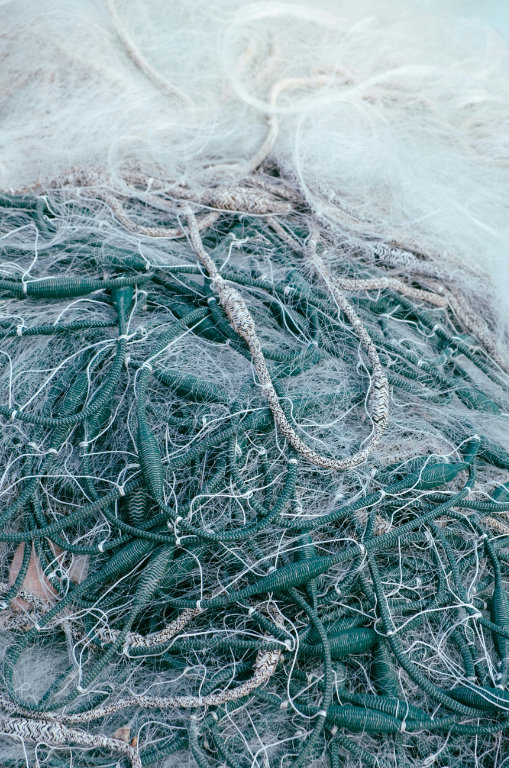} & 
\includegraphics[width=3cm]{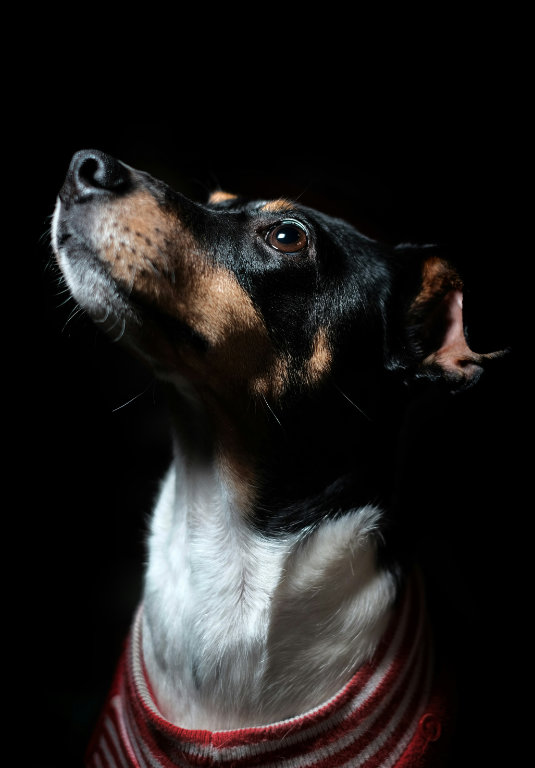} & 
\includegraphics[width=3cm]{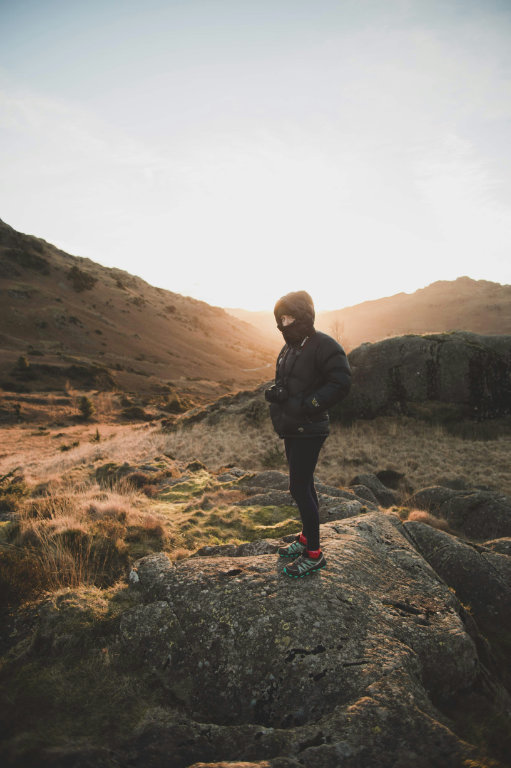} & 
\includegraphics[width=3cm]{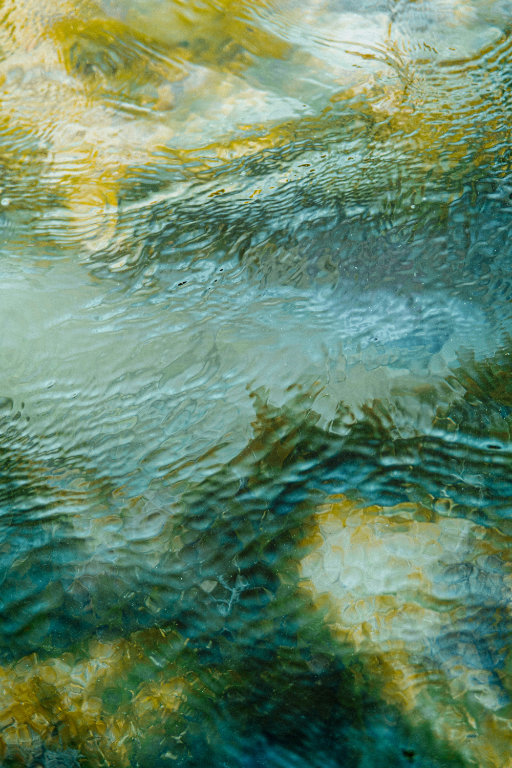} \\ % 

\rotatebox[origin=c]{90}{\tiny{\cite{lobashev2025color}}} & 
\includegraphics[width=3cm]{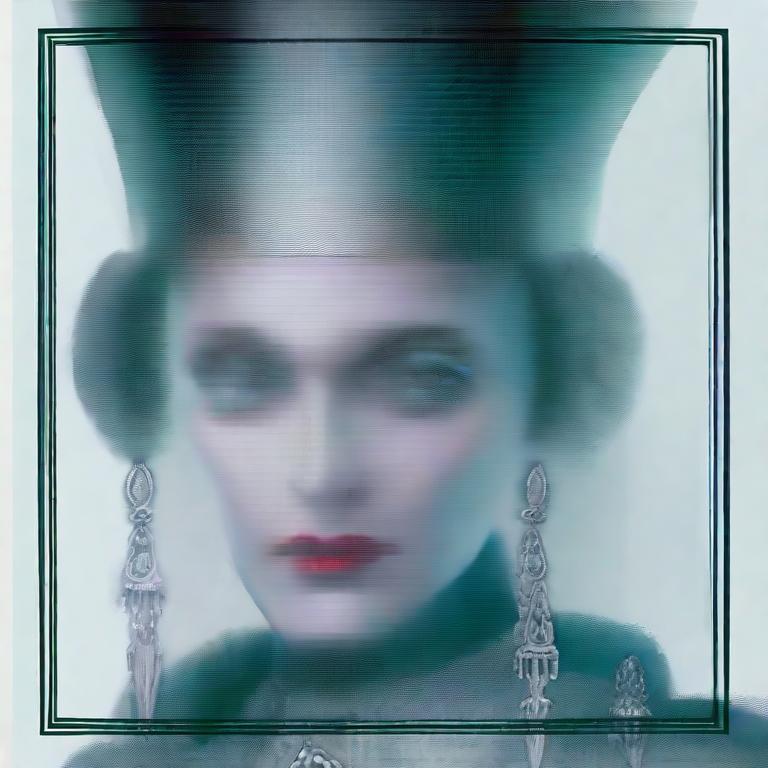} & 
\includegraphics[width=3cm]{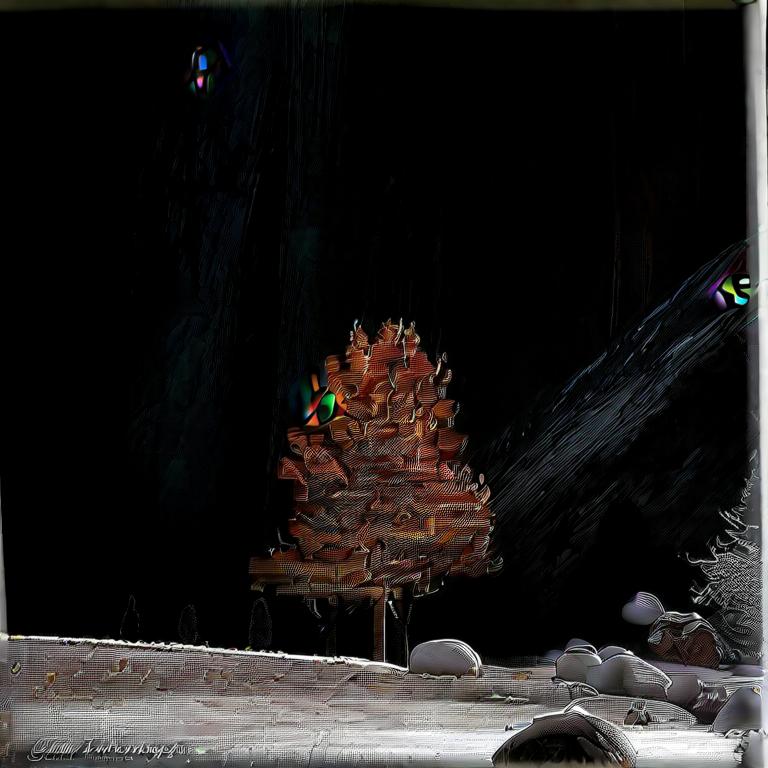} & 
\includegraphics[width=3cm]{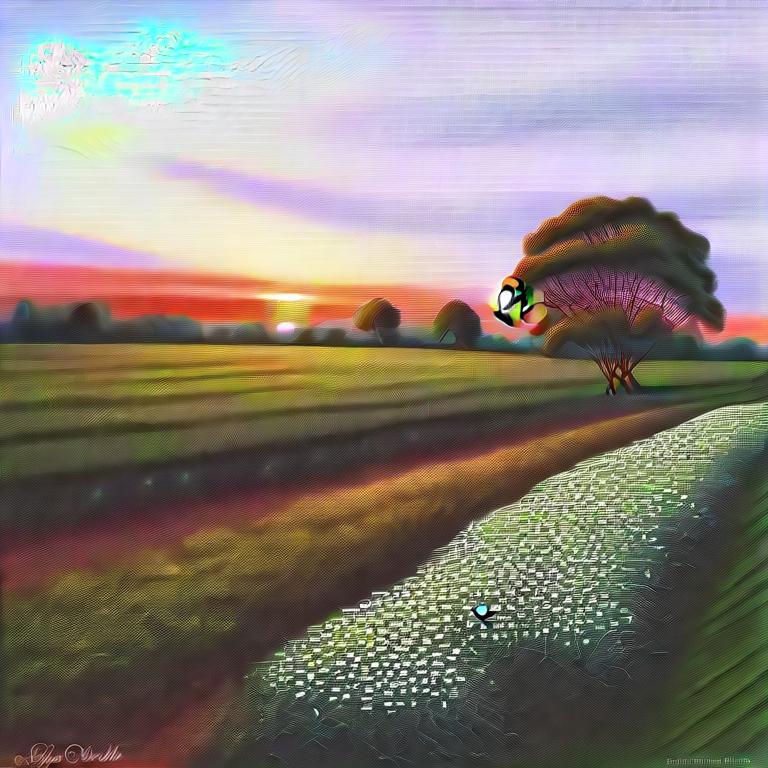} & 
\includegraphics[width=3cm]{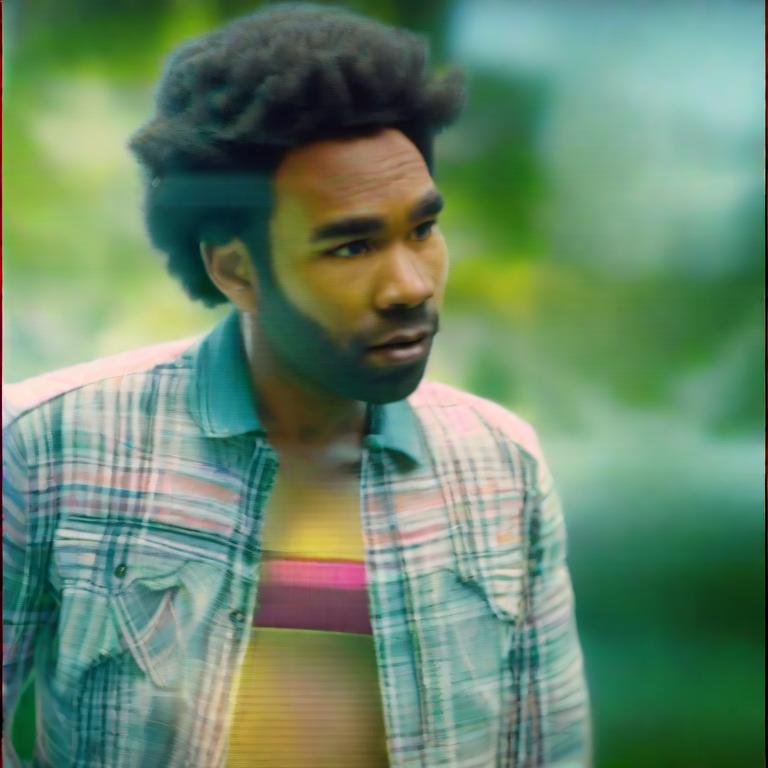} \\ %
\rotatebox[origin=c]{90}{\tiny{\textbf{\ours}}} & 
\includegraphics[width=3cm]{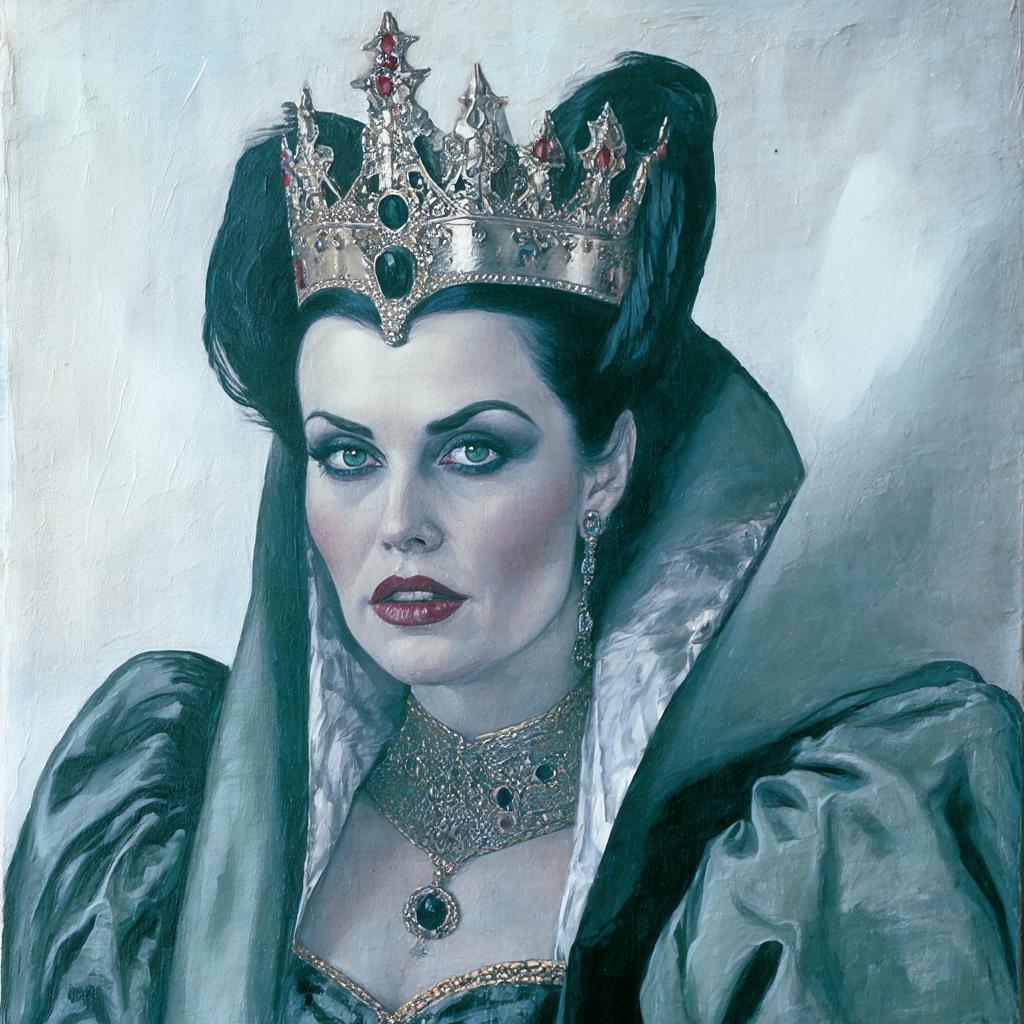} & 
\includegraphics[width=3cm]{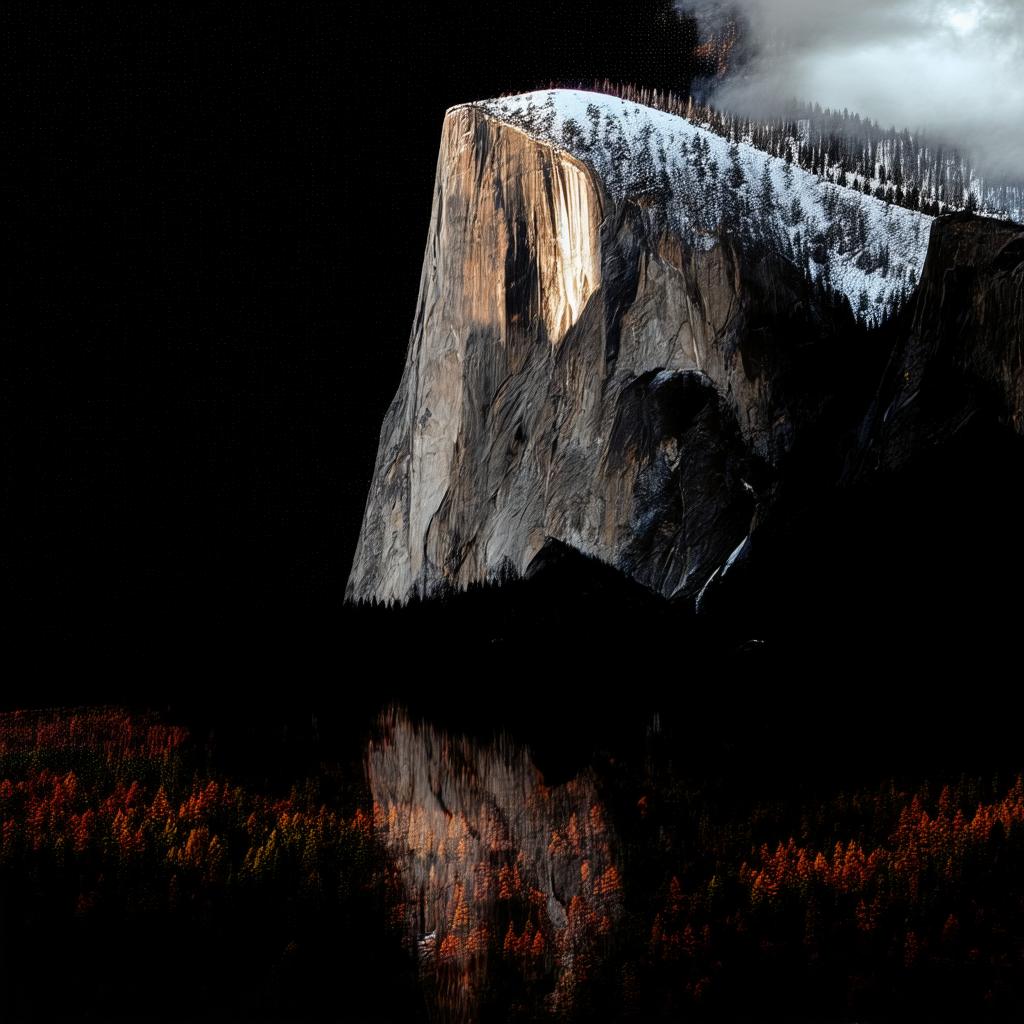} & 
\includegraphics[width=3cm]{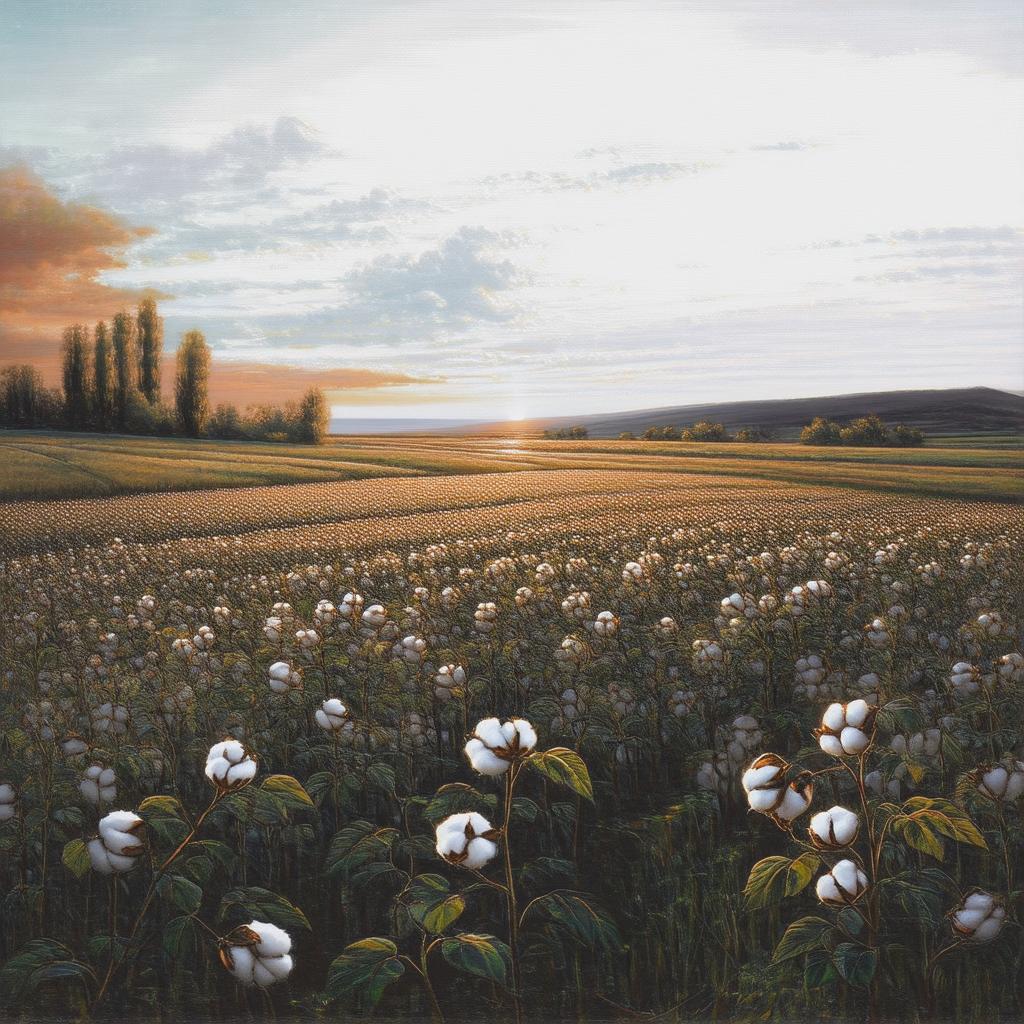} & 
\includegraphics[width=3cm]{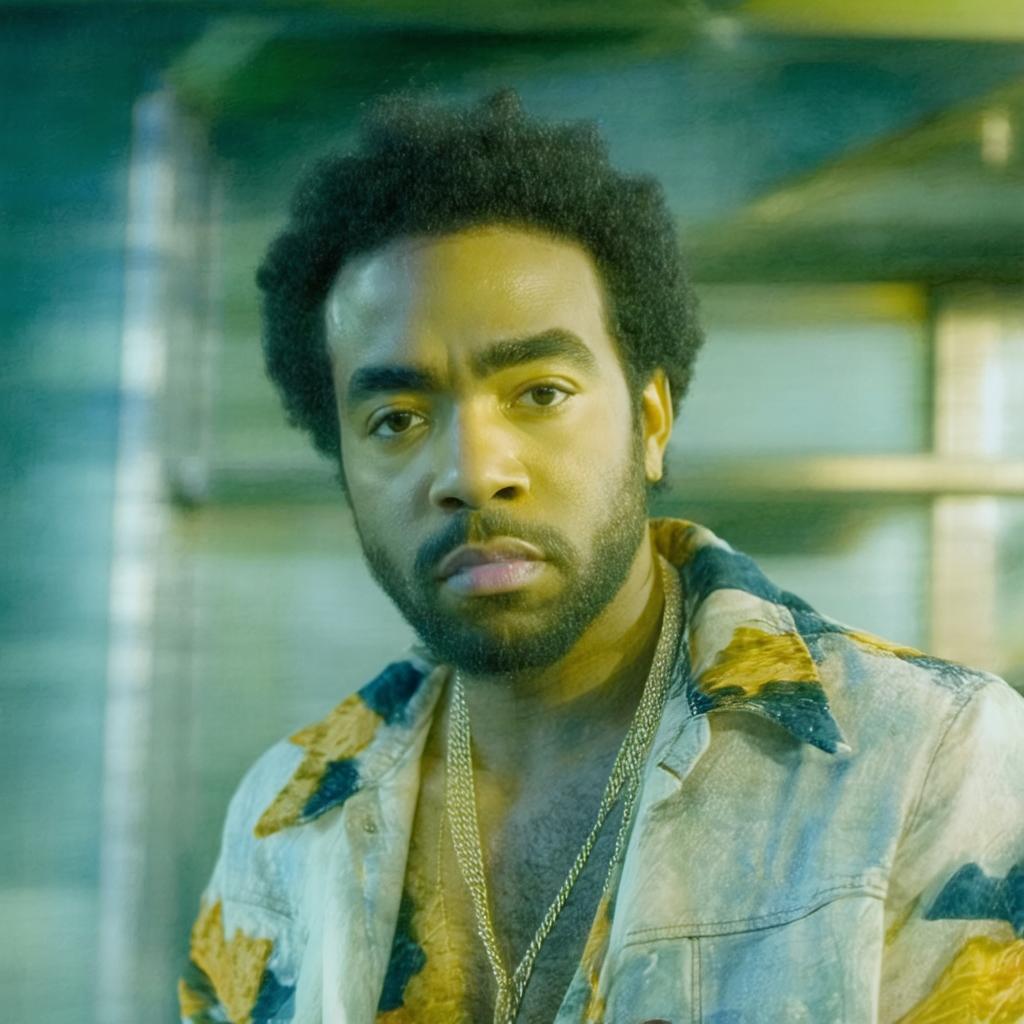} \\ % 
\end{tabular}

\titlecaption{Diffusion Guidance}{Further results on diffusion guidance with \citet{lobashev2025color}. Notice their approach often becomes blurry, collapses or contains artifacts.} 
\label{fig:guidance_suppl} 
\end{figure*}

In \fig{guidance_suppl} we present further results on diffusion guidance. \citet{lobashev2025color} proposed method and implementation often collapse, blur, or contain artifacts when the SWD guide enforces unlikely colors given the prompt. Our updated pipeline does not produce these issues.

\section{Color Matching}
\begin{figure*}[htb] 
\centering
\setlength{\tabcolsep}{3pt}
\renewcommand{\arraystretch}{0.8} 
\begin{tabular}{@{}p{0.05cm}*{3}{m{4.32cm}}@{}}
\rotatebox[origin=c]{90}{\fontsize{4pt}{4.2pt}\selectfont Source} & 
\includegraphics[width=4.3cm]{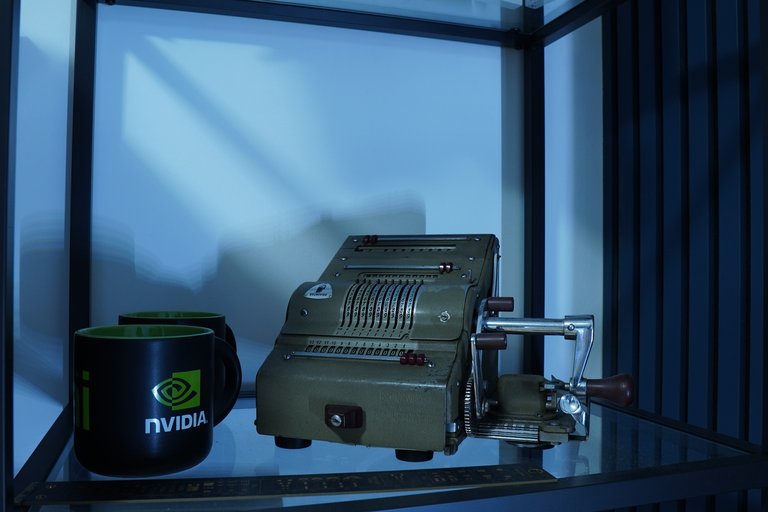} &
\includegraphics[width=4.3cm]{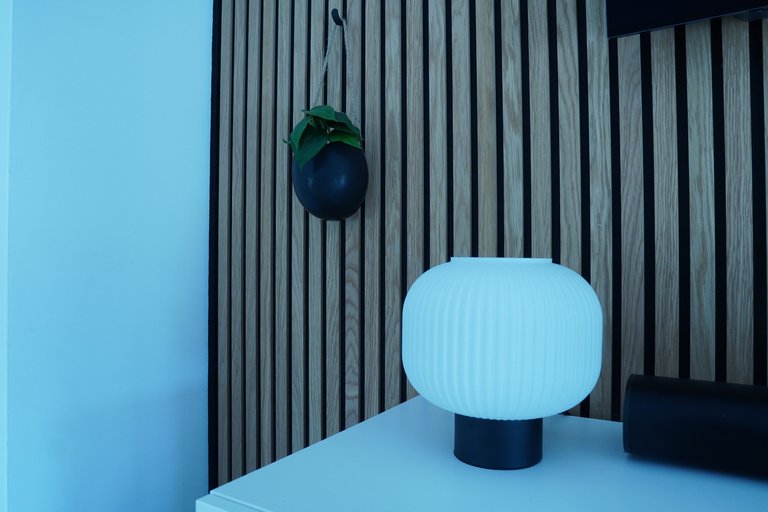} &
\includegraphics[width=4.3cm]{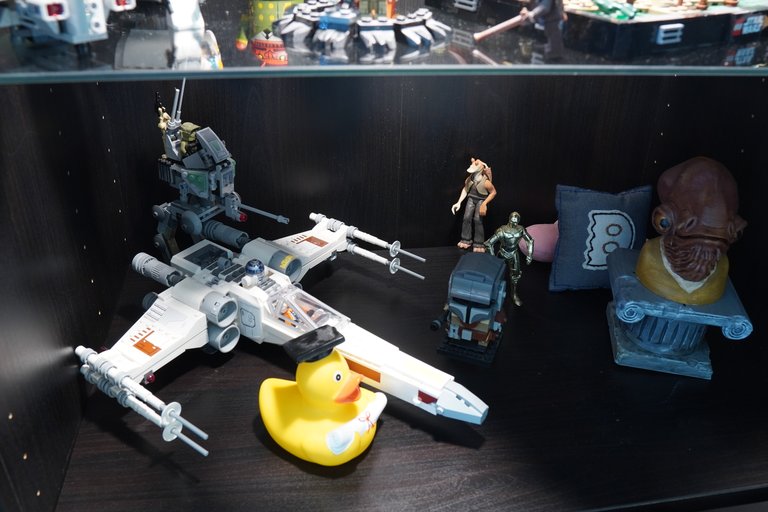} 
\\

\rotatebox[origin=c]{90}{\fontsize{4pt}{4.2pt}\selectfont Target} & %
\begin{tikzpicture}[spy using outlines={circle, magnification=2, connect spies}, inner sep=0.0cm]
    \node {\includegraphics[width=4.3cm]{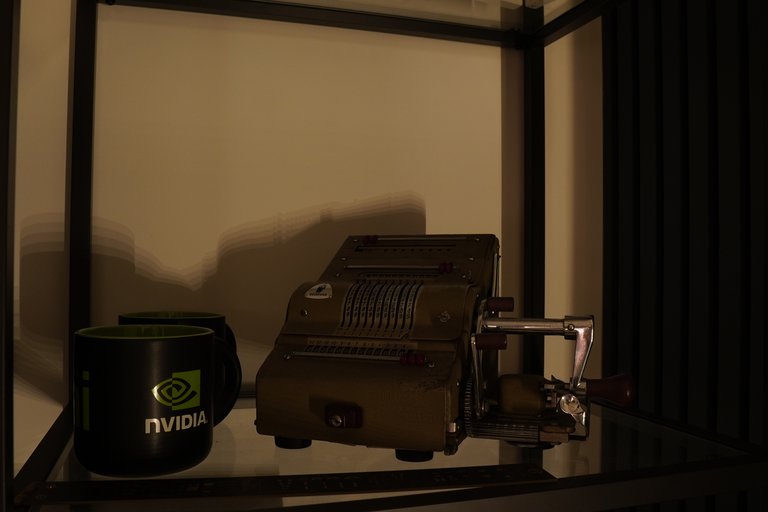}};
    \spy[size=1.0cm] on (-1.1, -0.65) in node at (-1.6, 0.8);
    \spy[size=1.0cm] on (-0.5, -0.45) in node at (1.6, 0.8);
\end{tikzpicture} &
\begin{tikzpicture}[spy using outlines={circle, magnification=2, connect spies}, inner sep=0.0cm]
    \node {\includegraphics[width=4.3cm]{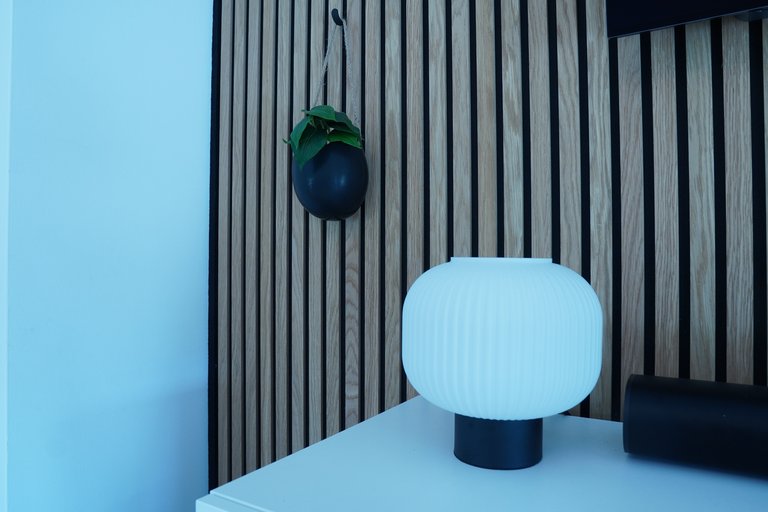}};
    \spy[size=1.0cm] on (-0.55, 0.5) in node at (-1.6, 0.8);
    \spy[size=1.0cm] on (0.5, -0.8) in node at (1.6, 0.8);
\end{tikzpicture} &
\begin{tikzpicture}[spy using outlines={circle, magnification=2, connect spies}, inner sep=0.0cm]
    \node {\includegraphics[width=4.3cm]{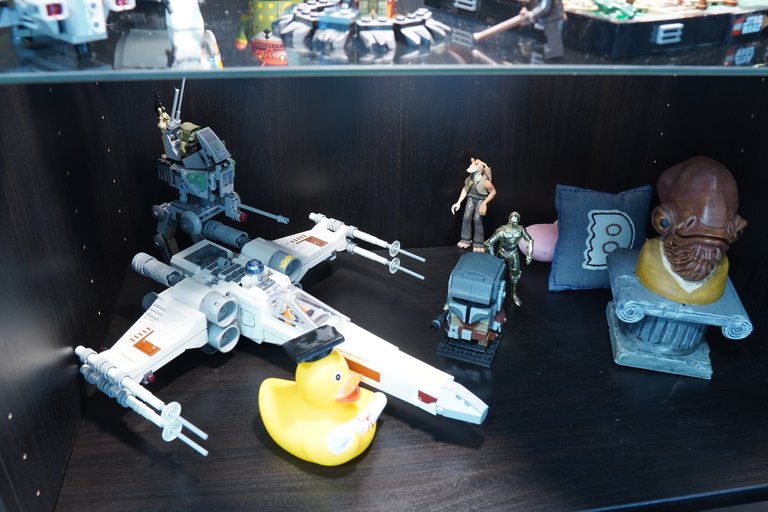}};
    \spy[size=1.0cm] on (-0.25, -0.9) in node at (-1.6, 0.8);
    \spy[size=1.0cm] on (0.6, -0.25) in node at (1.6, 0.8);
\end{tikzpicture} %
\\
\midrule

%%%
%Reinhard
%%%

\rotatebox[origin=c]{90}{\fontsize{4pt}{4.2pt}\selectfont\citet{reinhard2001color}} & 
\begin{tikzpicture}[spy using outlines={circle, magnification=2, connect spies}, inner sep=0.0cm]
    \node {\includegraphics[width=4.3cm]{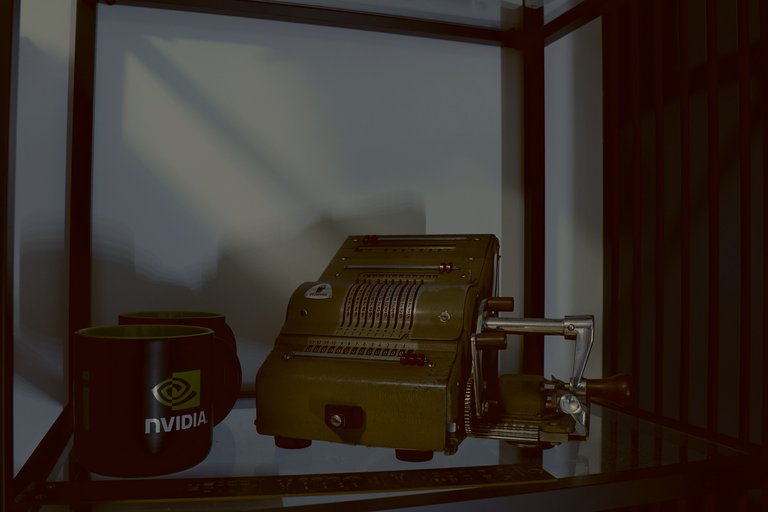}};
    \spy[size=1.0cm] on (-1.1, -0.65) in node at (-1.6, 0.8);
    \spy[size=1.0cm] on (-0.5, -0.45) in node at (1.6, 0.8);
\end{tikzpicture} &
\begin{tikzpicture}[spy using outlines={circle, magnification=2, connect spies}, inner sep=0.0cm]
    \node {\includegraphics[width=4.3cm]{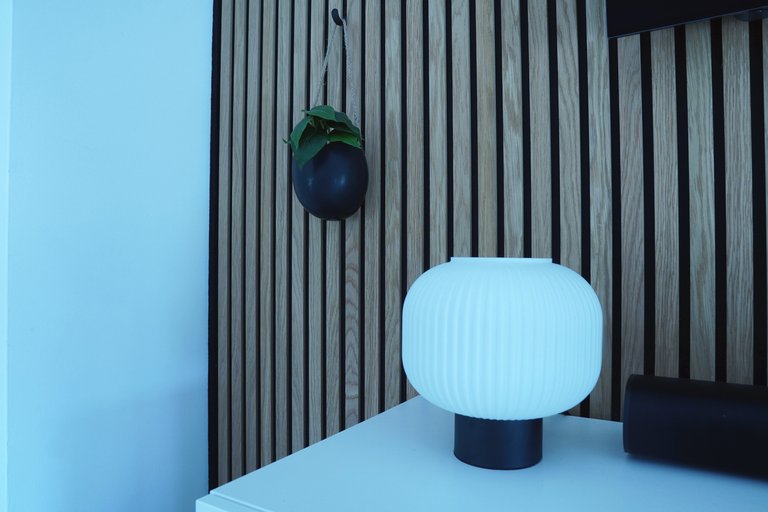}};
    \spy[size=1.0cm] on (-0.55, 0.5) in node at (-1.6, 0.8);
    \spy[size=1.0cm] on (0.5, -0.8) in node at (1.6, 0.8);
\end{tikzpicture} &
\begin{tikzpicture}[spy using outlines={circle, magnification=2, connect spies}, inner sep=0.0cm]
    \node {\includegraphics[width=4.3cm]{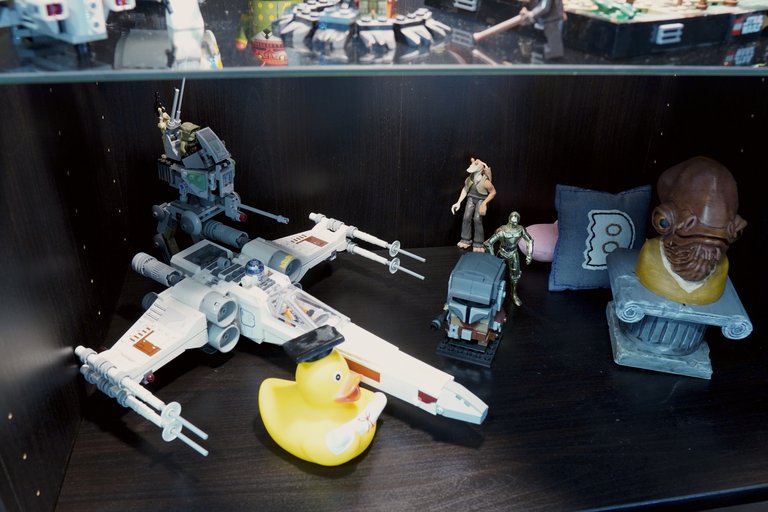}};
    \spy[size=1.0cm] on (-0.25, -0.9) in node at (-1.6, 0.8);
    \spy[size=1.0cm] on (0.6, -0.25) in node at (1.6, 0.8);
\end{tikzpicture} \\

%%%%%
% Nguyen
%%%%%
\rotatebox[origin=c]{90}{\fontsize{4pt}{4.2pt}\selectfont\citet{nguyenIlluminantAwareGamutBased2014}} & 
\begin{tikzpicture}[spy using outlines={circle, magnification=2, connect spies}, inner sep=0.0cm]
    \node {\includegraphics[width=4.3cm]{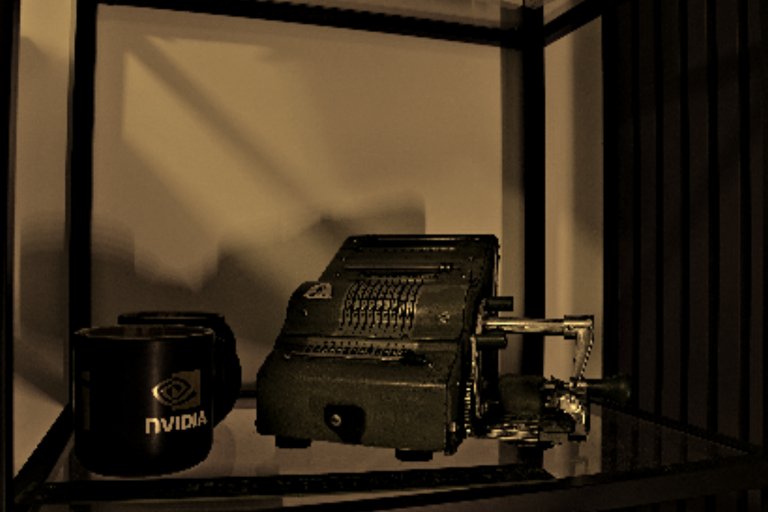}};
    \spy[size=1.0cm] on (-1.1, -0.65) in node at (-1.6, 0.8);
    \spy[size=1.0cm] on (-0.5, -0.45) in node at (1.6, 0.8);
\end{tikzpicture} &
\begin{tikzpicture}[spy using outlines={circle, magnification=2, connect spies}, inner sep=0.0cm]
    \node {\includegraphics[width=4.3cm]{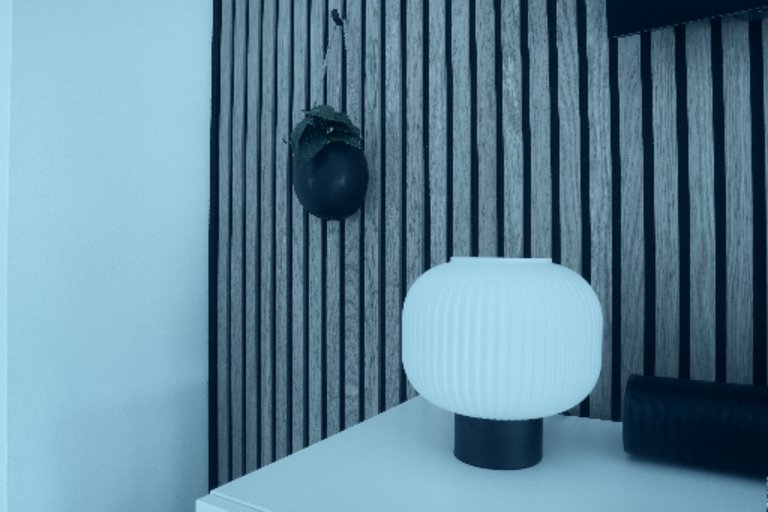}};
    \spy[size=1.0cm] on (-0.55, 0.5) in node at (-1.6, 0.8);
    \spy[size=1.0cm] on (0.5, -0.8) in node at (1.6, 0.8);
\end{tikzpicture} &
\begin{tikzpicture}[spy using outlines={circle, magnification=2, connect spies}, inner sep=0.0cm]
    \node {\includegraphics[width=4.3cm]{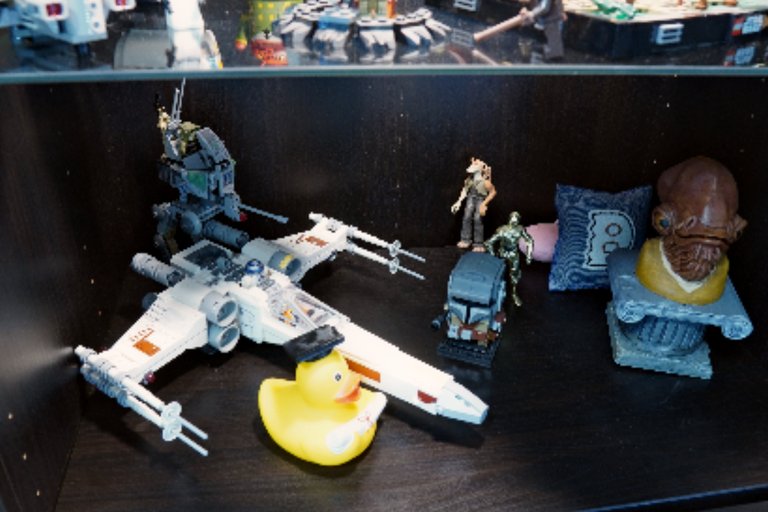}};
    \spy[size=1.0cm] on (-0.25, -0.9) in node at (-1.6, 0.8);
    \spy[size=1.0cm] on (0.6, -0.25) in node at (1.6, 0.8);
\end{tikzpicture} \\

%%%%%%
% Larchenko
%%%%%%

\rotatebox[origin=c]{90}{\fontsize{4pt}{4.2pt}\selectfont\citet{larchenko_color_2025}} & 
\begin{tikzpicture}[spy using outlines={circle, magnification=2, connect spies}, inner sep=0.0cm]
    \node {\includegraphics[width=4.3cm]{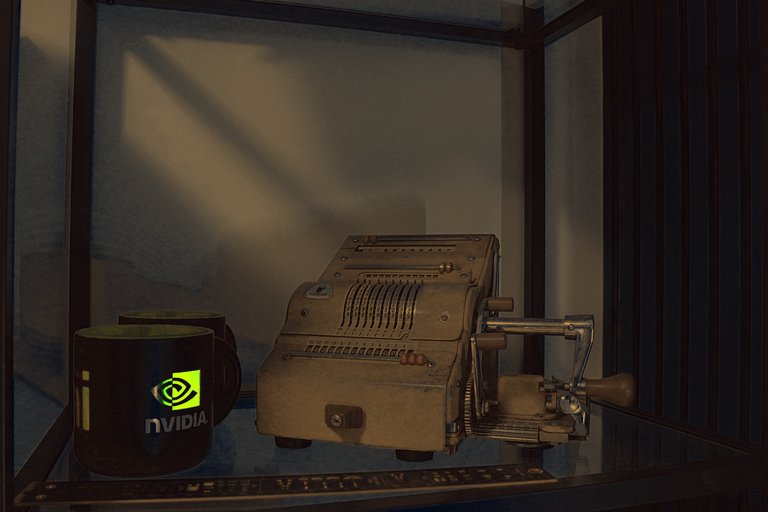}};
    \spy[size=1.0cm] on (-1.1, -0.65) in node at (-1.6, 0.8);
    \spy[size=1.0cm] on (-0.5, -0.45) in node at (1.6, 0.8);
\end{tikzpicture} &
\begin{tikzpicture}[spy using outlines={circle, magnification=2, connect spies}, inner sep=0.0cm]
    \node {\includegraphics[width=4.3cm]{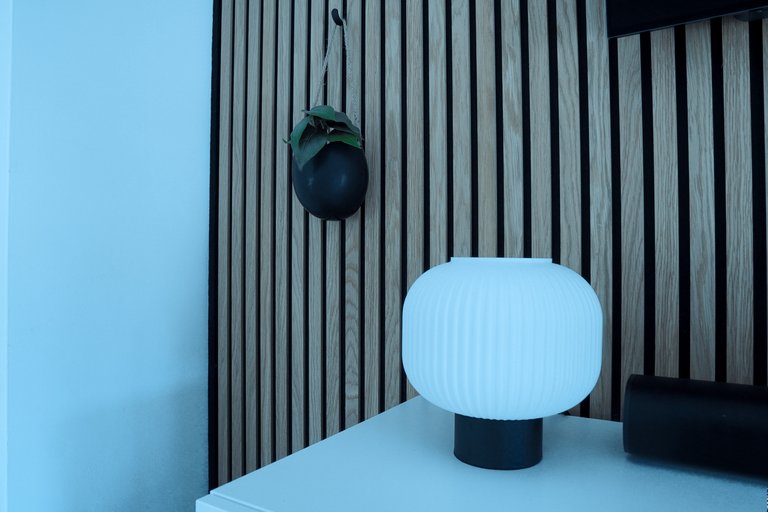}};
    \spy[size=1.0cm] on (-0.55, 0.5) in node at (-1.6, 0.8);
    \spy[size=1.0cm] on (0.5, -0.8) in node at (1.6, 0.8);
\end{tikzpicture} &
\begin{tikzpicture}[spy using outlines={circle, magnification=2, connect spies}, inner sep=0.0cm]
    \node {\includegraphics[width=4.3cm]{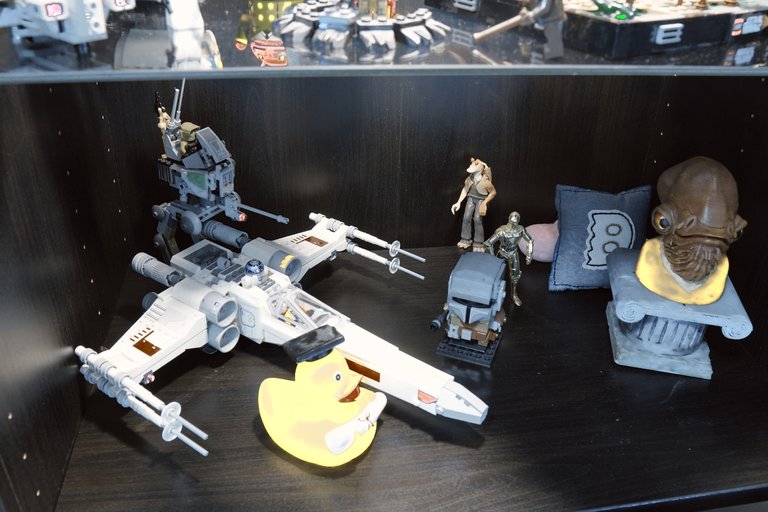}};
    \spy[size=1.0cm] on (-0.25, -0.9) in node at (-1.6, 0.8);
    \spy[size=1.0cm] on (0.6, -0.25) in node at (1.6, 0.8);
\end{tikzpicture} \\

%%%%%
% Ours
%%%%%

\rotatebox[origin=c]{90}{\fontsize{4pt}{4.2pt}\selectfont\textbf{\ours}} & 
\begin{tikzpicture}[spy using outlines={circle, magnification=2, connect spies}, inner sep=0.0cm]
    \node {\includegraphics[width=4.3cm]{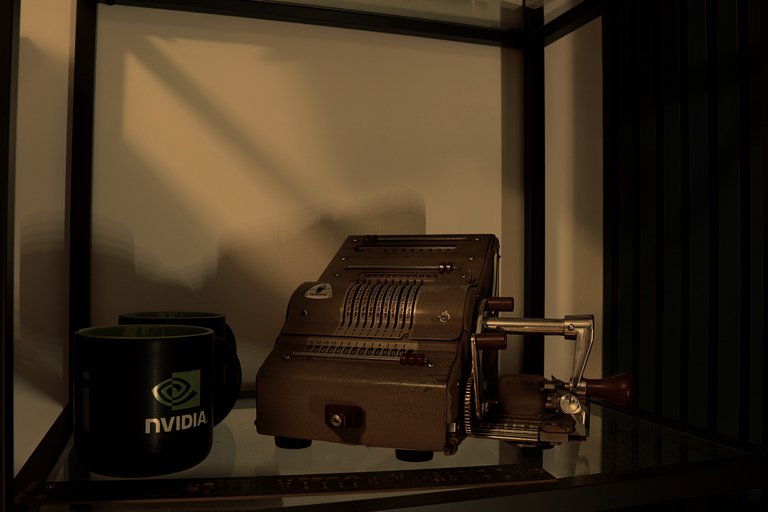}};
    \spy[size=1.0cm] on (-1.1, -0.65) in node at (-1.6, 0.8);
    \spy[size=1.0cm] on (-0.5, -0.45) in node at (1.6, 0.8);
\end{tikzpicture} &
\begin{tikzpicture}[spy using outlines={circle, magnification=2, connect spies}, inner sep=0.0cm]
    \node {\includegraphics[width=4.3cm]{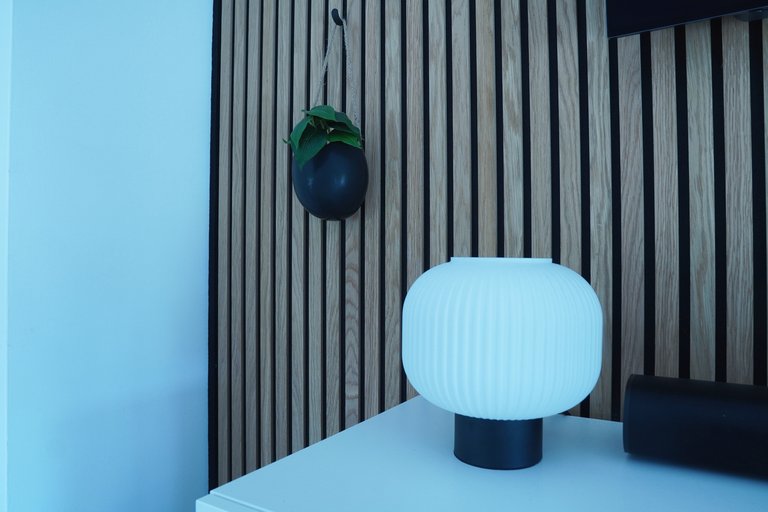}};
    \spy[size=1.0cm] on (-0.55, 0.5) in node at (-1.6, 0.8);
    \spy[size=1.0cm] on (0.5, -0.8) in node at (1.6, 0.8);
\end{tikzpicture} &
\begin{tikzpicture}[spy using outlines={circle, magnification=2, connect spies}, inner sep=0.0cm]
    \node {\includegraphics[width=4.3cm]{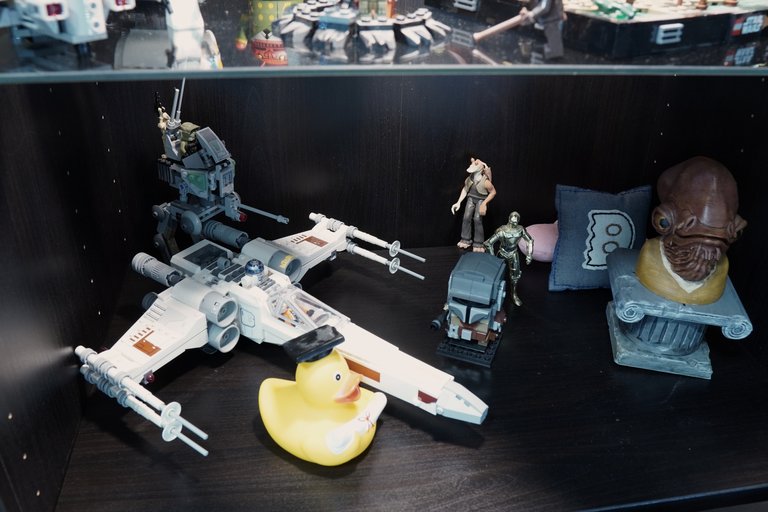}};
    \spy[size=1.0cm] on (-0.25, -0.9) in node at (-1.6, 0.8);
    \spy[size=1.0cm] on (0.6, -0.25) in node at (1.6, 0.8);
\end{tikzpicture} \\

\end{tabular}
\titlecaption{Color Matching}{More results on the color matching task. Notice the subtle changes and artifact free, high resolution color matching our method enables.} 
\label{fig:color_match_suppl} 
\end{figure*}

We show additional results in \fig{color_match_suppl} with our baselines. Notice that our results achieve consistent color matching performance and remain artifact-free. Our method also supports running at an arbitrary resolution.

\end{document}